\documentclass[nonatbib]{elsarticle}
\journal{Blockchain: Research and Applications}

\usepackage[utf8]{inputenc}

\usepackage[hyphens]{url}
\usepackage[hidelinks]{hyperref}
\hypersetup{breaklinks=true}
\usepackage{graphicx}
\graphicspath{{figures/}}
\usepackage{pgfplots}

\usepackage{booktabs}
\usepackage{tabularx}

\usepackage{natbib}

\begin{document}

\begin{frontmatter}
\title{A Systematic Literature Review of the Tension between the GDPR and Public Blockchain Systems}

\author[inst1]{Rahime Belen-Saglam}
\ead{R.Belen-Saglam-724@kent.ac.uk}
\affiliation[inst1]{organization={Institute of Cyber Security for Society (ICSS) University of Kent},
            addressline={Keynes College}, 
            city={Canterbury},
            postcode={CT2 7NP}, 
            country={United Kingdom}}

\author[inst1]{Enes Altuncu}
\ead{ea483@kent.ac.uk}
\author[inst2]{Yang Lu}
\ead{y.lu@yorksj.ac.uk}
\affiliation[inst2]{organization={School of Science, Technology and Health York, St John University},
            addressline={Lord Mayor's Walk}, 
            city={York},
            postcode={YO31 7EX}, 
            country={United Kingdom}}

\author[inst1]{Shujun Li\corref{cor1}}
\ead{S.J.Li@kent.ac.uk}
\cortext[cor1]{Corresponding author}

\begin{abstract}
The blockchain technology has been rapidly growing since Bitcoin was invented in 2008. The most common type of blockchain systems, public (permisionless) blockchain systems have some unique features that lead to a tension with European Union's General Data Protection Regulation (GDPR) and other similar data protection laws. In this paper, we report the results of a systematic literature review (SLR) on 114 research papers discussing and/or addressing such a tension. To be the best of our know, our SLR is the most comprehensive review of this topic, leading a more in-depth and broader analysis of related research work on this important topic. Our results revealed that three main types of issues: (i) difficulties in exercising data subjects' rights such as the `right to be forgotten' (RTBF) due to the immutable nature of public blockchains; (ii) difficulties in identifying roles and responsibilities in the public blockchain data processing ecosystem (particularly on the identification of data controllers and data processors); (iii) ambiguities regarding the application of the relevant law(s) due to the distributed nature of blockchains. Our work also led to a better understanding of solutions for improving the GDPR compliance of public blockchain systems. Our work can help inform not only blockchain researchers and developers, but also policy makers and law markers to consider how to reconcile the tension between public blockchain systems and data protection laws (the GDPR and beyond).
\end{abstract}

\end{frontmatter}

\section{Introduction}

Since Bitcoin was conceptualised in 2008, its underlying technology about blockchains (also known as distributed ledgers) has been considered as a breakthrough of secure computing without a centralised authority in an open environment. Its potential capabilities led many researchers and practitioners to consider that it is the next big revolutionising technology after the Internet~\citep{puthal2018blockchain}. Its applications have boomed in many sectors for various purposes and many researchers also started conducting research on this emerging technology. Although the blockchain technology has some built-in security and privacy mechanism by design, it has also introduced new security and privacy concerns, one of which is the conflict between the immutable nature of data on blockchain and the ``right to be forgotten'' (RTBF) of data subjects introduced in new data protection laws such as the European Union (EU) General Data Protection Regulation (GDPR) introduced in 2016~\citep{GDPR}. Such new concerns let many researchers, practitioners, policy makers and blockchain users to debate about legal compliance of blockchain systems and to explore ways to make blockchain systems more legally compliant with such new data protection laws and regulations. This paper aims at providing a comprehensive review of such efforts in the research literature.

After being passed by the European Parliament in 2016, the GDPR entered into force on 25 May 2018 in all EU member states. In addition, law markers in three non-EU member states of the European Economic Area (EAA), Iceland, Liechtenstein and Norway, also decided to adopt the GDPR. For the UK, after it left the EU, its law markers decided to keep the GDPR in its national law, but made some necessary changes to reflect the new status of the UK as a non-EU/EEA country, which led to the so-called UK GDPR~\citep{UK_GDPR}, a UK-specific version of the EU GDPR. In the rest of the paper, we will use the term GDPR in a broad sense to refer to the two different versions of the GDPR, since the differences are not essential for our discussions on the relationships between the blockchain technology and the relevant content defined in the GDPR.

In the context of the GDPR, legal compliance issues have been raised for a range of emerging technologies including IoT (Internet of Things), AI (artificial intelligence) and big data analytics, and also blockchains. One mostly discussed aspect of the tension between blockchains (especially public blockchain systems) and the GDPR is the following: the immutable nature of blockchains makes it impossible to delete personal information, therefore, it is not possible to exercise the RTBF (more formally known as the right to erasure) of data subjects as defined in the GDPR. Another aspect is about data sharing outside of the EU/EEA/UK: for a public blockchain system, it is normally the case that every node holds a full copy of all data, no matter where the node is physically located or even unknown.

Due to those GDPR-compliance challenges, many researchers looked at the tension between the GDPR and blockchains in recent years and some also attempted to propose solutions to address some of the challenges. In a 2018 report~\citep{EUBlockchainObservatoryForum2018report}, the EU Blockchain Observatory \& Forum stated that ``Public, permissionless blockchains represent the greatest challenges in terms of GDPR compliance''. Despite the active research on this very important topic, to date we have noticed only two systematic literature reviews (SLRs) covering related research progress, both published in 2021. In one SLR, \citet{haque2021gdpr} identified 39 papers covering this topic by searching into two databases (IEEE and Scopus), and in the other SLR, \citet{suripeddi2021blockchain} identified 41 papers for their review by searching into three databases (Science Direct, ACM and IEEE). Both SLRs are not sufficiently comprehensive due to the limited databases and keywords they used and the over-strict inclusion criteria. We also noticed another literature review paper following a different review technique (Levy and Ellis' narrative review of literature methodology), which used a forward and backward search technique to posit a framework for adopting a blockchain that follows the GDPR~\citep{al2020designing}. This non-systematic literature review also suffers from having a very limited number of papers covered -- just 39.

For our SLR, we expanded the databases searched to Scopus, WoS (Web of Science) and Google Scholar, which allowed us to access gray literature as well. Our SLR therefore led to a much more comprehensive coverage with 114 research articles, making it possible to draw a much bigger picture of relevant research work. We also decided to limit our scope to public blockchains only considering the statement in the EU Blockchain Observatory \& Forum's 2018 report~\citep{EUBlockchainObservatoryForum2018report}. This allowed us to focus on blockchain systems with more essential challenges in terms of the GDPR compliance.

Compared with past reviews on the same topic, our SLR makes a number of new contributions due to our larger coverage of related research papers and a more in-depth analysis of the included papers. First of all, we have considered different types of personal data that can be stored and processed on a blockchain and identified both challenges and proposed solutions for each data type. Our findings also cover limitations and consequences of proposed solutions as well as contradicting opinions that will allow our readers to get a better idea about the current state of the art. Second, we considered different roles and responsibilities in the blockchain data processing ecosystem, provided perspectives at the network and application levels, and categorized discussions in the research literature accordingly, all of which have been largely overlooked in other literature reviews. Finally, we reviewed the covered research papers by considering a broader scope of GDPR-related elements, which allowed a much more in-depth and precise representation of the literature.

For our SLR, we followed the PRISMA protocol widely used in many disciplines~\citep{liberati2009prisma}. Our results revealed that the tension between the GDPR and public blockchains has been studied around three main issues: (i) difficulties in exercising data subjects' rights such as the RTBF due to the immutable nature of public blockchains; (ii) difficulties in identifying roles and responsibilities in the public blockchain data processing ecosystem (particularly on the identification of data controllers and data processors); (iii) ambiguities regarding the application of the relevant law(s) due to the distributed nature of blockchains. Our work also led to a better understanding of GDPR-compliance related solutions proposed in the literature, e.g., those around assuring the RTBF using hashing, and the use of smart contracts to manage consent. The results of our SLR can help inform blockchain researchers and developers, policy makers and law markers to consider how to reconcile the tension between public blockchain systems and the GDPR. Note that our results are not limited to the GDPR since many other data protection laws and regulations share similar data protection principles with the GDPR.

The rest of the paper is organised as follows. In Section~\ref{sec:background}, important background information about the blockchain technology and the GDPR is given. Section~\ref{sec:researchmethodology} explains our research methodology. Our detailed analysis of the covered papers is given in Section~\ref{'sec:results_findings}. Then, we summarise the results into three main areas (GDPR-compliance, proposed solutions, roles and responsibilities) in Section~\ref{sec:discussion}. The final section concludes the paper.

\section{Background}
\label{sec:background}

\subsection{Blockchain Technology}
\label{sec:blockchaintechnology}

From a technical perspective, a blockchain is a distributed database that is formed as a chain of data blocks and offers a solution through decentralising storage and processing of data. Each block in a blockchain normally composes two parts: transactional data and metadata. The metadata typically contains, inter alia, a timestamp, hash value of the block, and a hash value of the previous block. All hash values are computed using a cryptographically hard one-way function. This allows blocks to be linked to each other to form a chronological database. This very nature of the blockchains results in any modification of data to be detected by other participants of the network as the hash of the next block would not correspond to the data on the modified one. This feature is called ``immutability'' and leads public blockchains to be regarded as tamper-proof.

As another important feature of public blockchains, a full copy of a distributed database is stored at each node that is part of the blockchain system. Since there is no central authorities, trust is achieved via the distributed storage (i.e., a distributed ledger) and a distributed consensus mechanism. The latter is needed to ensure different nodes will converge to the same distributed ledger, rather than all nodes produce different ledgers therefore leading to inconsistency in the system. The distributed consensus mechanism determines how new data blocks are added into a blockchain and how all nodes agree which branch to follow if there are multiple chain branches. There are several consensus algorithms used by different blockchain systems, and Proof of Work (PoW) used by Bitcoin is so far the most widely used one, in which new blocks are added to the chain by nodes who compete against each other by solving a mathematical puzzle (normally defined by a cryptographic hashing function). The node who first solves the puzzle creates a new block and a longer chain for others to follow. Such nodes are called miners. Miners need to spend a lot of computational power on solving mathematical puzzles and are incentivised by being awarded coins for being the first puzzle solver. Those algorithms are used to confirm consensus of the current state of the ledger and to ensure that all nodes have the same copy.

The blockchain technology also utilises asymmetric cryptosystems, mainly for verifying authenticity of a transaction and its sender and receiver. Each user in a blockchain network has their own private key and public key. The private key is used by a transaction sender to sign a transaction using a digital signature algorithm, which can then be verified by other users using the sender's public key.

Blockchain systems can be classified into three broad categories: public (permissionless) blockchains, consortium (permissioned) blockchains and private blockchains. Public blockchains are open to anyone and allow any participants to join the network and read, send, or receive data on the blockchain. In contrast, there are constraints on consortium blockchains and normally write permissions are granted to a pre-selected set of participants only. When only one participant has such a privilege, then we have a private blockchain system.

Finally, smart contracts are another associated technology based on the blockchain technology, which can fully automate self-enacting electronic contracts. They allow a distributed protocol (such as a set of business rules) to be executed and enforced automatically.

\subsection{The GDPR}
\label{sec:gdpr}

In order to pursue the objective of protection of fundamental rights and to protect personal data of individuals, the GDPR strengthens the protection of individuals' personal data primarily by defining principles and the lawful bases for processing their personal data, and also specifying rights for individuals.

In this section, we will give the definition of a relevant subset of these elements which are important to understand the GDPR compliance issues of public blockchain systems.

\subsubsection{Personal Data and Data Subjects}

Two core concepts, personal data and data subjects, are at the core of the GDPR. The GDPR defines ``personal data'' in Article 4 as:
\begin{quote}
\small
``\textit{Any information relating to an identified or identifiable natural person (`data subject')}''
\end{quote}
Here, the definition of ``identifiable natural person'' (i.e., ``data subject'') is given as:
\begin{quote}
\small
``\textit{One who can be identified, directly or indirectly, in particular by reference to an identifier such as a name, an identification number, location data, an online identifier or to one or more factors specific to the physical, physiological, genetic, mental, economic, cultural or social identity of that natural person.}''
\end{quote}

This definition is expanded under Article~4(1) and it is stated that it is possible to define information that in itself would not be considered personal data but, when combined with other information, it can be considered personal data. Pseudonymised data can be given as an example here. The GDPR defines pseudonymisation as:
\begin{quote}
\small
``\textit{the processing of personal data in such a manner that the personal data can no longer be attributed to a specific data subject without the use of additional information, provided that such additional information is kept separately and is subject to technical and organisational measures to ensure that the personal data are not attributed to an identified or identifiable natural person}''
\end{quote}

The ability to identify a person based on additional information in pseudonimisation technique leads pseudonymised data to be considered as personal data. This opinion is based on Recital~26, which states that
\begin{quote}
\small
``\textit{personal data which have undergone pseudonymisation, which could be attributed to a natural person by the use of additional information should be considered to be information on an identifiable natural person.}''
\end{quote}
In the research literature, there have been different opinions on whether pseudonymisation could render data anonymous or not.
One example of pseudonymous data is encrypted data~\citep{ICO_Encryption}, which is mentioned in Article~32 of the GDPR for ensuring the security of personal data. In its essence, encryption is a mathematical function which uses a secret value (the key) to encode data so that only users with access to that key can read the information. The holder of the key has the ability to re-identify individuals through decryption of that data. It is not denoted as a mandatory technique for the GDPR compliance but given as an essential data protection measure to mitigate the risk of data processing activities and a convenient way for data controllers to demonstrate compliance with the GDPR.

Unlike pseudonymised data, anonymised data is entirely excluded from the GDPR in its Recital~26. Regarding what constitutes anonymised data, Recital~26 defines anonymisation as follows,
\begin{quote}
``\textit{the principles of data protection should therefore not apply to anonymous information, namely information which does not relate to an identified or identifiable natural person or to personal data rendered anonymous in such a manner that the data subject is not or no longer identifiable}''.
\end{quote}
It is not an explicit definition, however, an opinion given by the Article~29 Data Protection Working Party~\citep{Article29WP2014OpinionAnonymisation} in relation to the EU DPD (Data Protection Directive) 1995~\citep{EU_DPD1995}, the predecessor of the GDPR, is still widely used as a general guidance. The opinion sets a very high standard and requires that the identification must be prevented irreversibly.


\subsubsection{Data Controllers and Data Processors}

In addition to personal data and data subjects, there are two other very important roles defined in the GDPR: data controllers and data processors. The definition of ``data controller'' is given in Article~4(7) as:
\begin{quote}
\small
``\textit{the natural or legal person, public authority, agency or other body which, alone or jointly with others, determines the purposes and means of the processing of personal data; where the purposes and means of such processing are determined by Union or Member State law, the controller or the specific criteria for its nomination may be provided for by Union or Member State law}''.
\end{quote}
The definition suggests that the controller is responsible for the processing of personal data, imposing several legal responsibilities for the controller. In Article~4(8), ``data processor'' is defined as follows:
\begin{quote}
\small
``\textit{a natural or legal person, public authority, agency or other body which processes personal data on behalf of the controller.}.''
\end{quote}
The personal data used in both definitions can be explained as any personal data related to a data subject.

The first part of the definition of the `data controller' implies that no one, not even a natural person, is excluded from responsibility when it comes to the processing of personal data. The second part given as ``that jointly or alone'' deepens the definition to include joint responsibility for the processing of personal data. Finally, the third part given as ``determines the purposes and means of the processing of personal data'' exemplifies that the one who has the decision-making power, not the factual power over the processing, is considered as the controller.

The definition of the ``data processor'' clarifies that a controller must exist for a processor to exist. Furthermore, it should be a separate legal entity with regard to the controller to be classified as a (non-controlling) processor. Controllers delegate the task to processors who process data as separate legal entities within the means and purposes of the controller's own agenda.

\subsubsection{Data Protection Principles}

The GDPR sets specific criteria for data controllers and processors to assure that personal data is processed in a fair and lawful way. For this goal, its Article~5 sets seven key data protection principles: 1) \emph{lawfulness, fairness and transparency}; 2) \emph{purpose limitation}; 3) \emph{data minimisation}; 4) \emph{accuracy}; 5) \emph{storage limitation}; 6) \emph{integrity and confidentiality (security)}; and 7)  \emph{accountability}. For instance, gaining the data subjects' consent is an example of \emph{lawful processing} since it is a valid ground under Article~6(1)(a) of the GDPR for collecting and processing personal data. Using personal data in a \emph{fair} way refers to not processing the data in a way that is unduly detrimental, unexpected or misleading to the data subject~\citep{ICO_Fairness}. As its name suggests, \emph{transparency} requires to be clear, open and honest to people from the beginning about how their personal data is being processed. The second principle, \emph{purpose limitation}, requires that personal data be ``collected for specified, explicit and legitimate purposes and not further processed''. The principle of \emph{data minimisation} is given in Point~(c) of Article~5 as
\begin{quote}
``\textit{personal data shall be adequate, relevant and limited to what is necessary in relation to the purposes for which they are processed}''.
\end{quote}
Another related principle, \emph{storage limitation} requires that the period for which the personal data is stored is limited to a very strict minimum. It should not be longer than it is necessary for the purposes for which the personal data are processed and data controllers must delete personal data when it is no longer needed. The principle of \emph{accuracy} dictated that data must be kept up to date and inaccurate data must be deleted. The principle of \emph{integrity and confidentiality} ensures that the personal data is processed and stored in a fashion that appropriate security measures are put in place to protect the personal data. Lastly, the GDPR requires a party to exist that is responsible under the principle of \emph{accountability}. This party is expected to take the responsibility for what is done with personal data and to have appropriate measures and records in place to be able to demonstrate the compliance.

\subsubsection{Lawful Basis for Processing}

According to the GDPR, it is required to have a valid lawful basis in order to process personal data. Obtaining explicit consent from the data subject for the processing of any personal data is one of the most commonly used bases for lawful processing. Explicit consent implies freely given, specific, informed and unambiguous indication of the data subject's preferences about the processing of their personal data. Article~7 of the GDPR provides three fundamental principles or rules for obtaining consent from the data subjects: controllers are responsible for demonstrating consent was given, a data subject has the right to withdraw consent at any time, and finally written requests for consent must be clear. Exercising the right to withdraw consent is expected to be as easy as giving consent. Article~22 also notes that the data subject has the right to not be subjected to automated decision making unless this kind of processing is based on the data subjects' explicit consent. The controller/processor has to stop all automated processing of the data if an explicit consent is not gathered. However, they can continue such processing if they are able to demonstrate another compelling legitimate ground. Article~6(1)(f) gives controllers and processors a lawful basis for processing where interest of processing outweighs the data subjects' rights and freedoms. It stated that, personal data can be processed without gathering explicit consent when:
\begin{quote}
\small
``\textit{processing is necessary for the purposes of the legitimate interests pursued by the controller or by a third party except where such interests are overridden by the interests or fundamental rights and freedoms of the data subject which require protection of personal data, in particular where the data subject is a child.}''
\end{quote}

\subsubsection{Data Protection by Design and Default}

The GDPR introduces explicit requirements for the protection of personal data concerning data protection by design and by default in Article~25. Data protection by default requires data controllers to implement technical and organisational measures that are designed to ensure that the processing of personal data meets the GDPR's requirements and otherwise to ensure protection of data subjects' rights. Data protection by design requires data protection requirements to be considered in all phases of system development and appropriate measures to be fulfilled to built-in data protection measures.




\subsubsection{Data Subject's Rights}

There are a number of rights granted to the data subjects in the GDPR. Both controllers and processors must fulfil certain requirements and duties towards such rights of data subjects to comply with the GDPR.

\paragraph{Right to erasure (Right to be forgotten/RTBF)}
The right to erasure (also known as the right to be forgotten/RTBF) mandates that controllers delete data in certain cases. According to Article~17, data subjects are granted with the right to request removing all related personal data.
According to Article~6(1), when data is no longer necessary for the purposes for which it was collected, it must be erased. If the processing is based on consent and the data subject withdraws it, actions must be taken to erase the data as long as there is no other ground for processing (Article~7). It is also possible for a data subject to object to processing and if there is no overriding reason to continue storing, it must be deleted (Article~21). Otherwise, as long as a lawful means for processing exists, the data can continue to be stored.

\paragraph{Right to rectification}
Under Article~16 of the GDPR, data subjects have rights to make a request to have their inaccurate personal data rectified, or completed if it is incomplete. Here, rectification means that data is updated to be accurate. Thus, this right has close links to the accuracy principle of the GDPR explained before.

\paragraph{Right to be informed and right to access}
Right to be informed requires data controllers to provide information to the data subjects regarding the processing and storage of their personal data. This right is expanded by the right of access, through which individuals can make access request to their personal data and gain in-depth information regarding the lawfulness of processing and how their personal data is handled.

\paragraph{Right to object and automated decision making}
The right to object enables data subjects to object to the processing of their personal data in certain circumstances. The controller can continue such processing if and only if they are able to demonstrate a compelling legitimate ground and that their interests of processing outweighs the data subjects' rights and freedoms. Article~22 of the GDPR sets additional rules to protect individuals against automated decision-making that has legal or similarly significant effects on them. Automated decision-making means making a decision solely by automated means without human involvement. Under Article~22, data subjects have the right not to be subject to a decision solely based on automated processing.

\paragraph{Right to data portability}
The right to data portability allows data subjects to access and move, copy or transfer their personal data easily from one electronic processing system to another in a safe and secure way, without affecting its usability. Under this right, data subjects have the right to request their personal data in a common and easy-to-read computer format or to request that a controller transmits this data directly to another controller.

\paragraph{Right to restrict processing}
Data subjects have the right to request the restriction or suppression of their personal data in certain situations: if the data subjects contests the accuracy of their personal data; if the processing is unlawful; if the data subject needs them to establish, exercise or defend a legal claim; and finally if data subjects have objected to processing their data.

\section{Research Methodology}
\label{sec:researchmethodology}

The overall aim of our research is to understand how researchers have studied the tension between public blockchain systems and the GDPR. To achieve this aim, we formulated the following research questions (RQs):
\begin{itemize}
\item \textbf{RQ1)} What issues public blockchain systems can lead to in relation to data subjects' rights and data protection principles provided by the GDPR?

\item \textbf{RQ2)} What solutions have been proposed in the research literature to address the tension between public blockchain systems and the GDPR?

\item \textbf{RQ3)} How researchers have considered legal roles and responsibilities of different stakeholders of public blockchain systems, e.g., who should be considered as data controllers and processors in public blockchain systems?
\end{itemize}

\subsection{Identifying Data Items}

\begin{figure}[!htb] 
\includegraphics[width=0.8\linewidth]{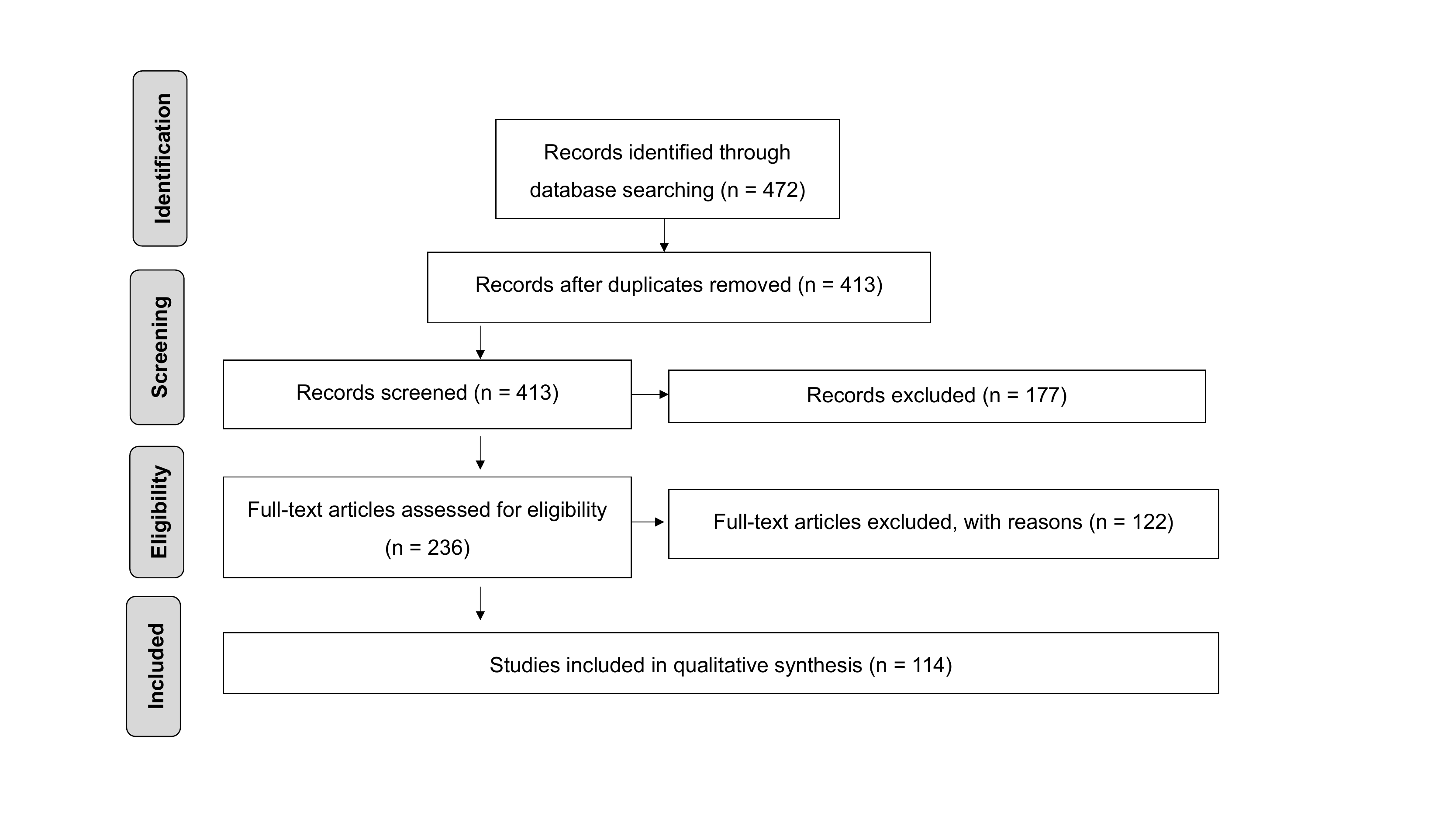} 
\caption{The diagram of the SLR procedure we used following the PRISMA protocol and the results of different steps of data item identification.}
\label{fig:Prisma}
\end{figure}

To conduct the SLR, we needed to first identify relevant data items -- research papers for our study. To this end, we utilised the PRISMA (Preferred Reporting Items for Systematic reviews and Meta-Analyses) protocol widely used by researchers for SLRs in multiple disciplines~\citep{liberati2009prisma}. The protocol involved a number of steps as shown in Figure~\ref{fig:Prisma}.

The first step is to select databases for searching for relevant research papers. We decided to use three scientific databases: Elsevier's Scopus\footnote{\url{https://www.scopus.com/}}, Clarivate's Web of Science (WoS)\footnote{\url{https://www.webofknowledge.com/}}, and Google Scholar\footnote{\url{https://scholar.google.com/}}. These databases were the most widely used databases with a very comprehensive coverage of research papers collectively. We did not use specific publishers' own databases because they are largely covered by the above three general databases. For all three databases, we used the same search query (note that all searches are case insensitive):
\begin{quote}\small
\texttt{( ( blockchain* OR Bitcoin OR cryptocurrenc* OR "distributed ledger*")\\
AND ( GDPR OR "General Data Protection Regulation" ) )}
\end{quote}
For Scopus and WoS, we searched into the metadata, i.e., titles, abstracts and keywords. For Google Scholar, there were only two options for the searches: title and fulltext. When we attempted searching into fulltext, Google Scholar returned too many candidate data items, so we decided to search into titles. Because Google Scholar had a relatively simplistic search syntax, we split the above search query into four sub-queries and then merged the results. All searches into the three databases were completed on 21th December, 2021.

The initial set of data items returned from multiple searches were merged, which gave us 472 papers. Then, the results were de-duplicated, leading to 413 papers. After that, we followed the following exclusion and inclusion criteria to screen all candidate papers.

\textbf{Exclusion Criteria}
\begin{itemize}
\item Books, theses, book chapters and other data items that are not research papers were excluded because such items either have not been properly peer reviewed or their fulltexts are hard to obtain, and many parts of their content are often published as separate research papers.

\item Non-English articles were excluded.

\item The article that were not peer reviewed were excluded.

\item The articles covering private or consortium blockchains only were excluded.

\item The articles in which the GDPR compatibility of public blockchains are covered only in the literature review sections with no further original discussions or research contributions relevant for our RQs were excluded.

\item The articles that discuss blockchain or the GDPR in general and lack discussions regarding the GDPR compliance issues were excluded.
\end{itemize}

\textbf{Inclusion Criteria}
\begin{itemize}
\item The articles that include some discussions on different GDPR-compliance issues of public blockchain systems were included.

\item The articles that propose one or more methods to help manage the GDPR compliance of public blockchain systems were included.
\end{itemize}

The screening process was conducted by the first author, and it involved reading titles and abstracts to exclude papers (leading to 236 papers) and then reading fulltext to make the final selection (leading to 114 papers selected).

\subsection{Encoding Data Items}

After obtaining the relevant papers, the first author followed a thematic approach to qualitatively analyse all papers to develop an encoding theme. The encoding process was done using NVivo\footnote{\url{https://www.qsrinternational.com/nvivo-qualitative-data-analysis-software/}}, one of the most widely used software tools for qualitative analysis. During the qualitative analysis of all papers, the first author identified discussions related to one or more research questions identified for the SLR, and incrementally defined codes to capture such discussions. Generated codes have been reviewed regularly and adjusted where necessary. The encoding scheme was reviewed and validated by the second and third authors, each of whom reviewed 25 randomly selected papers and checked the encoding results. Their feedback was considered by the first author to finalise the encoding scheme and make necessary changes to the encoding results. The last author participated in the general discussion on the encoding scheme and reviewed the final version to approve it.

\section{Results and Findings}
\label{'sec:results_findings}

This section describes the results from the SLR.

\subsection{General Statistics}

\begin{figure}[!htb]
\begin{tikzpicture}
\begin{axis}[
    every tick label/.append style={font=\small},
    ybar stacked,
	bar width=15pt,
	width=0.7\linewidth,
	height=7cm,
	nodes near coords,
    enlargelimits=0.13,
    legend style={at={(0.5,1.2)},
      anchor=north,legend columns=-1},
    ylabel={Number of papers},
    y label style={font=\small},
    symbolic x coords={2016, 2017,2018, 2019, 2020, 2021, 2022},
    xtick=data,
    ]

\addplot coordinates {
    (2016,1) 
    (2017,4) 
    (2018,25) 
    (2019,32)
    (2020,32)
    (2021,19)
    (2022,1)
};
\end{axis}
\end{tikzpicture}
\caption{Distribution of papers per year (2016-22)}
\label{fig:papersAmongYears}
\end{figure}
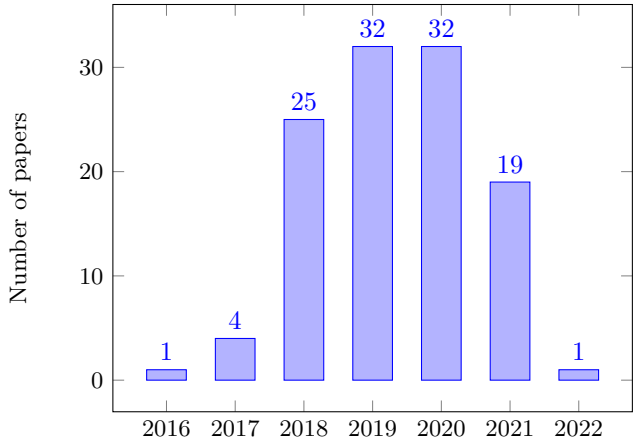

The distribution of the articles among years can be seen in Figure~\ref{fig:papersAmongYears}. As displayed in the figure, the interest into the GDPR compliance issues of public blockchain systems gained pace in 2018 and received the most attention from researchers in 2019 and 2020. As the saturation point has been reached in 2020, a decline in the number of papers has been observed in 2021. There is also one paper published in 2022 because that paper became searchable in December 2021 but was included in a 2022 issue. A majority of the studies have been conducted by researchers in computer science and related disciplines, but some were conducted by law researchers (which is not surprising given the half of the topic is about data protection law).

\subsection{Encoding Scheme}

Due to the broad scope of discussions in the literature, we ended up with several categories of codes as seen in the encoding scheme given in Table~\ref{tab:encodingScheme}. The codes that received the most attention (the ones that were used in more than 10 articles) can be seen in Figure~\ref{fig:popularCodes}. The codes are sufficiently self-explanatory so we do not include lengthy explanations to them in Figure~\ref{fig:popularCodes}. More detailed discussions on all themes and codes are summarised in the following subsections: Section~\ref{subsec:PersonalDataOnBlockchain} covers personal data on blockchains, Section~\ref{subsec:RolesInBlockchainDataProcessing} covers GDPR-related roles and responsibilities in blockchains, Section~\ref{subsec:TechnicalSolutionsforDataProtection} covers research papers proposing solutions to address the GDPR compliance issues of public blockchain systems, Sections~\ref{subsec:RTBF}--\ref{subsec:RightToRestrictProcessing} cover different data subject rights defined in the GDPR in the context of public blockchain systems, Section~\ref{subsec:LawfulnessFairenessTransparency} covers the first data protection principle on lawfulness, fairness and transparency, Section~\ref{subsec:PrinciplesOthers} covers other data protection principles, and Section~\ref{subsec:Others} covers two other topics: data protection by design and by default, and the territorial scope (i.e., data sharing beyond the EU/EEA/UK).

\begin{table}[htb]
\centering
\caption{Encoding scheme}
\label{tab:encodingScheme}
\begin{tabularx}{\linewidth}{p{0.3\linewidth}X}
\toprule
\textbf{Category/Theme} & \textbf{Codes}\\
\midrule
Personal data in blockchains & PublicPrivateKeys, PersonalDataOfOthers\\
\midrule
Roles and responsibilities in blockchains & WhoIsProcessorOrController, DataSubjectIsDataController, WhoHasLegalResponsibility\\
\midrule
Solutions for protection of personal data in blockchains & ZeroKnowledgeProof, ChamelonHash, RingSignatures, Salting, MerkleTrees, SecureMultiPartComputation, PseudonymisedDataIsPersonalData, PseudonymisedDataIsAnonymizedData, DoNotStorePersonalDataOnChain, RisksOfQuantumComputers, DoNotReusePublicKeys, AnonymizationIsGDPRCompliant, AnonymizationIsNotGDPRCompliant, AnonymizationIsIllegal, SensitiveDataStorageOnChain\\
\midrule
Data subject's rights & RTBF (ImmutabilityIsAProblem, ImmutabilityIsNotAProblem, HashingOut, RemoveSecretKey, ConsensusToDelete, DisableAccess, Prunning, MainChainSideChain), RightToRectification, RightToBeInformed, RightOfAccess, RightToObject, AutomatedDecisionMaking,  RightToDataPortability, RightToRestictProcessing\\
\midrule
Lawfulness, fairness and transperancy & ConsentManagementViaBlockchain, AccessControlViaSmartContracts, LegitimateUse, UseCasesForLegitimateUse, Transperancy, Lawfulness, DataBreaches, DataBreachNotification\\
\midrule
Other principles & DataMinimisation, StorageLimitation, Security\\
\midrule
Other topics & ProtectionByDesignDefault, TerritorialScope\\
\bottomrule
\end{tabularx}
\end{table}

\pgfplotsset{
barcharts axis style/.style={
xbar,
bar width=8pt,
xmin=0,
enlarge x limits=0,
enlarge y limits=0.05,
axis x line=none,
tick style={draw=none},
axis lines*=left,
y dir=reverse,
ytick=data,
reverse legend,
legend style={font=\small},
legend cell align=left,
nodes near coords,
nodes near coords align={horizontal}
},
}

\begin{figure}[!htb] 
\centering
\begin{tikzpicture}
\begin{axis}[
width=0.7\linewidth,
height=8cm,
symbolic y coords={HashingOut, ImmutabilityIsAProblem, WhoIsProcessorOrController, ConsentManagementViaBlockchain, DataMinimisation, AccessControlViaSmartContracts, PseudonymisedDataIsPersonalData, ProtectionByDesignDefault, DoNotStorePersonalDataOnChain, ZeroKnowledgeProof, RemoveSecretKey, Security, PublicPrivateKeys, StorageLimitation, Transperancy, RightToAccess, LegitimateUse},
barcharts axis style]
\addplot
coordinates {(58,HashingOut) (53,ImmutabilityIsAProblem) (43,WhoIsProcessorOrController) (26,ConsentManagementViaBlockchain) (24,DataMinimisation) (23,AccessControlViaSmartContracts) 
(19,PseudonymisedDataIsPersonalData) (18,ProtectionByDesignDefault)  (18,DoNotStorePersonalDataOnChain) (17,ZeroKnowledgeProof) (16,RemoveSecretKey) (16,Security) (16,PublicPrivateKeys) (14,StorageLimitation) (14,Transperancy) (12,RightToAccess) (10,LegitimateUse)};
\end{axis}
\end{tikzpicture}
\caption{Most popular codes}
\label{fig:popularCodes}
\end{figure}
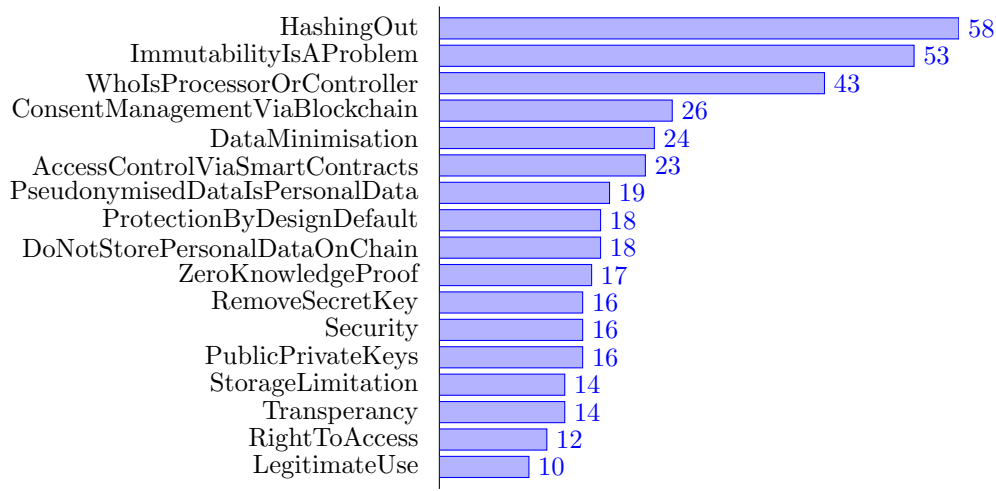

\subsection{Personal Data in Public Blockchain Systems}
\label{subsec:PersonalDataOnBlockchain}

In order to assess whether personal data may be processed legitimately on public blockchain systems, this subsection is dedicated to understand different types of data on the blockchain and their key components.

\subsubsection{Transactional Data}
\label{subsubsec:TransactionData}

Transactional data is the most common data type in all types of blockchain systems. Depending on the underlying use case, the content of a transaction tends to include personal data such as personal identifiers, financial or medical information relating directly or indirectly to data subjects. In case of public blockchain systems that cover smart contracts, executions of smart contract functions are also held in the transactions.

Transactional data can appear in three forms in blockchain systems: plain, encrypted, or hashed. Keeping data in plain text is problematic from a data protection perspective, especially for public blockchain systems. Therefore, it is often the case that some public blockchain systems choose to keep data in an encrypted or hashed form. However, encrypted or hashed data is still personal data, as it falls under the category of pseudonymised data defined by the GDPR as explained before. As mentioned before, the GDPR defines pseudonymisation as ``the processing of personal data in such a manner that the personal data can no longer be attributed to a specific data subject without the use of additional information''. Here, the important question is to define this additional information in the blockchain context and identify how the personal data can be revealed. When data is stored in an encrypted form, it can still be decrypted with the correct key which makes the key the additional information to reveal the personal data.

It is discussed in the literature that there is a linkability risk when data stored in a hashed form: the possibility to link a particular piece of data and a hash value can still be found or a hash value might be used to infer personal information, when the same hash value is stored multiple times~\citep{erbguth2019five, duarte2019introduction}. Therefore, it is not surprising that we observed a general consensus in the literature, which highlights that transactional data pseudonymised via encryption or hash functions should still be considered personal data~\citep{giordano2021blockchain, Politou2019, wirth2018privacy, molina2021design, giannopoulou2018distributed, kolan2020medical, wilford2021digital, Ferrari2018, duarte2019introduction, teperdjian2020puzzle, dutta2020blockchain, schmelz2018towards, karasek2021reconciliation, alessi2019decentralized, duarte2019introduction}.

However, even though not many, there are different opinions proposed in the literature. The GDPR makes it clear that pseudonymisation of data does not equal to anonymisation\footnote{The GDPR, supra note 291, Recital~26}, however, it does not make the distinction between the two methods very clearly. Recital~26 of the GDPR specifies that data becomes anonymous if it is ``reasonably likely'' that no identification of the data subject can be derived, which led to different understandings among researchers and even different national data protection authorities. For example, according to the Irish data protection authority (DPA) Data Protection Commission (DPC), the data has to be rendered ``irreversibly'' anonymous and the criterion of irreversibility is linked to the absence of reasonable likelihood of identifiability~\citep{giannopoulou2020data}. The French data protection authority CNIL (Commission nationale de l'informatique et des libertés) takes a similar position and acknowledges that anonymisation tends to make identifiability ``practically impossible''~\citep{Martin-Bariteau2018}. However, Spanish DPA (AEPD, Agencia Española de Protección de Datos) provides a more absolute approach regarding hash functions and reported that whether to consider hashed data as anonymised or pseudonymised depends on a variety of factors ranging from the entities involved to the type of the data at hand~\citep{giannopoulou2020data}. Similarly, the UK's DPA Information Commissioner's Office (ICO) once advised on its website (on 20th October 2017) that\footnote{\url{https://ico.org.uk/media/for-organisations/data-protection-reform/overview-of-the-gdpr-1-13.pdf} (Page 4)}:
\begin{quote}
``\textit{Personal data that has been pseudonymised -- e.g. key-coded -- can fall within the scope of the GDPR depending on how difficult it is to attribute the pseudonym to a particular individual.}''
\end{quote}
However, this moderate definition seems to have changed as now the ICO's website clearly states that\footnote{\url{https://ico.org.uk/for-organisations/guide-to-data-protection/guide-to-the-general-data-protection-regulation-gdpr/what-is-personal-data/what-is-personal-data/\#pd4}}
\begin{quote}
``\textit{However, pseudonymisation is effectively only a security measure. It does not change the status of the data as personal data.}''
\end{quote}
In the context of the GDPR's predecessor the EU DPD 1995, a similar opinion was provided by Article 29 Data Protection Working Party~\citep{Article29WP2014OpinionAnonymisation}, which acknowledged that even though pseudonymisation reduces the linkability to original identity, it does not eliminate the risk of the data subject being identified, e.g., decrypting an encrypted piece of data via brute-force attack without the decryption key.

We observed various discussions on this topic in papers we covered in the SLR. For instance, it was noted by~\citeauthor{erbguth2019five} that when only the data subject has the key and nobody else can get hold of it, it is doubtful if the GDPR is meant to protect the data subject from the risk of decrypting the data itself~\citep{erbguth2019five}. \citeauthor{giordano2021blockchain} pointed to the same issue and emphasised that, since the key usually remains in the exclusive ownership of the user himself, there is no intermediary nor central body has knowledge of the link between the key and its user. Therefore, he proposed that the nature of the information and of data ﬂows affected by the blockchain technology is still far from being defined~\citep{giordano2021blockchain}. In a similar study, \citeauthor{rampone2018data} differentiated the two roles and proposed that key-coded data is personal data only for the owner of the list of correspondence that links the codes and the data subject's identities and for those who can reasonably gain possession of it. However, she did not consider key-coded data personal for those who do not have the list of correspondence and are not allowed to have access to it~\citep{rampone2018data}.

In a relevant study, \citeauthor{guggenmos2020develop} conducted a participatory action based research and they held workshops to pinpoint blockchain systems' GDPR compatibility issues. Their study covered developers, regular stand-ups and management meetings. As a reflection from their legal analysis, they concluded that the prototype under analysis in the study did not comply with the GDPR as the use of an identifier made all data on the blockchain personal data. However, surprisingly it was added that the legal opinion indicated that a pseudonymisation solution would resolve this problem~\citep{guggenmos2020develop}. Perhaps more surprisingly, \citeauthor{stan2019new} argued that the health data, which is considered sensitive by the GDPR, could be stored as a hash on a block, without violating the GDPR as the data cannot be returned to its original state and it is therefore sufficiently anonymous~\citep{stan2019new}. Similarly, \citeauthor{Politou2019} suggested to protect sensitive data in the long term by using symmetric algorithms with long key lengths~\citep{Politou2019}. However, researchers also stated that such a choice would have a severe impact on the storage requirements of the designed blockchain systems. \citeauthor{eichler2018blockchain} pointed to off-chain storage solutions in the same context and stated that if the data linking the hashed data to a data subject was kept off-chain and was later erased, the hashed data should once again be considered sufficiently anonymous~\citep{eichler2018blockchain}.

Among those discussions in the literature, the main challenge is given on the fact that data is stored for an \emph{unlimited} period of time in public blockchain systems. Thus, potential future technological development is frequently suggested to be considered when assessing the reliability of current techniques to protect privacy of personal data on blockchains. It is commonly emphasised in the literature that even though it is unlikely for the state-of-the-art methods to link encrypted personal data back to the data subject at the moment, the same cannot be guaranteed in the future due to the rapid advancement in the technology~\citep{eichler2018blockchain, giannopoulou2018distributed, schellinger2022yes, erbguth2019five, wirth2018privacy}.
For instance, advancements in quantum computing were mentioned to pose risks to public-key cryptography as available quantum computers may soon be powerful enough to derive private keys used to encrypt personal data today~\citep{shahaab2020managing}.

To summarise, there have been different opinions regarding whether pseudonymisation techniques used by blockchain systems could render data sufficiently anonymous and there is no observed consensus on what techniques are sufficient to anonymise personal data to the point where the resulting output can potentially be stored in a blockchain system in a GDPR-compliant way.

\subsubsection{Metadata}
\label{subsubsec:metadata}

Metadata is another set of data stored in blockchains that may qualify as personal data. Blockchain technologies rely on public-key cryptography where public keys are used for validating transactions. Those keys are essential elements of the metadata and must be publicly available on the blockchain to enable validation of transactions.

We observed a consensus in the literature that public keys serve as the type of identifiers mentioned in Recital~30 of the GDPR, since those keys are often used to identify the origin of transactions and when associated with other information they constitute personal data~\citep{Martin-Bariteau2018, Finck2018a, buocz2019bitcoin, Politou2019, dekhuijzen2019call, molina2021design, ahmed2020towards, jaccard2018gdpr, kolan2020medical, Ferrari2018}. 

The French DPA CNIL considered the risk of identification of individuals via use of additional information and noted that blockchain applications should implement solutions to ensure that any additional personal data is not stored on the blockchain in clear text~\citep{Martin-Bariteau2018}. This was stressed with highlighting the fact that public keys are essential to the blockchain's proper functioning and their retention periods are aligned with the those of the blockchain's lifetime~\citep{Martin-Bariteau2018}.

\citeauthor{Finck2018a} explained the potential of public keys to identify individuals with the following examples~\citep{Finck2018a}. 1) If someone uses the blockchain to transfer ownership of a house that will be public due to the nature of blockchain, then, if the person's neighbour would know that such a transfer took place, he/she could associate the public key to the transfer that was made and link the public key with the house owner. 2) Some users may prefer to share their public keys online intentionally to receive donations, which may link their address to their real-world identities. 3) Additional information that might be gathered in accordance with regulatory requirements, such as where cryptoasset exchanges perform ``Know Your Customer'' (KYC) and ``Anti-Money Laundering'' (AML) duties, can lead to disclosures of real-world identities behind the public keys (this scenario was also mentioned in~\citep{duarte2019introduction}).

In addition to the above simple scenarios where identification of data subjects behind public keys can happen, more advanced pattern analysis is also given in the literature as a risk. It has been argued that patterns may emerge if the same public key is used by the same natural person in several transactions, which can be used to re-identify them~\citep{schmelz2018towards, duarte2019introduction}.

Even though not many, there are researchers who expressed opposite opinions in the context of considering public keys as personal data. \citeauthor{rampone2018data} argued that the definition of personal data, albeit in the form of pseudonymous data, given in the GDPR does not apply to public keys used in blockchain systems~\citep{rampone2018data}. His argument is around public keys being used to solve a technical problem about facilitating trust in a peer-to-peer network and not actually designed to allow for revealing personal identities. Therefore, he suggested them to be considered neither personal nor pseudonymous data even though they could be used to carry out advanced digital forensic searches to track down the identity of the private key holders. He noted that a public key is not always associated with a natural person and it may be used by a legal entity, which makes equating public keys and pseudonymous data wrong. He also added that, due to the lack of a correspondence list that maps public keys to personal IDs, and such a correspondence list cannot be easily obtained in normal conditions, public key is nothing but a piece of information indicating a certain credit availability. In payment situations where the debtor and the creditor know each other, it would be a contingent correspondence related only to a given transaction in progress, however, \citeauthor{rampone2018data} noted that this could not be extended to other transactions.

With similar arguments, \citeauthor{eichler2018blockchain} proposed that public keys were not expected to be personal data in two circumstances: when a key does not belong to a natural person or was not created on behalf of a natural person; and when a key could not be linked to a natural person by reasonable means and is therefore truly anonymous~\citep{eichler2018blockchain}. As done by \citeauthor{rampone2018data}, highlighting that public keys are unavoidable component of the blockchain technology, \citeauthor{eichler2018blockchain} noted that the law must acknowledge a new way to think about public keys.

A more moderate approach was followed by \citeauthor{koscina2021blockchain} who recognised public keys as personal data, however, using them in blockchains is interpreted as the maximum minimisation of information (Article~5(c) of the GDPR)~\citep{koscina2021blockchain}. Similarly, \citeauthor{giannopoulou2018distributed} also argued that combined with necessary privacy enhancing mechanisms, public keys could fulfil the data minimisation requirements of the GDPR~\citep{giannopoulou2018distributed}.

In brief, to make a truly GDPR-compliant blockchain system, public keys are one of the biggest challenges as they are an essential component of the blockchain technology and cannot be moved to be off-chain like other data. However, there are techniques provided in the literature for anonymisation of public keys such as ring signatures and zero-knowledge proofs, which will be explained later in Section~\ref{subsec:TechnicalSolutionsforDataProtection}.

\subsection{Roles and Responsibilities in Blockchain Data Processing}
\label{subsec:RolesInBlockchainDataProcessing}

One of the most common debates in the literature in the context of GDPR compliance of public blockchain systems is around identification of data controllers and processors. As explained before, the GDPR identifies a data controller as an entity that jointly or alone determines the purposes and means for the processing of the personal data. The key component in this definition is being able to have the decision-making power over the processing. On the other hand, a processor is an entity who processes personal data on behalf of a controller.

For data controllers of public blockchain systems, there are three different opinions among researchers as discussed in the papers covered by our SLR: nodes (lightweight nodes, full nodes, miners, all nodes as joint controllers), designers/developers, and users. In contrast, for data processors public blockchain systems, there is more consensus among researchers: users and nodes on the blockchain network. In this subsection, we summarise such discussions and highlight conflicting interpretations.

The majority of the articles covered in our SLR provided a network perspective and are based on the assumption that data subjects are participants of the network and add personal data to the blockchain themselves. However, some other researchers considered the scenario where a user of a blockchain-based application adds personal data to the blockchain on behalf of a data subject. The identification of roles and responsibilities in these two types of scenarios differ greatly, as summarised in the following two subsections from two different perspectives (the blockchain network and application domains).

\subsubsection{Perspective of Blockchain Network}
\label{subsec:networkPerspective}

As stated above, a majority of studies covered in our SLR highlighted different roles and responsibilities of different actors who contribute to the functioning of the blockchain network differently. A blockchain network can consist of full nodes (nodes that maintain a full local working copy of the whole blockchain)~\citep{bitcoinwiki_fullnodes} and lightweight nodes (those who do not download the entire blockchain but only the block headers)~\citep{bitcoinwiki_lightweight}. Lightweight nodes rely on full nodes to access the full content of the network. Both lightweight and full nodes can request to create new transactions in the network by broadcasting such request to all nodes. Some full nodes are miners, who write to the blockchain network by investing processing power into solving a cryptographic puzzle so that they can create new blocks and get rewarded for each new block created. A blockchain system uses a distributed consensus protocol to allow miners to create new blocks and for full nodes to jointly decide which branch of a blockchain network will become the main chain. From the GDPR's perspective, this protocol determines the purposes of the blockchain system and means of data processing. The rules in a distributed consensus protocol and a blockchain system are created by developers who are another type of actors responsible for functioning of the network. In some studies, researchers did not differentiate the different types of nodes and stakeholders and argued that the owner of each node should be considered as a joint ``controller'' of the processing of personal data according to the GDPR~\citep{fabiano2017IoT, schmelz2018towards}. However, other researchers recognised the different types of node and stakeholders. Their opinions are summarised below.

\paragraph{Lightweight nodes}
According to \citeauthor{buocz2019bitcoin}, lightweight nodes of the Bitcoin network cannot be considered controllers since they can only request to create transactions and determine only the input address, the output address, and the transferred amount~\citep{buocz2019bitcoin}. They argued that lightweight nodes' ability to request creation of transactions would be an excessive interpretation of the term ``controller'' since they only define ``what'' (the transactions) but not ``why'' and ``how'' of the processing like a real controller should do.

\paragraph{Full nodes and miners}
\citeauthor{buocz2019bitcoin} noted the essential contributions of full nodes to the functioning of the blockchain network, however, as they cannot change the protocol by themselves or choose a different protocol within the respective software, they were not considered controllers. However, \citeauthor{jaccard2018gdpr} argued that when a number of full nodes form more than 50\% of all mining power, they should qualify as joint controllers~\citep{jaccard2018gdpr}. Considering the activity of full nodes similar to Internet hosting, they also claimed that every full node and perhaps every miner would qualify as a data processor under the GDPR. They further added that following a similar logic lightweight nodes might also qualify as processors. Responsibilities of the miners were found insufficient as controllers by some researchers due to the lack of their power in determining the purposes or means of the processing~\citep{buocz2019bitcoin, ramos2019privacy, giordano2021blockchain, schellekens2020conceptualizations, eichler2018blockchain}. \citeauthor{schellekens2020conceptualizations} added that miners are unlikely qualified as processors because users neither know the miners nor do they have a contractual relation with them~\citep{schellekens2020conceptualizations}. On the contrary, miners were considered to be data processors by some other researchers~\citep{duarte2019introduction, suripeddi2021blockchain}.

It has also been acknowledged by some researchers that, in certain cases, nodes and miners could define their own purposes and set up their own means via accessing the public database stored on the blockchain to collect personal data for commercial purposes, or changing the rules of the blockchain-based platforms by creating a fork in the chain~\citep{duarte2019introduction}. In such cases, one could argue that nodes and miners become joint controllers.
 
Even though the lack of capacity to qualify as a controller dominates research papers covered in our SLR, we also identified some opposite opinions. For instance, \citeauthor{ibanez2018blockchains} argued that miners could be considered as controllers since they determine why and how their own local version of the block is processed~\citep{ibanez2018blockchains}. Some other researchers also agreed that every miner on a public blockchain network could qualify as a controller in theory~\citep{hofman2019margin, herian2018regulating}.

\paragraph{All nodes as joint controllers}
Holding all nodes responsible is an alternative opinion in many papers covered~\citep{buocz2019bitcoin, Fabiano2018a, wirth2018privacy, molina2021design, giannopoulou2020data, karasek2021reconciliation, alessi2019decentralized}. It has been argued that either all nodes collectively -- as a partnership -- is the controller within the meaning of Article~4(7) of the GDPR or all individual nodes are joint controllers under Article~26 of the GDPR~\citep{buocz2019bitcoin, wirth2018privacy}. However, some researchers highlighted that joint controller-ship requires that controllers, by means of arrangements between them, determine (and thereby divide) their respective obligations, but this does not correspond to nodes in public blockchain systems~\citep{dekhuijzen2019call, ahmed2020towards}. \citeauthor{campanile2020privacy} added that lawful processing of personal data requires all nodes holding personal data to be known by the users as joint controllers, which is not possible for public blockchain systems~\citep{campanile2020privacy}.

\paragraph{Designers/Developers}
The discussions regarding responsibilities of designers and developers are even more diverse. While examining the responsibilities of the developers, \citeauthor{buocz2019bitcoin} started the discussions with the following question~\citep{buocz2019bitcoin}: who determines the content of the code (``governance of infrastructure'')? This question is important since it defines the responsibility of the data controller(s) in a blockchain system. \citeauthor{buocz2019bitcoin} highlighted the dynamic nature of open-source project teams where a dynamic group makes proposals and offers inputs to improve the code. They added that despite this dynamic nature, development and maintenance of the code ultimately relies on a small number of core developers who play a key role in the design of the platform. \citeauthor{schellekens2020conceptualizations} had the view that even though the core developers determine the content of the protocol proposal, they cannot decide whether it will actually be the way that data will be processed within a blockchain network~\citep{schellekens2020conceptualizations}. Therefore, some researchers concluded that developers could not be considered as controllers~\citep{buocz2019bitcoin, eichler2018blockchain, giannopoulou2018distributed, schellekens2020conceptualizations}. \citeauthor{eichler2018blockchain} added that the capacity of developers is limited to developing tools and it is up to participants of the blockchain system to decide how those developed tools are used~\citep{eichler2018blockchain}. System administrators, for instance, are mentioned as an important type of actors who decide whether to adopt the code or not.

Interestingly, \citeauthor{jaccard2018gdpr} suggested that the first designers be considered as the first data controllers and to be held liable for certain damages and responsible for the respect of certain obligations. Keeping the data protection by design and by default principles in mind, they added that authorities and private parties could be tempted to hold those designers liable as data controllers.

We also observed discussions on the responsibilities of developers of smart contracts. Some researchers considered smart contract developers as processors since they process personal data on behalf of data controllers according to Article~28 of the GDPR~\citep{ramos2019privacy, erbguth2019five, dutta2020blockchain}. \citeauthor{ramos2019privacy} argued that the same applied to miners as they followed the data controllers' instructions for checking whether a transaction meets set technical criteria~\citep{ramos2019privacy}. \citeauthor{dutta2020blockchain} noted that both smart contract developers and smart contracts themselves could also be considered as data processors~\citep{dutta2020blockchain}.

\paragraph{Participants with special power}
Some researchers claimed that~\citep{guggenmos2020develop, campanile2021risk}, data on a blockchain network is pseudonymised, and it only qualifies as personal data to those participants who possess certain additional information that allows attribution of the data to a data subject. Based on that, they argued that only those participants who possess the additional information (e.g., decryption keys) required for attribution qualify as controllers. Related opinions were proposed in~\citep{tatar2020law, karasek2021reconciliation}, where it was reported that in off-chain storage solutions, the party who controls the off-chain storage could be counted as a data controller as they have the power of determining the purposes and means of data processing.

\paragraph{Users}
Another type of actors, users, were also discussed as controllers by some researchers since they decide what information is included in a transaction, and by this means, determine the details of processing~\citep{wrigley2019people, ramos2019privacy, fabiano2017IoT, al2020designing, kondova2020self, giannopoulou2020data, walters2019privacy, Ferrari2018, erbguth2019five}. This interpretation raises the interesting point that individual users can be both a data controller and a data subject, which makes many of the data processing requirements unnecessary. It also raises a difficult question~\citep{millard2018blockchain}: might they be exempt from regulation considering that they are processing data in the course of a purely personal or household activity?

The strongest argument regarding the user's role is that if a user chooses to use a blockchain or blockchain-based application, they makes themself determine the ``purposes'' and ``means''~\citep{schellekens2020conceptualizations, duarte2019introduction}. For instance, \citeauthor{duarte2019introduction} claimed that when a user chooses to use a blockchain network even though there are different types of payment methods and different platforms, they determines the ``means'' and making a transaction would mean determining the ``purpose''~\citep{duarte2019introduction}. On the other hand, \citeauthor{buocz2019bitcoin} argued exactly the opposite~\citep{buocz2019bitcoin}: although a user has the factual power over the processing as they could choose to connect to the blockchain and leave whenever they want, it is unclear if they have the power to determine the means and the purpose of the processing, and thus, they could not be labelled as a controller in a public blockchain system.

\citeauthor{schellekens2020conceptualizations} noted that it could be desirable to designate the user as a joint controller together with the administrators of nodes (and also with core developers of the blockchain system). He explained that this would create a clear addressing point for a data subject seeking to exercise their rights~\citep{schellekens2020conceptualizations}. However, considering the administrators as joint controllers together with the core developers was given as the strongest argument in the same study. \citeauthor{schellekens2020conceptualizations} also noted that even though a user could be considered as a data controller, they may not be able to fulfil the responsibilities of a controller, which would include making binding contracts with processors, exercising the necessary control over the full nodes, and deleting data from the blockchain. In this context, \citeauthor{al2020designing} recommended the use of a contract which would include the terms and conditions to be agreed upon whenever a user, a node or a miner first uses a blockchain system~\citep{al2020designing}. This approach can help define the use case and then the role of a user as a data controller, a data processor or a data subject.

\paragraph{Enforcement of responsibilities}
Some researchers also raised concerns on legal responsibilities of controllers. In summary, it was reported that problems of enforcement would remain unclear due to the distributed nature of blockchain systems and the network lacks identifiable managing partners, and clear and transparent allocation of responsibilities~\citep{buocz2019bitcoin, Finck2018a, lima2018blockchain, riva2020happens, jaccard2018gdpr, teperdjian2020puzzle}. Holding a collective responsible that does not have any statutory representatives would cause ambiguities for legal enforcement bodies. It is hard to answer how the responsible parties should take care of blockchain security, for instance, in case of a data breach~\citep{teperdjian2020puzzle}. \citeauthor{teperdjian2020puzzle} also emphasised that data subjects require a contact person to exercise their rights such as the right of access, the right to object to the processing of data and to automated processing, however, decentralised and automated blockchain systems have no single point of contact to make these requests~\citep{teperdjian2020puzzle}.

\citeauthor{buocz2019bitcoin} argued that legal responsibilities could be allocated to all peers in a blockchain network, however, he also added that identifying the individuals behind the network nodes could be very complicated in practice since they are constantly changing~\citep{buocz2019bitcoin}. Due to those difficulties, \citeauthor{tatar2020law} recommended imposing the obligation to identify a person or an entity as the representative of the whole users in a given blockchain network prior to joining the system~\citep{tatar2020law}. \citeauthor{wrigley2019people} stated that even this would most likely require a significant amount of processor agreements in practice, and in theory, it is certainly not unfeasible~\citep{wrigley2019people} to make joint agreements between all participants and parties that run the nodes.

Another concern was given regarding the governing law. \citeauthor{herian2018regulating} stated that, to identify the jurisdiction whose law should be applied, we need to know where a data controller is physically located~\citep{herian2018regulating}. This can be difficult to achieve for public blockchain systems.

\paragraph{Governmental positions}
In addition to researchers' opinions we observed in the research papers covered in our SLR, it is useful to compare them with positions of relevant national authorities in the EU/EEA/UK. CNIL, the French DPA, noted in a 2018 report~\citep{CNIL2018} that participants who have a right to write on the chain and who decide to submit data for validation by miners can be considered as data controllers. Therefore, CNIL considers any legal person who registers personal data, on behalf of a natural person, in a blockchain system as a data controller. However, natural persons, outside a professional or commercial activity, are not considered as data ``controllers'' due to the principle of domestic exception defined in Article~2 of the GDPR. CNIL did not evaluate developers or creators as a whole, but noted that designers of smart contract algorithms may be qualified as processors or controllers, depending on their role in determining the purposes. They considered miners as processors and suggested creating a contract between miners and the controller, specifying the obligations of each party and incorporating the provisions of Article~28 of the GDPR. However, CNIL did not consider miners as controllers due to their lack of power on intervening in the purpose of the transactions.

On the other hand, the Hungarian NPA, National Authority for Data Protection and Freedom of Information, adopted a different position and considered each user who adds data to a blockchain as a data controller~\citep{hunagrianDPO}. EU Blockchain Observatory and Forum, a semi-governmental body in the EU, also highlighted the ambiguities in a 2018 report~\citep{EUBlockchainObservatoryForum2018report}, which stated that in situations where application developers or consortia act as intermediaries between individual users and a blockchain network, they would most likely be considered as data controllers. However, it was also covered in the report that there are cases where it was difficult, and perhaps impossible, to identify a data controller, particularly when blockchain transactions are written by data subjects themselves.

\subsubsection{Perspective of Application Domain}
\label{subsec:applicationDomainPerspective}

The discussions given in Section~\ref{subsec:networkPerspective} mainly depend on the assumption that the data subject is a participant of the blockchain system and puts personal data of their own (see Use Case 1 in Figure~\ref{fig:UCs}). However, there is another use case, which is more common in some applications, where data subjects use an application where a blockchain system is used as a service. In the latter case, service and application providers that determine the purpose and means of personal data processing are argued to be data controllers~\citep{moerel2018blockchain, erbguth2019five, Ferrari2018}. It is noted that, only when nodes and miners have a more active role in processing the data for their purposes, they qualify as data controllers~\citep{al2020designing}, not while processing data on behalf of user without any impact on the algorithm used.

\subsection{Technical Solutions for Protection of Personal Data in Public Blockchain Systems} \label{subsec:TechnicalSolutionsforDataProtection}

Blockchain systems do not necessarily mean to keep personal data, however, the possibility of personal data being kept in a blockchain system cannot be ruled out. In this subsection, we first discuss data anonymisation methods for blockchains, and then, summarise technical solutions proposed in the literature to protect personal data in those systems so the blockchain system can be more GDPR-compliant.

\subsubsection{Anonymisation}

Some researchers investigated the GDPR compliance issues of anonymised data in public blockchain systems. Anonymisation services were considered fully-compliant with the GDPR to protect privacy, however, there were concerns about de-anonymisation attacks and whether the GDPR's threshold for anonymisation is currently reachable on public blockchain systems~\citep{biryukov2014deanonymisation, meiklejohn2013fistful, karasek2021reconciliation}. On the other hand, \citeauthor{karasek2021reconciliation} claimed that if the data is anonymised, then the linkage of data on a public blockchain with a data subject would be impossible for any controller without the use of additional information possessed by that data subject. She considered it practically impossible for data controllers to find the user in possession of additional information needed to identify the data subject~\citep{karasek2021reconciliation}.

Another concern regarding anonymised data is the lack of legal certainties and untraceable payment transactions that contradict the KYC and AML laws~\citep{schmelz2018towards}. Therefore, even though the use of data anonymisation services might solve issues with the GDPR, it will likely raise other legal issues.

It is important to underline that data anonymisation techniques are mainly applied to transactional data, however, as discussed before personal data in public blockchain systems are not limited to transactions (e.g., public keys can be considered personal data). In the rest of this section, we summary data protection techniques proposed for public blockchain systems, highlighting the type of personal data that they can protect.

\subsubsection{Hashing Out}

The most common technique proposed for protecting personal data in transactions is the ``hashing out'' technique. It is achieved by storing hashes of data on-chain and keeping the actual data off-chain by using a local database, eliminating several concerns raised by the distributed nature of public blockchains. This also allows to store more data on the blockchain as the size of hashes is much smaller than that of actual data.

\subsubsection{Zero Knowledge Proof}

Another proposed solution, which received the second most attention research papers covered in our SLR, is zero-knowledge proof (ZKP). ZKP is a cryptographic technique used to ensure privacy without damaging transparency~\citep{quiniou2019blockchain}. It allows the entire blockchain network to agree on the validity of a transaction without revealing the content of the transaction and is recognised by several researchers as an effective privacy-enhancing technology that can lower the risk of liability for GDPR violations~\citep{hasselgren2020, molina2020, erbguth2019five, damian2019applying, molina2021design, eichler2018blockchain, jaccard2018gdpr, walters2019privacy, Ferrari2018, dutta2020blockchain, suripeddi2021blockchain}. \citeauthor{schellinger2022yes} recommended this technique if the verification of important information such as balances, coordinates, or signatures is required~\citep{schellinger2022yes}. Some researchers also added that this technique should be considered from the very beginning of the development cycle, i.e., it was recommended as a privacy-by-design solution~\citep{moerel2018blockchain, mannan2019gdpr, pagallo2018chronicle, giannopoulou2021putting}. This technique was also seen as a solution to comply with the RTBF~\citep{poelman2021investigating}. Although it is a prominent solution used in many applications, its main drawback was reported as the high computational workload~\citep{Schwerin2018}.

\subsubsection{Merkle Trees}

Similar to ZKP, Merkle trees were recommended by some researchers to assure data integrity. \citeauthor{schellinger2022yes} claimed that ZKPs or Merkle trees can be used to achieve a privacy-preserving record of data on the blockchain that does not fall within the scope of the GDPR~\citep{schellinger2022yes}. Merkle trees were also recommended in the scope of implementing privacy by design~\citep{schmelz2018towards}.

\subsubsection{Ring Signatures}
Another type of cryptographic methods proposed for addressing the GDPR compliance issue are ring signatures. Ring signatures refer to digital signatures performed by a group, each member of which has a private key to sign a given transaction. While it is known that one of those members initiate the signing, it is not possible to know which member it is. Thus, this provides a strong protection to personal data. Therefore, some researchers adopted or recommended the use of ring signatures for GDPR compliance~\citep{al2020designing, jaccard2018gdpr, cruz2019privacy, walters2019privacy, dutta2020blockchain}. \citeauthor{giannopoulou2021putting}, however, noted that ring signatures are not yet subject to standardisation processes by neither the developer communities nor any formal standardisation bodies. In addition, it also remains unclear if they reach the GDPR required anonymisation threshold~\citep{giannopoulou2021putting}.

Like ZKPs, ring signatures also rely on advanced cryptography, which makes it harder to integrate into blockchain protocols~\citep{Schwerin2018}. Moreover, the issues on ZKPs regarding high computational workload are also valid for ring signatures~\citep{Schwerin2018}.

\subsubsection{Secure Multi-Party Computation}

Secure multi-party computation (SMPC) is another technical solution proposed~\citep{jaccard2018gdpr}, which aims to provide privacy by allowing multiple parties to perform computations over encrypted data without revealing their input to each other. This enables hiding content of a transaction while still allowing validation of the content. In SMPC, during the processing of personal data, each user's input is split into multiple pieces and distributed randomly to other users. Therefore, each user can only see some meaningless portion of the original data. However, it does not seem to be possible to implement SMPC at scale since it requires a vast amount of system resources like ZKPs and ring signatures~\citep{Politou2019}.

\subsubsection{Other Solutions}

In addition to the solutions summarised above, stealth addresses, one-time keys and adding noise to data are other technical solutions mentioned to hide addresses and transactions~\citep{al2020designing}. It was also recommended to use a salt in the hash function as a means to reduce the probability of obtaining the input value~\citep{molina2021design}. Third-party mixing services of public blockchain transactions were also proposed to help users mitigate risks of re-identification~\citep{walters2019privacy}. This technique helps mitigate the risk of identification preventing ``linkage attacks'' to find out connections between transaction inputs and outputs to identify users~\citep{walters2019privacy}.

Avoiding reusing of the public key was another (less technical) solution suggested as it becomes more difficult to de-anonymise a data subject when a unique public key is used for every transaction~\citep{shahaab2020managing, jaccard2018gdpr}. This was also suggested in the context of smart contracts~\citep{wirth2018privacy}.

\subsubsection{Avoiding Blockchains}

Despite the variety of solutions proposed, some researchers claimed that storing personal data in blockchains conflicts with the GDPR and should simply be avoided~\citep{guggenmos2020develop, sarier2021comments, sarier2021efficient, erbguth2019five, peel2019gdpr, zichichi2020efficiency, bernabe2019privacy, Ferrari2018}. Off-chain solutions are encouraged by some researchers~\citep{guggenmos2020develop, sarier2021comments, sarier2021efficient, peel2019gdpr}. For instance, to achieve the GDPR compatibility, \citeauthor{alessi2019decentralized} proposed to develop modules that allow to store personal data in a centralised cloud environment and preserve only business logic in a blockchain network~\citep{alessi2019decentralized}.

There are some domain-specific discussions as well. \citeauthor{kolan2020medical} argued that personal medical data should not be stored directly on blockchains~\citep{kolan2020medical}. \citeauthor{zheng2018blockchain} also preferred not to store health information in blockchains in their proposed solution~\citep{zheng2018blockchain}. Similarly, \citet{ma2018nudging} focused on personal data collected and processed by banking systems and provided a data privacy classification~\citep{ma2018nudging} for data storage. For such systems, they proposed to allow only public information on chain without any restriction. Most sensitive information is not put on chain as a default setting. For sensitive information owned by customers, they are allowed to put them on chain if they so wish. Finally, sensitive information owned by banks, which is mainly confidential information required for the operation of banks, is given as bank's decision to put on the chain or not.

\subsection{Right to Erasure (Right to Be Forgotten/RTBF)}
\label{subsec:RTBF}

Due to the nature of data immutability, public blockchain systems provide a high standard regarding data integrity. However, as mentioned before, this feature is in direct conflict with the data subjects' ``right to erasure'' under the GDPR. This conflict is the most discussed topic in 53 papers (nearly half of all papers covered in this SLR), e.g., in~\citep{Schwerin2018, guggenmos2020develop, tatar2020law, zemler2019concepts, rampone2018data, moerel2018blockchain, hasselgren2020, berberich2016blockchain, ramos2019privacy, mannan2019gdpr, van2018blockchain, erbguth2019five, giordano2021blockchain}. Highlighting the natural outcome that any attempt to change or manipulate data stored in a block would distort the whole blockchain's consistency, some researchers noted that once a systems based on the blockchain technology fulfils the request of data erasure, this would be accomplished at the expense of blockchain consistency, which would be detrimental to reliability and trust~\citep{tatar2020law, moerel2018blockchain}.

An ideal scenario given in the papers covered by this SLR is the consensus of all participants of a blockchain on the joint execution of requests to delete personal data from the decentralised ledgers~\citep{wirth2018privacy, Politou2019, moerel2018blockchain, kadena2018security, hillmann2020selective}. It requires the agreement of a majority of all nodes. For instance, in an opaque blockchain whereby one party has the power over 50\% of all nodes, the erasure of data can be feasible: the majority of nodes would erase the data, and all other nodes would subsequently erase the data as well. Similarly, the alteration of personal data can be resolved by changing the stored data for a majority of all nodes. In the case that data processors (not controllers) run nodes, it can be solved by making joint agreements between all participants and parties that run the nodes.

However, it was noted that this scheme adds significant performance overhead~\citep{Politou2019}. In addition, in public blockchains, no single node can efficiently eliminate a set of personal data requested for erasure or inform the network about such a request. Considering the difficulties in reaching a consensus to delete personal information in public blockchains, researchers proposed several other techniques to overcome this problem. In the following, we summarise those techniques and their consequences according to papers this SLR covers.

\subsubsection{Hashing out}

Hashing out or off-chain storage is the most discussed solution for GDPR-compliant processing of personal data in blockchain with the aim of implementing the RTBF. As explained before, it falls under a new class of personal data created by the GDPR, which is pseudonymised data. In common understanding, pseudonymisation does not prevent personal data from being personal, but it gives the organisation more leeway for its processing as the corresponding risks are lower. Researchers proposed to use this technique in two ways to implement RTBF.

Some of the studies suggested erasing any off-chain data once a request to erasure is received, e.g., as proposed in~\citep{eichler2018blockchain, guggenmos2020develop, faber2019bpdims, makhdoom2020privysharing, sarier2021comments, shahaab2020managing, moerel2018blockchain, molina2020, campanile2021risk, moslavac2020consent, berberich2016blockchain, kondova2020self, mannan2019gdpr, chavez2020securing, molina2021design, pagallo2018chronicle, naik2020your, casino2019immutability, rotondi2019distributed, wulff2020right, jaccard2018gdpr, dauden2021lightweight}. In this type of solutions, off-chain repositories are used to store personal data of users and in the blockchains only a hash value pointing to the storage location of the personal data stored on the off-chain repository. In this way, once a data subject requests deletion of their personal data, the personal data on the off-chain repository can be deleted, which makes the immutable hashed data pointer stored in blockchain become null and void, and thereby, the system becomes GDPR compliant. A main advantage of this approach is that the information on the blockchain can be used to validate data stored in local repositories~\citep{erbguth2019five, damian2019applying}, which helps ensure integrity. \citeauthor{barati2019developing} proposed to combine this technique with smart contracts and keep a Boolean value in a smart contract that determines whether the service provided by an actor enables users to erase their data at any time~\citep{barati2019developing}. In some studies, researchers enriched this solution by specifying the type of information that should be stored off-chain. For instance, \citeauthor{walters2019privacy} suggested to store all personally identifiable information off-chain~\citep{walters2019privacy}. In another study, \citeauthor{zheng2018blockchain} proposed a solution where they preferred off-chain repositories for continuous dynamic data to make it easier to update data over time~\citep{zheng2018blockchain}. They also preferred to hash personal demographic data such as name, address, person identifier, etc. Similarly, \citeauthor{Ferrari2018} argued that only transactional data should be recorded in blockchains, and any credential or identifying information should be stored off-chain~\citep{Ferrari2018}.

In addition to the first type of solution, some researchers proposed to use the hashing-out technique with encryption to implement RTBF. In this way, a private key is transmitted to the data subject to encrypt the hashed value and deleting the key is argued to amount to erasure for the purposes of the GDPR. Key destruction allows the service provider to erase the `linkability' of the blockchain hash pointer to the data located in distributed off-chain repositories~\citep{lima2018blockchain, campanile2021risk, zieglmeier2021gdpr, erbguth2019five, dejanovic2019using, guggenmos2020develop}. Unlike outright erasure, the encrypted hash values will still exist on-chain but can only be accessed by the data subject through the exclusive control of the private key. With a single point of storage, it is possible to delete the link between the blockchain and the data storage. Moreover, it gets difficult to apply reverse engineering to restore data from the encrypted hash. It is only important to ensure that the key information required to link hashed data to off-chain data can be shared securely, and deletion can be done reliably~\citep{campanile2021risk}.

Another approach similar to disabling access to hashed data is storing the encrypted data in blockchains and deleting encryption keys once a data erasure request is received~\citep{stan2019new, dejanovic2019using, el2020examining, jaccard2018gdpr, campanile2020privacy, Ferrari2018,  tatar2020law}. However, French data protection authority CNIL explained that this approach is not an actual erasure according to the GDPR~\citep{CNIL2018}, which was also confirmed by other researchers~\citep{campanile2020privacy, karasek2021reconciliation}. Karasek-Wojciechowicz noted that if, after key destruction, the controller is still able to link the data stored on the ledger with the natural person which could be achieved via analysing the content of public ledgers and external data that the controller could access, then, this process could not be considered as erasure under the GDPR~\citep{karasek2021reconciliation}. It was also noted in the literature that today's encryption algorithms might no longer be considered secure in the future which might make it possible to decrypt the data without the knowledge of the original encryption key~\citep{ibanez2018blockchains, eichler2018blockchain}. Another difficulty of this solution was given as managing the decryption keys among many parties that need access to the data~\citep{Politou2019}. \citeauthor{pagallo2018chronicle} also noted that the destruction of data or keys did not eliminate the possibility to re-identify individuals~\citep{pagallo2018chronicle}.

Even though hashing out is the most discussed solution to implement RTBF, it was also considered as a ``betrayal'' to the decentralisation principle of blockchains since a certain degree of control of data remains in the hands of a single centralised party~\citep{ibanez2018blockchains}. This leads to reintroduction of a trusted third-party, which would contradict to the motivation behind using blockchains~\citep{Finck2018a, tatar2020law}. Researchers provide potential solutions to design a GDPR-compliant off-chain storage solution that does not need a trusted third-party. Such solutions include challenge-response patterns, off-chain signature patterns, delegated computing patterns, low-contract footprint patterns, and content addressable storage patterns~\citep{Finck2018a}. However, these solutions also introduce new vulnerabilities, reversing the benefits of storing data in a blockchain database in an immutable, tamper-proof, secure, and transparent way~\citep{tatar2020law}. Data stored in off-chain repositories are prone to be compromised, and \citeauthor{cruz2019privacy} argued duplicating the off-chain data with two different hashes so that if one set of data is compromised, the personal data does not become lost~\citep{cruz2019privacy}. However, this technique was also argued to introduce complexity and delays~\citep{Politou2019} and does not solve the problems completely as transactional metadata saved on the chain may still be considered personal data and so subject to the GDPR~\citep{al2020designing, zemler2019concepts, Politou2019}. For instance, when combined with the other information, hashed data may reveal sensitive personal information and can become a target to dictionary attacks~\citep{Politou2019}. Dissimilar to those opinions, however, \citeauthor{stan2019new} argued that if all information about a data item is stored only as a hash on chain, the hashes do not violate the GDPR as they are sufficiently anonymous~\citep{stan2019new}.

\subsubsection{Pruning Techniques}

Pruning is another technique used or suggested to overcome conflicts with the RTBF~\citep{Finck2018a, farshid2019design, moerel2018blockchain}. Particularly, old transactions and blocks are deleted after a predefined amount of time, whereas old block headers containing the hashed version of the removed block data are maintained that ensures the integrity and security of the data~\citep{Politou2019}. Therefore, pruning techniques were considered by some researchers to serve regulatory requirements, allowing the old transactions to be forgotten from the network~\citep{farshid2019design, geraud2017twisting}. In addition, it was argued that it can also offer an increased level of user privacy since old transactions might not be locatable~\citep{Politou2019}.

However, it was also argued that blockchain pruning meets scalability and privacy requirements at the expense of security~\citep{Finck2018a}. \citeauthor{Politou2019} added that pruning might add an expensive overhead, leading further inconsistencies and scalability issues when the blockchain's state is verified~\citep{Politou2019}. On the other hand, there is no guarantee that all nodes would choose not to store the full chain in public blockchains~\citep{Politou2019}. \citeauthor{dutta2020blockchain} focused on state tree pruning and smart contract self-destruct in Ethereum in particular and added that data removal did not depend on participants' demands~\citep{dutta2020blockchain}. The only way to remove code from the blockchain was reported as a contract at that address performing the ``selfdestruct'' operation which leads the storage and code to be removed from the state. However, after this operation, it is still part of the history of the blockchain, and therefore, this process cannot be considered the same as deleting data from a hard disk~\citep{dutta2020blockchain}.

\subsubsection{Chameleon Hash}

Another solution proposed is to use chameleon-hashed blockchains that allow a trusted authority to rectify, amend or overwrite the content of the blockchain~\citep{ateniese2017redactable}. Chameleon hash functions work like any other hash functions with the difference that they have a trapdoor that can be used to generate collisions allowing to alter a data item without changing the corresponding hash value of the data, and therefore, being able to maintain the connection to its successor in a blockchain~\citep{ateniese2017redactable}. The knowledge of the trapdoor key allows to find collisions and thus to replace the content of a given block. So, the users among whom the trapdoor key is secretly shared or a centralised authority holding the key can redact the blockchain content in specific and exceptional circumstances. This functionality was listed a potential solution for implementing RTBF by some researchers~\citep{moerel2018blockchain, Politou2019, pagallo2018chronicle, wulff2020right, dutta2020blockchain}. 

\citeauthor{al2020designing} criticised the approach as it defeats the purpose of blockchain by requiring third-parties and/or a centralised authority~\citep{al2020designing}. \citeauthor{ibanez2018blockchains} also highlighted that adding redactability to an existing blockchain is not possible because the decision for this concept has to be made before a network is set up~\citep{ibanez2018blockchains}. Besides, old copies of blockchain would still contain the redacted data~\citep{ateniese2017redactable, Finck2018a}, which makes it highly questionable for this technique to be considered as deletion of personal data under the GDPR. Cutting the immutability of blockchains was also seen as an security risk as it opens an additional door for hackers~\citep{wulff2020right}.

\subsubsection{Truncated Hashing}

\citeauthor{lee2019modifiable} proposed to use truncated hash values to address the RTBF, which allow to modify transactions by making truncated hash values of modified versions equal to their original target values~\citep{lee2019modifiable}. A hierarchical multi-blockchain model was used to improve the efficiency of such transaction modifications. It was reported that the method did not sacrifice any of the core benefits of the blockchain technology including transparency, security, and traceability. This method however has not been sufficiently scrutinised by other researchers.

\subsection{Right to Rectification}
\label{subsec:RightToRectification}

While many research papers covered in this SLR discussed the RTBF, a much smaller number of papers looked at another highly related right, the right to rectification. Among those, a majority of the studies concluded that technical characteristics of public blockchains are in direct conflict with the right to rectification~\citep{sim2019blockchain, daoui2019gdpr, haque2021gdpr, duarte2019introduction, schmelz2018towards}. The main reason for this is that, similar to the issues with deletion, information in public blockchain systems cannot be corrected, but changed by adding a new block to the chain. In addition to the immutable nature of blockchains, \citeauthor{al2020designing} reported one more technical barrier for exercising this right: the difficulty or impossibility of addressing all the full nodes of the network to make necessary updates~\citep{al2020designing}. In this sense, \citeauthor{duarte2019introduction} added that even if it is possible for a data subject to identify all the nodes or to identify enough nodes (over 50\%) to rectify their personal data, coordination among them would be extremely difficult to ensure~\citep{duarte2019introduction}.

There are a couple of solutions proposed regarding rectification. \citeauthor{al2020designing} assumed that this right could be granted by providing a supplementary statement, which explains the fact that a transaction could be amended by publishing a new transaction with the new or correct data without the need to delete the previous one completely~\citep{al2020designing}. \citeauthor{dejanovic2019using} designed a process where as the hash value of a new data block is added to the blockchain, the encrypted key for the old data block is deleted and therefore made inaccessible~\citep{dejanovic2019using}.

\subsection{Right of Access}

As mentioned above, right of access gives data subjects the right to access their personal data held by any service provider subject to compliance with the GDPR. Dissimilar to other rights such as the RTBF or the right to rectification, the right of access has been reported to be entirely compatible with the blockchain technology by some studies since the data is available to all members of the network~\citep{Martin-Bariteau2018, molina2021design, poelman2021investigating, kolan2020medical, schmelz2018towards}.

On the other hand, \citeauthor{al2020designing} argued that since the controllers in a blockchain system only handle the encrypted or hashed versions of data but not the actual form of it, in order to comply with this right, policies and user agreements should be provided to data subjects that explain technical details of how the network functions~\citep{al2020designing}. Similarly, \citeauthor{duarte2019introduction} stated that it is difficult for nodes to know exactly which data is stored on a blockchain so that they can provide data subjects with information concerning the processing of their personal data~\citep{duarte2019introduction}.

Providing another perspective, the EU Blockchain Observatory and Forum pointed out that, since the data subjects are not provided with a contact person from whom they could request whether their data were being processed and for what purpose, etc., it is problematic to enforce the right of access in public blockchain systems~\citep{EUBlockchainObservatoryForum2018report}. \citeauthor{riva2020happens} was in complete agreement with this opinion~\citep{riva2020happens}. Moreover, \citeauthor{giordano2021blockchain} added that even if it is possible to identify the specific node as the data controller, that node might not have the requested information~\citep{giordano2021blockchain}.

\subsection{Right to be Informed}

Even though not many, there are also some concerns regarding the right to be informed which requires data controllers to provide the data subject with information on the period for which their personal data will be stored. Due to the immutable nature of blockchains, this requirement was stated to be difficult to fulfil~\citep{daoui2019gdpr}. However, facilitating the exercise of this right was asserted not to represent a serious issue to organisations by some researchers~\citep{kolan2020medical, cruz2019privacy}. \citeauthor{cruz2019privacy} suggested having privacy information on chain and enforcing it to be verified by all participants as an acknowledgement of having read it~\citep{cruz2019privacy}. She added that this information should include how data subjects could exercise their rights, and who is responsible for dealing with such a request when the occasion arises.

\subsection{Right to Object}

Right to object is one of the GDPR requirements which is not easy to meet for public blockchains due to their permanent nature. However, this right is overlooked in majority of the discussions in the literature, and issues related to immutability are mainly centred around the RTBF. 

One study that covers difficulties in exercising the right to object highlighted that knowing the controller of the blockchain is a prerequisite, and therefore, this right might be difficult to comply with~\citep{poelman2021investigating}. In their proposed study, \citeauthor{dauden2021lightweight}, however, claimed that smart contracts could be used to exercise right to object as they could be implemented to allow to revoke consent~\citep{dauden2021lightweight}.

\subsection{Rights related to Automated Decision Making including Profiling}

Another theme we identified is about concerns related to automated decision making. The GDPR requires explicit consents of the data subject to allow automated processing of their personal data. \citeauthor{herian2018regulating} asserted that smart contracts engender risks contravening the fundamental rights of the data subject under Article~22 of the GDPR~\citep{herian2018regulating}. The EU Blockchain Observatory and Forum also noted that the difficulties in identifying data controller prevented data subjects to oppose to automated decision making~\citep{EUBlockchainObservatoryForum2018report}.

Highlighting the lack of a specific definition of what must be considered to be a decision, \citeauthor{riva2020happens} argued that it is not clear whether automated data processing management through blockchains fall into the GDPR regime or not~\citep{riva2020happens}. She hypothesised that the solution might be to allow the data subject to make the decision in real time, however, she also noted that this would undermine the benefits of automating the whole set of data processing. This was also agreed by \citeauthor{poelman2021investigating}~\citep{poelman2021investigating}. Considering data subjects' set-up options as the ``human intervention'' required by the law was given as a potential solution by \citeauthor{riva2020happens}, however, it was also stated that this might not satisfy the requirement as the human intervention had to occur at the end of the process and could not be general and preventive~\citep{riva2020happens}. Based on this, \citeauthor{riva2020happens} claimed that some parts of the GDPR is neither up to date nor able to correctly tackle current socio-technological needs.

Unlike other researchers, \citeauthor{Ferrari2018} claimed that the protection from automated decision making is not a challenge for blockchain-based automation of transactions as smart contacts generally provided greater auditability and transparency compared to other methods of executing algorithmic-based transactions~\citep{Ferrari2018}.

\subsection{Right to Data Portability}

Data portability is one of the GDPR requirements that aims to give data subjects more control over data, allowing them to obtain and reuse their personal data for their own purposes across different services~\citep{ICOPortability}.

Discussions similar to those related to the right to be informed and the right of access were also found regarding the right to data portability. Researchers considered the exercise of these right as compatible with the technical properties of the blockchain based on the fact that data written to a public blockchain is available to the general public~\citep{eichler2018blockchain, kolan2020medical, Ferrari2018, schmelz2018towards}. CNIL agreed that there is not a problem with exercising data portability in blockchain systems~\citep{CNIL2018}.

However, there are opposite opinions stated. For instance, some researchers noted that this right requires the interoperability between different blockchains, and this does not exist in practice yet due to the lack of standardisation of blockchain systems~\citep{jaccard2018gdpr, Kusber2020}. \citeauthor{giordanengo2019possible} provided another perspective and asserted that moving data between providers would imply the deletion of data held by the old provider which was not possible in public blockchains~\citep{giordanengo2019possible}. Finally, the lack of precise and identified data controllers was seen as another barrier for this right as it disallows data subjects to forward their data portability request~\citep{riva2020happens}.

\subsection{Right to Restrict Processing}
\label{subsec:RightToRestrictProcessing}

Smart contracts are the only technique identified in the limited number papers that cover the right to restrict processing. According to \citeauthor{poelman2021investigating}, the right to restrict processing could be executed by implementing smart contracts to limit the use of data when necessary~\citep{poelman2021investigating}. It was noted that the first step for this solution is to establish which nodes have access to personal data. Such a solution was implemented by \citeauthor{dauden2021lightweight} where the restriction was performed by the consent smart contract which could limit the personal data that could be collected~\citep{dauden2021lightweight}.

\subsection{Lawfulness, Fairness and Transparency}
\label{subsec:LawfulnessFairenessTransparency}

The decentralised nature of blockchain networks challenges the principle of lawfulness of processing, in particular for the use cases where personal data is processed based on consent~\citep{van2018blockchain}. Hereby, there is a considerable amount of research conducted on how to manege consent in public blockchain systems in a lawful way. In this subsection, we provide summary of such studies and another lawful basis ``legitimate interest'' which was identified as another theme for lawful processing. Due to the very nature of public blockchains, unsurprisingly, transparency is a third theme emerged as a closely related topic. We put all the three themes together and more discussions are summarised at the end of this subsection.

\subsubsection{Consent Management}

Smart contracts have been proposed by many researchers as an ideal solution to manage consents of the data subjects~\citep{Schwerin2018, Rantos2018, kaiser2018towards, makhdoom2020privysharing, moslavac2020consent, Neisse2018, Schwerin2018, erbguth2019five, sharma2020blockchain, wirth2018privacy, bowles2021blockchain, pagallo2018chronicle, faber2019bpdims, molina2021design, loukil2020patriot}. The idea behind using smart contracts for consent management is based on the power of translating privacy preferences into automated rules in smart contracts. Then, those contracts can check the validity of a data access request by a third party and allow individuals to verify who can access what part of their personal data~\citep{shahaab2020managing, molina2020, kolan2020medical, silva2021our, riva2020happens}.

This solution has been proposed in different contexts or applications, e.g., for protecting healthcare data~\citep{bowles2021blockchain, kolan2020medical} and financial data~\citep{ma2018nudging}, two special categories of personal data according to the GDPR. Education is another context identified where contracts can be used to transfer personal data related to educational and professional personal records among educational stakeholders~\citep{alkouz2019eppr}. Consent management problems in online social networks is the focus of another study, where smart contracts were proposed as the essential component of a solution~\citep{ahmed2020towards}.

Even though it has been utilised in different contexts, the strategy behind using smart contracts to manage consents of individuals in a GDPR-compliant way is pretty much the same in all the studies. The aim of all such studies is to prevent undesired data access or to provide proofs for privacy violations. \citeauthor{neisse2017blockchain} provided a preventive mechanism, which disallows undesired behaviours of data controllers including misuse or exchange with third-parties~\citep{neisse2017blockchain}. They identified three possible models with different contracts in relation to the number of data subjects and controllers: data subject contract for a specific controller; data subject contract for a specific data item; and controller contract for multiple data subjects. In the first model, privacy preferences (usage control policies) of data subjects are embedded in specific smart contracts deployed in the blockchain for each controller or processor receiving their data. In the second model, smart contracts are created for each data item to be shared with multiple data controllers, allowing control at data item level. Finally, in the third model, each controller expresses their privacy conditions in a smart contract with an interface allowing users to join (give consent) or leave the contract (withdraw consent). The first and the second models were implemented in a follow-up study~\citep{Neisse2018}, and after running some experiments, it was reported that for more sensitive data with less frequent exchanges (e.g., medical data) the first model is more adequate. On the other hand, the third model was reported to be ideal for more dynamic data with more frequent exchanges and strict scalability and performance requirements~\citep{Neisse2018}.

\citeauthor{barati2019developing} used smart contracts in a broader context, and in addition to consent management, they translated different GDPR rules into contracts including encryption for preventing unauthorised access to sensitive data, erasure of data upon request and restriction of personal data to be transferred outside EU/EEA/UK~\citep{barati2019developing}. The first contract developed by the researchers is a GDPR-compliance contract and it allows users to identify what operations could be executed on their personal data. Secondly, user consent contract was developed to enable users to give a vote as a consent or negation for the execution of operations already claimed through a GDPR-compliant contract. Via this contract, users can retrieve the corresponding blockchain records and accept or reject the execution of each operation. Two more contracts were provided in the study for submission and verification purposes. Verification contracts aim to identify data privacy violations when an executed operation of an actor does not get user consent or some personal data processed by the operation is different from those already claimed by an actor via the GDPR-compliance contract~\citep{barati2019developing}.

Some other GDPR elements were considered by~\citeauthor{BaratiPrivacy2021}, in another study of them, where compliance with three GDPR obligations for cloud providers, namely data protection, data minimisation, data transfer and data storage, were implemented via smart contracts~\citep{BaratiPrivacy2021}. In a similar study, \citeauthor{barati2019enhancing} used the blockchain technology again to provide the audit trail of IoT devices under GDPR rules~\citep{barati2019enhancing}. As done in~\citep{barati2019developing}, those rules are translated into smart contracts to protect personal data in a transparent and automatic way and to facilitate the automatic verification of smart objects whose roles are a data controller or a data processor. The abstract model and business processes proposed in the study were reported to show how the integration of GDPR and blockchain could appear in the design patterns of IoT devices to achieve a greater transparency of privacy~\citep{barati2019enhancing}. Those solutions were improved in another study where~\citeauthor{barati2020gdpr} formally examined the verification of GDPR rules on IoT devices at the design time prior to the usage or manipulation of users' personal data~\citep{barati2020gdpr}. Finally, in their recent study, \citeauthor{barati2021privacy} provided a reactive mechanism so that the cloud providers who have violated the GDPR rules can be detected via developed smart contracts~\citep{barati2021privacy}. Highlighting the increasing number of cloud providers, they noted that even though some of the providers might be directly visible to a user, some others might not be, which was reported to raise data privacy concerns violating the transparency of the data processing. They proposed an architecture based on blockchains and smart contracts to address this requirement.

The privacy of users in the IoT ecosystem was also focused in a study conducted by~\citeauthor{Rantos2018}~\citep{Rantos2018}. The user-centric solution they proposed allows data subjects to manage their consents regarding their personal data in the IoT ecosystem and to exercise their rights defined by the GDPR. Additionally, the blockchain technology is used to support the consent integrity, non-repudiation and versioning in a publicly verifiable manner. Another personal data sharing management system for IoT was developed by~\citeauthor{alessi2019decentralized}, where smart contracts are proposed for the same purpose~\citep{alessi2019decentralized}. The application was designed to delete data referenced by it from the cloud storage when the user withdrew a consent.

A similar system architecture was proposed by~\citeauthor{marikyan2021privacy} to manage agreement between data subject and a data controllers (cloud service providers) before service delivery and any data usage~\citep{marikyan2021privacy}. The verification process managed by the smart contracts involved ensure that user consent had been obtained and that sharing of data with external cloud providers is undertaken in a transparent way.

In addition to smart contracts developed to verify consent of the data subjects, \citeauthor{dauden2021lightweight}~\citep{dauden2021lightweight} provided a purpose smart contract. In their design, once the consent is validated via consent smart contracts, the data controller creates a new purpose smart contract which allows data subjects to decide whether to agree on the processing purpose. The data processor is allowed to request data subject's personal data to process it if this contract is also validated.

\citeauthor{heiss2020put} focused on detecting consent violations in a publicly verifiable way and reporting them~\citep{heiss2020put}. Their smart contract based solution was designed to support service providers to fulfil three particular obligations: establishing an auditable archive of consent policies, providing an appropriate technical measure to ensure and to be able to demonstrate the legally valid processing of personally identifiable information, and finally reporting of consent violations to the supervisory authority. Similarly, in the architecture they proposed, \citeauthor{deSousa2018feasibility} suggested that evidence of the data subject's consent would be stored in the blockchain to enable the regulator to traverse the blockchain whenever a consent needs verifying~\citep{deSousa2018feasibility}. In another paper published by the same authors, storing proofs of consent was reported to assure the integrity of persisted consents and evaluations~\citep{deSousa2019blockchain}. The advantage of recording user consents and updates in blockchain immutably was covered in another study~\citep{pei2020udpp}, where making the consents sticky to the data in order to empower data privacy was given as the main contribution of the study. Comparing their solution with traditional methods in text mode, it was noted that their proposal translates consents into switchable right buttons, making it very easy for users to express their consents by switching on/off the button preference~\citep{pei2020udpp}. \citeauthor{wirth2018privacy} followed a slightly different approach where smart contracts had access to a securely hosted decryption function~\citep{wirth2018privacy}. In the proposed framework, the data subject is the single source who has the key used for decryption which enables data subject to be notified whenever their personal data is accessed.

\subsubsection{Legitimate Use}

Legitimate interests is one of the lawful bases for processing personal data, which gives flexibility to data controllers and processors if personal data is used in expected ways, with a minimal privacy impact, or where there is a compelling justification for the processing~\citep{ICOLegitimateUse}. There is a considerable amount of research on legitimate use of a blockchain system, which was generally proposed as a counter-argument of the need to meet the RTBF. As mentioned above, the RTBF is not an absolute request and only applies when data is no longer necessary for the purposes for which it was collected. In the literature, some researchers proposed counter-arguments based on this point. For instance, \citeauthor{daoui2019gdpr} argued that when deciding to participate in the blockchain, the data subject would know that their personal data would be processed for the duration of the blockchain, which would mean until the last server is destroyed. They noted that as long as data subjects are informed that the duration period would be infinite and that personal data of each member of the blockchain is necessary for the purpose of data processing, the RTBF would not apply~\citep{daoui2019gdpr}. A very similar opinion was given by~\citeauthor{mannan2019gdpr}, who argued that once users gives their consent to have their data permanent on the blockchain, the irreversibility of opt-in mechanism does not conflict with the GDPR~\citep{mannan2019gdpr}.

Another perspective focused on the flexible definition of deletion under the GDPR and it was argued that, considering the functioning principle of blockchains, data stored in blockchains is still necessary for the processing purpose, as those systems are immutable  by design~\citep{berberich2016blockchain, mannan2019gdpr}. According to these studies, such arguments give blockchains a legal ground for processing personal data permanently.

In the same context, \citeauthor{giordano2021blockchain}~\citep{giordano2021blockchain} focused on Paragraph~2 in Article~17 of the GDPR, which provides the following:
\begin{quote}
``\textit{Where the controller has made the personal data public and is obliged pursuant to paragraph 1 to erase the personal data, the controller, taking account of available technology and the cost of implementation, shall take reasonable steps, including technical measures, to inform controllers which are processing the personal data that the data subject has requested the erasure by such controllers of any links to, or copy or replication of, those personal data.}''
\end{quote}
Based on this definition, he argued that since the erasure of personal data is not technically feasible or would require disproportionate efforts, data controllers should not be obliged to a result they could not realistically guarantee. However, he also added that data subjects would still be entitled to receive compensation for any damage they received.

Similarly, \citeauthor{walters2019privacy} argued that a single data subject's privacy request should be overridden by the legitimate interests of all independent users -- to ensure network functionality and the sanctity of information~\citep{walters2019privacy}. Of course, it was highlighted that this argument could only stand if the design of the public blockchain is such that certain privacy standards are met. \citeauthor{zemler2019concepts}, however, highlighted that due to lack of judgements on this field, these legal arguments should be treated with prudent care~\citep{zemler2019concepts} as these approaches could be considered illegal by a law court in the future.

Protecting public interest and protecting legal interest were argued to be considered to overcome conflict with the RTBF in another study~\citep{labancz4conflict}. With a similar motivation, different use cases were recommended in the literature where permanent storage would have a legal basis. For instance, \citeauthor{cruz2019privacy} argued that the use of blockchain-based systems would be convenient where existence of that personal information needs to be proved~\citep{cruz2019privacy}. Similarly, \citeauthor{dutta2020blockchain} recommended the use of blockchain systems for data governance in the areas of transparency and data provenance~\citep{dutta2020blockchain}. \citeauthor{erbguth2019five} suggested providing a permanent justification to use a blockchain-based system and provided the publication of election results as an example~\citep{erbguth2019five}.

In another paper, \citeauthor{subramanian2020blockchain}~\citep{subramanian2020blockchain} summarised results from a panel discussion, which include a recommended decision path to determine when to use blockchain technologies considering the privacy issues raised by the GDPR, derived from \citeauthor{Pedersen2019BlockchainDecisionPath}'s work in \citeyear{Pedersen2019BlockchainDecisionPath}~\citep{Pedersen2019BlockchainDecisionPath}. According the proposed decision path, the following 11 questions should be considered to decide if a blockchain system and which type should be used: 1) Need for a shared common database? 2) Multiple parties involved? 3) Involved parties have conflicting incentives/trust issues? 4) Participants can/want to avoid a trusted third party? 5) Governing rules vary between participants? 6) Rules of transactions remain predominantly constant? 7) Need for an immutable log? 8) Governance allows public network access? 9) Are transactions to be public? 10) Where is consensus determined? and 11) Is off-chain data storage required?

\subsubsection{Transparency}

The GDPR requires data processing to be clear, open and honest with data subjects from the start about identity of the data controllers, and how and why their personal data is being used. Data processing is transparent in public blockchains, as data is stored in the chain and all participants have the information of the system transactions. Therefore, several researchers claimed that this principle is met by blockchains by design~\citep{sim2019blockchain, molina2021design, dagostini2017gdpr, el2020examining}. In addition, according to some of the respondents interviewed in a study conducted by \citeauthor{poelman2021investigating}, due to the transparency of the data in public blockchain systems, this technology gives individuals a strong position in regard to the unlawful exchange of data as they provide footprints which could be linked to consents given~\citep{poelman2021investigating}.

\subsection{Other Data Protection Principles}
\label{subsec:PrinciplesOthers}

\subsubsection{Data Minimisation and Purpose Limitation}

Data minimisation is another requirement of the GDPR which could be argued to contradict the nature of blockchains~\citep{schmelz2018towards, Finck2018a, Fabiano2018a, erbguth2019five, arisi2020blockchain, Ferrari2018}. While one reason for this is that replication of the data across all nodes in a public blockchain prevents a system to store minimum amount of data~\citep{schmelz2018towards, sim2019blockchain}, another reason was given as that the append-only nature of the blockchain systems prevents data to be deleted even though data minimisation principle requires data not to be further processed in a manner that is incompatible with legitimate purposes~\citep{Finck2018a, dekhuijzen2019call, naik2020your, poelman2021investigating, el2020examining}.

Even though not many, there are some arguments for the possibility of adapting data minimisation in blockchains. Off-chain solutions were recommended by several researchers as they allow performing modifications and minimisation~\citep{Finck2018a, al2020designing, mannan2019gdpr, dekhuijzen2019call}. In this context, \citeauthor{Finck2018a} highlighted the difficulty of storing pseudonymous public keys off-chain as they could not be retroactively removed from the ledger and argued that existence of public keys could be a challenge for data minimisation~\citep{Finck2018a}. Unlike \citeauthor{Finck2018a}, \citeauthor{koscina2021blockchain} interpreted storing public keys in blockchains as the maximum minimisation of information and argued that public keys could fulfil the data minimisation requirements of the GDPR~\citep{koscina2021blockchain}.

As another possible solution, \citeauthor{barati2020gdpr} drew attention to adapting smart contracts and proposed a solution in which a contract verifies the data minimisation via checking the scope of personal data processed by the system~\citep{barati2020gdpr}. Pruning is another technique suggested in the literature as it allows removing unused blocks from a blockchain~\citep{moerel2018blockchain}. In addition to smart contracts and pruning techniques, storing data hashes was given as another solution to minimise the processing and storage of personal data~\citep{schmelz2020securing}. Finally, implementation of anonymisation procedures was suggested to ensure that personal data is stored when it is strictly necessary, and therefore, to comply with data minimisation~\citep{cruz2019privacy}.

There are also some debates in the literature regarding the legitimate use in discussions around data minimisation. For instance, \citeauthor{tatar2020law} stated that it is the choice of users what to share in blockchain systems, and therefore, they argued that users could limit the data share to the minimum scope necessary to achieve the specific purpose~\citep{tatar2020law}. \citeauthor{walters2019privacy} provided another counter-argument emphasising that the principle of data minimisation depends on what the ``purpose'' of the information stored is~\citep{walters2019privacy}. He asserted that when the purpose of storing information is to ensure the security and accuracy of the data, the immutable nature of public blockchains might not violate the data minimisation principle~\citep{walters2019privacy}. A government land registry was given as an possible valid use case for this purpose.

\subsubsection{Storage Limitation}

Storage limitation requires not to keep personal data longer than it is needed. As expected, this principle stands in tension with immutability which has been confirmed by multiple researchers~\citep{van2018blockchain, giordano2021blockchain, arisi2020blockchain, sim2019blockchain, el2020examining, zemler2019concepts, walters2019privacy, suripeddi2021blockchain}. The techniques suggested for complying with storage limitation are the same as those for data minimisation. For instance, prunning is one of those approaches, proposed by \citeauthor{moerel2018blockchain}~\citep{moerel2018blockchain}. Implementing smart contracts that can keep the total time taken for data processing and the period of time during which personal data will be kept on the storage of an user is another technique followed in another study~\citep{barati2019developing}. Finally, off-chain storage was also given as a potential solution as it allows erasing of personal data when necessary or at a specified time period~\citep{naik2020your}.

\subsubsection{Integrity and Confidentiality (Security)}

Despite the variety of concerns reported in the literature regarding conflicts between characteristics of blockchains and different GDPR elements, for security principle, there is almost a consensus among researchers that blockchains offer multiple opportunities as a tool for enhanced security~\citep{arisi2020blockchain, naik2020your, poelman2021investigating}. First of all, assuring the integrity of personal data was highlighted in several studies as the blockchain technology is generally noted for its resiliency and the absence of a single point of failure~\citep{arisi2020blockchain, haque2021gdpr, poelman2021investigating, campanile2020privacy}. It was highlighted that distributed management and storage of those systems prevents a single point of failure and attacks on such a single point~\citep{naik2020your}. In addition, cryptography used in blockchains is considered to help to preventing unauthorised parties to alter, delete or steal data~\citep{naik2020your, haque2021gdpr, poelman2021investigating, campanile2020privacy}. Thus, blockchain-based systems are argued to be preferably used in a GDPR-compliant environment in terms of implementing the security principle.

However, even though not many, we observed a couple of concerns raised in the literature regarding the security principle. Security of the encryption and decryption keys is one of them as highlighted by \citeauthor{erbguth2019five}~\citep{erbguth2019five}: due to the immutable nature of public blockchains, changing the passwords for the encrypted data on such systems is not possible. Considering the fact that those passwords need to be changeable when there is a risk that a password has been compromised, he argued that putting encrypted sensitive data on a blockchain might violate information security standards.

Another point highlighted in the literature was ambiguities in terms of responsible units for enforcing security policies, the use of appropriate encryption methods and training for users when required~\citep{dekhuijzen2019call}. Hereby, it was noted that there might be an increase in the risk of personal data breaches. Foreseeing similar risks, \citeauthor{duarte2019introduction} suggested off-chain repository solutions, arguing that data breaches could be detected more easily since any unauthorised alteration made could be easily spotted~\citep{duarte2019introduction}. Additionally, \citeauthor{hasselgren2020} focused on data breach notifications required by the GDPR and suggested the use of smart contracts to ensure automated and imitate notifications to relevant authorities upon data breach~\citep{hasselgren2020}. This was given as especially beneficial to avoid severe penalties.

Similarly, the confidentiality principle was found not aligned with the nature of public blockchains according to some studies~\citep{arisi2020blockchain, Kusber2020}. \citeauthor{Kusber2020} asserted that confidentiality could only be ensured with off-chain storage of affected data~\citep{Kusber2020}. Besides, they argued that the use of hash algorithms, which can become weak, also deteriorated integrity. However, it is noteworthy that there are also opposite opinions in the literature regarding confidentiality. Particularly, cryptography used in blockchain systems was seen as a tool to support confidentiality by some other reesarchers~\citep{naik2020your, haque2021gdpr, rotondi2019distributed, bernabe2019privacy}.

\subsection{Other Topics}
\label{subsec:Others}

\subsubsection{Data Protection by Design and Default}

The GDPR defines requirements for data protection by design and data protection by default in an abstract way, which led researchers to interpret those requirements differently and follow different approaches to meet them. Even though there is a clear understanding that data protection requirements should be considered from the very beginning of the development cycle, a variety techniques have been argued to be a solution to be inline with data protection by design and default due to those different concerns and understandings.

The researchers focusing on privacy of personal data mainly recommended cryptographic techniques like zero-knowledge proofs~\citep{schellinger2022yes, moerel2018blockchain, mannan2019gdpr, pagallo2018chronicle, giannopoulou2021putting}. Off-chain solutions~\citep{molina2021design} and not storing personal data on-chain were also recommended for the same purpose~\citep{moerel2018blockchain, poelman2021investigating}. Another concern is regarding the immutable and distributed nature of blockchains. In this context, solutions like pruning~\citep{moerel2018blockchain} or limiting ledger storage by storing the entire ledger on one (or a few) instances only to enable deletion~\citep{moerel2018blockchain} were recommended. In addition, decentralised control and distributed storage was seen as a major conflict with privacy by design by some others~\citep{walters2019privacy}.

There are also studies that argued that public blockchains satisfy those requirements due to its very nature. For instance, immutability and decentralised data storage is stated to ensure the integrity and accuracy of the data, and hence data protection by design and default~\citep{tatar2020law, haque2021gdpr}. It was highlighted that the individuals received the highest level of control over their personal data via blockchains acquiring the ability to selectively share to any service provider of their choice~\citep{sim2019blockchain}. Based on this idea, privacy by design and by default were argued to be fulfilled~\citep{sim2019blockchain}. Encryption techniques and hash functions were also considered to meet those requirements~\citep{berberich2016blockchain, molina2021design, haque2021gdpr, schmelz2020securing, giannopoulou2021putting, alessi2019decentralized}. However, using hash values and public key cryptography alone was not seen to be able to guarantee privacy by design by some researchers~\citep{Schwerin2018}.

\subsubsection{Territorial Scope}

According to the territorial scope defined by in Article~3 of the GDPR, the regulation applies to personal data processing of any data subjects if the data controller or processor has an establishment in EU/EEA/UK or if the data processing is for delivering a good or service or behavioural monitoring while a subject is in the EU/EEA/UK. Regarding transfer of personal data to outside of EU/EEA/UK, the GDPR requires the receiving country or organisation to provide an adequate level of data protection or the data controllers/processors provide appropriate safeguard.

The distributed nature of public blockchains allows anyone to join the network as a node and create transactions. Therefore, there is no geographical border to its network which could pose the issues regarding the localisation and the application of the law. It is almost impossible for data controllers to be aware of where the participants of the blockchain are located and to ensure compliance~\citep{zemler2019blockchain}. Given that, several studies that investigated the GDPR compliance of public blockchain systems reported this as a major conflict~\citep{daoui2019gdpr, zemler2019blockchain, herian2018regulating, Ferrari2018, duarte2019introduction, dutta2020blockchain, schmelz2018towards, karasek2021reconciliation}. This was also mentioned by CNIL, where the constraints regarding territorial scope was found difficult to implement and permissioned blockchains were recommended instead of public ones~\citep{CNIL2018}. EU Blockchain Observatory and Forum provided an opinion inline with CNIL's, where international transfers of personal data were found problematic for the GDPR compliance.

Even though solutions to overcome this issue are limited compared to other conflicts reported in the literature, a couple of studies proposed some strategies. \citeauthor{al2020designing} suggested having a detailed data security and protection agreement prepared among participants of the network as a way to address the problem, following mechanisms such as the standard contractual clauses allowed by the European Commission~\citep{al2020designing}. In some other studies, technical solutions based on smart contracts were provided that controlled the country of each actor receiving personal data, and if it is a country that is outside EU/EEA/UK, the algorithm sets the GDPR compliance to false~\citep{barati2019developing, barati2019enhancing, barati2021privacy}.

\section{Further Discussions}
\label{sec:discussion}

In this section, we summarise and discuss our key findings for our research questions and highlight limitations of research work reviewed in our SLR.

\subsection{GDPR Compliance Issues of Public Blockchain Systems (RQ1)} \label{subsec:DiscussionComplianceIssues}

Our first research question (RQ1) aims to investigate the issues that the public blockchain technology can lead to in relation to data subjects' rights and data protection principles provided by the GDPR. Our findings demonstrate that difficulties in exercising the right to erasure (RTBF) is the most commonly cited conflict in the literature. It is followed by the right to rectification and the data minimisation principle, which are both at odds with the immutable nature of public blockchains. Finally, territorial scope is another major theme identified.

It should be noticed that the conflicts regarding rights granted to data subjects by the GDPR or the data processing principles are tightly connected to the debates regarding anonymisation and allocation of responsibilities on blockchains. As discussed before, different techniques, including encryption, deletion of encryption keys, and off-chain storage, have been argued to correspond to anonymisation of data even though opponent opinions are more common in the literature. Those discussions are important as once such an assumption could be made where a technique is considered to satisfy the threshold set by the GDPR for anonymisation, the personal data storage in public blockchains via those techniques become GDPR-compliant by design. However, we did not identify any consensus regarding the existence of such a technique, and it remains to be seen as the potential advancements in technology make it impossible to claim the current approach to be \%100 robust. Advancements in the future cannot be ruled out in those discussions since the data is stored in public blockchains permanently.

In addition, the question about allocation of responsibilities on blockchains is crucial in determining who should comply with the obligations, be held responsible for any violation and to whom the data subjects could reach out for their requests. Difficulties in identifying data controllers and processors is the second most commonly discussed conflict (see Section~\ref{subsec:DiscussionlegalRolesResponsibilities}). However, it is noteworthy to briefly mention here that this issue was not only discussed in isolation but also the difficulties it causes in exercising rights guaranteed by the GDPR, such as the RBTF, the right to be informed, the right to access, or the right to data portability, were also highlighted by many researchers. Although it is of critical importance, it is not possible to report a dominant opinion regarding identification of data controllers. Therefore, we can speculate that techniques that will be provided for GDPR compliance of public blockchains will remain insufficient as long as the rules and definitions are not updated in the GDPR to make them clearer for blockchains and other similar decentralised solutions.

As stated at the beginning of this subsection, data minimisation and purpose limitation is the third most common conflict covered in the literature. Two main challenges were highlighted in those discussions: replication of the data across all nodes in a public blockchain; and the append-only nature of the blockchain systems that prevents data to be deleted. Surprisingly, only a few studies discussed public keys in the context of data minimisation and two opponent opinions were reported. In one of those studies, public keys were considered a challenge for data minimisation~\citep{Finck2018a}. However, some others interpreted the use of public keys as the maximum minimisation of information and argued that they could fulfil the data minimisation requirements of the GDPR~\citep{koscina2021blockchain, giannopoulou2018distributed}.

Finally, territorial scope is another major theme identified in the literature in the sense of the GDPR compliance issues of public blockchains. The constraints regarding territorial scope defined by the GDPR was found difficult to implement by several researchers due to the lack of geographical border of public decentralised systems. We believe that similar to the definitions made for data processors and controllers by the GDPR, this is another area that should be handled in a clearer way in the regulations considering the inevitable increase in globalisation of information-centric businesses.

\subsection{Solutions Proposed for GDPR Compliance (RQ2)}

As a response to our second research question (RQ2), our results revealed three major research areas in the literature: techniques proposed to overcome problems regarding deletion of personal data; management of consent and privacy preferences; and techniques to secure data and ensure data privacy.

Regarding the first research area, hashing out is far most the most common technique proposed in the literature with the aim of implementing the RTBF. In this approach, personal data is stored in off-chain repositories and only the hash data pointing to offline storage is stored in the distributed ledgers. This allows off-chain data to be deleted once a request to erasure is received. In the second approach, a private key is transmitted to the data subject which is used to link the blockchain hash pointer to the data located in distributed off-chain repositories and erasure is accomplished via key destruction. Even though both of the techniques were claimed to be effective in means of assuring RTBF in many studies, they fall short in maintaining the decentralisation principle of blockchains and reintroduce several security problems such as trusted third parties.

\begin{figure}[!htb] 
\includegraphics[width=\linewidth]{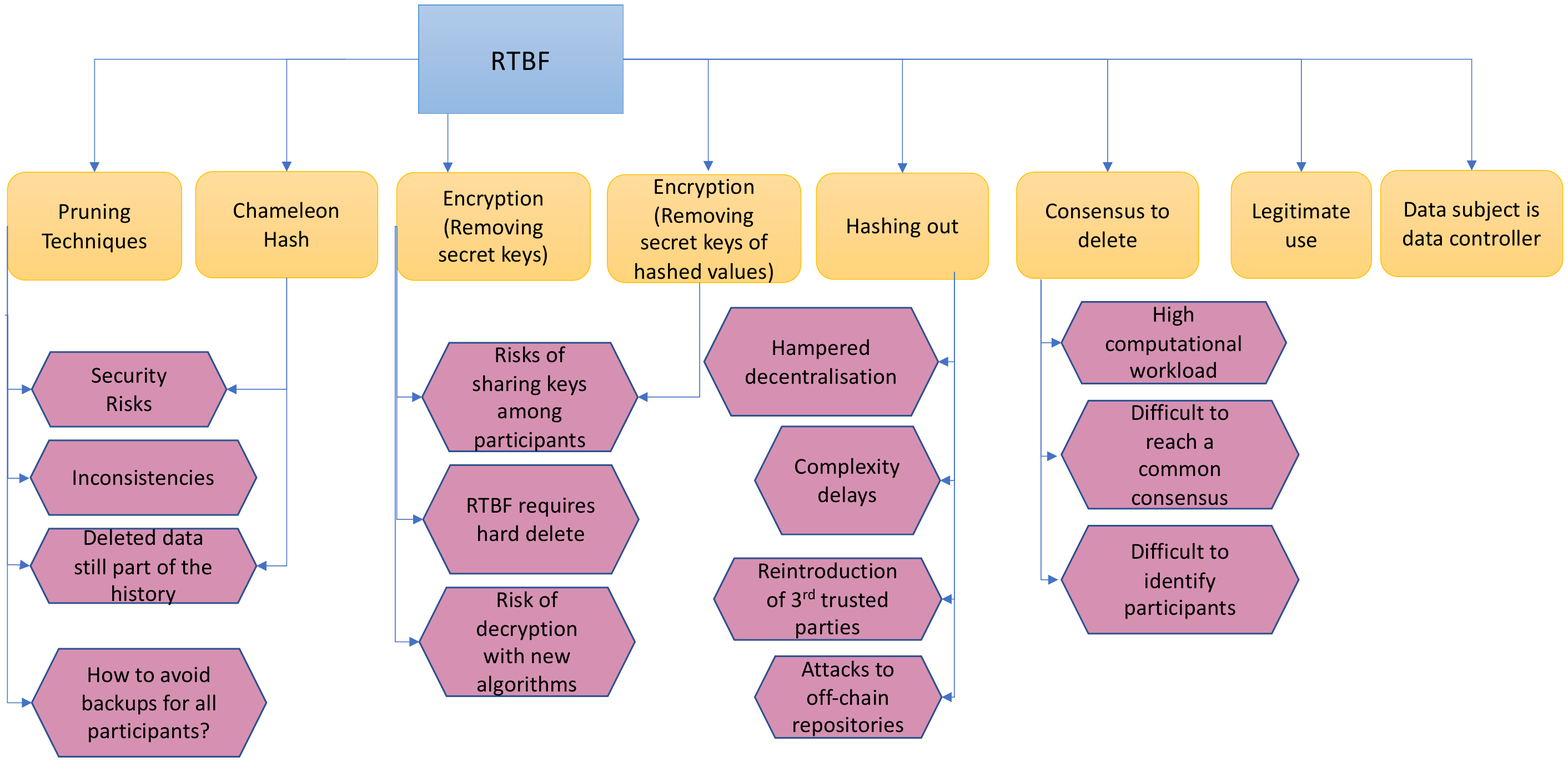}
\caption{Solutions proposed for addressing the RTBF}
\label{fig:RTBF}
\end{figure}

In Figure~\ref{fig:RTBF}, we summarised major solutions proposed in the literature for addressing the RTBF problem, where red hexagons represent limitations argued for each solution. Here, it is important to underline that none of those solutions can be used to delete metadata in blockchains even though it is recognised as personal data by majority of the studies, and therefore, fall in the scope of the GDPR.

One interesting detail for this figure is perhaps the case where data subject is data controller. Once this assumption is made, many of the data processing requirements including the RTBF become unnecessary, and therefore, this is added as a solution in Figure~\ref{fig:RTBF}.

In the second research area, smart contracts were heavily discussed as a solution to managing consent and controlling privacy preferences. They mainly enable to translate privacy preferences into automated rules which can check the validity of a data access request by a third party and handle it transparently for all parties involved. This approach has two main use cases: preventing unauthorised data access and providing proofs for privacy violations. As this process is fully automated, it allows detection of consent violations in a publicly verifiable way and can demonstrate legally valid processing of personal information. The immutable nature of blockchains also enables to assure the consent integrity, non-repudiation and versioning in a publicly verifiable manner. However, it was also noted in the literature that in an ideal sense, data subjects should be able to obtain human interventions and challenge the decision after the smart contract has been executed, which is not possible via smart contracts~\citep{Martin-Bariteau2018}. We identified one study in the literature that recognised this problem and proposed to use smart contracts through a securely hosted decryption function that enables the data subject who is the single source with the key used for decryption to be notified whenever their personal data is accessed~\citep{wirth2018privacy}. Therefore, designing blockchain solutions that can support human interventions could be reported as a potential research area.

\begin{figure}[!htb] 
\includegraphics[width=\linewidth]{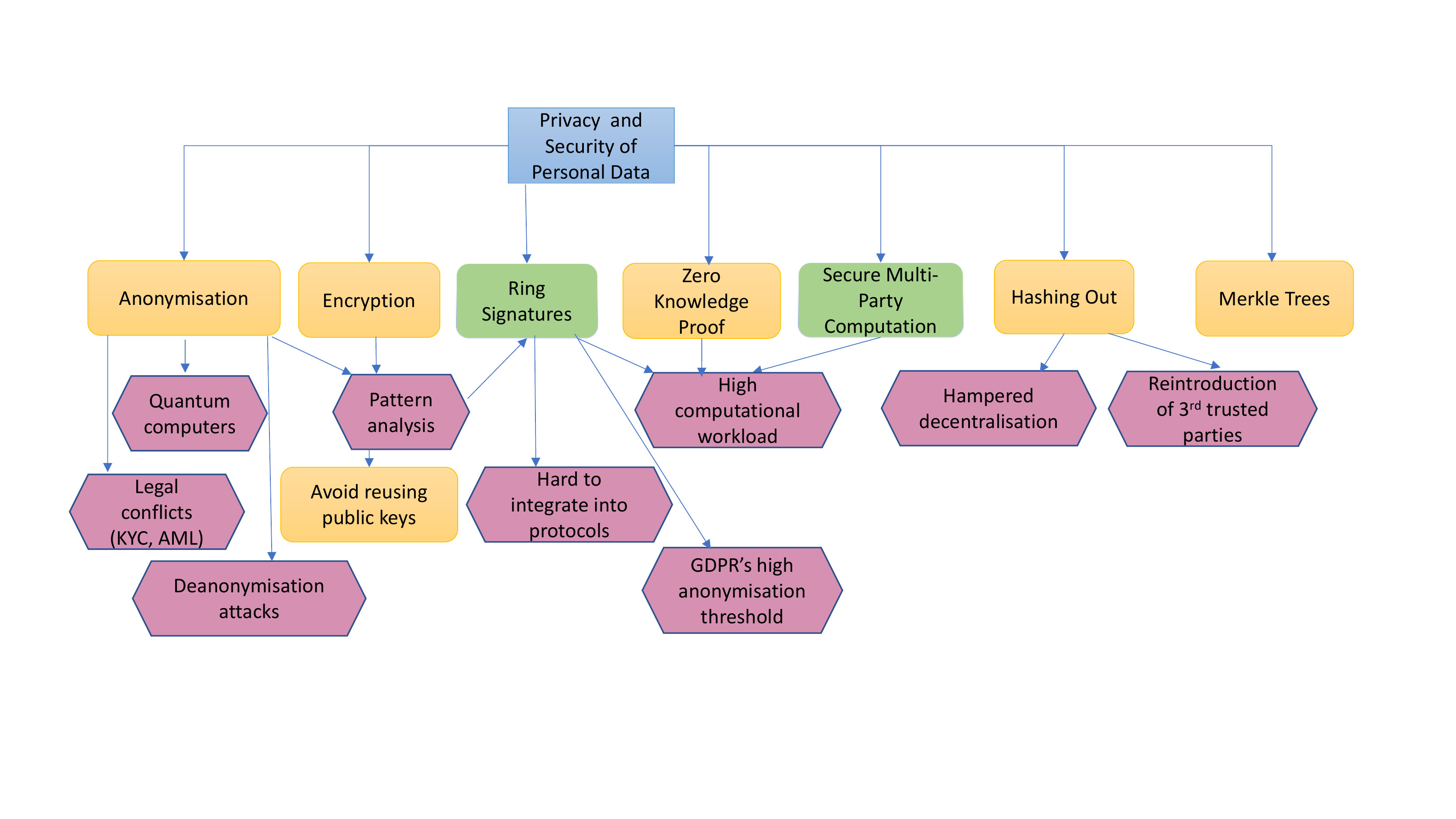}
\caption{Solutions proposed to assure data security and privacy}
\label{fig:PrivacyDataTypes}
\end{figure}

Regarding techniques to assure data security and privacy, there is a variety of techniques proposed with two main goals: assuring data integrity, and assuring data privacy or data confidentiality. ZKP is the most commonly cited technique that has been proposed to assure data confidentiality. Even though not received as much attention as ZKP, Merkle trees have also been proposed in the literature. Ring signatures, SMPC and tumbling techniques are the emerging themes identified, all of which can be used to assure data integrity or confidentiality of personal data on the chain. A full list of solutions and their consequences reported in the literature can be seen in Figure~\ref{fig:PrivacyDataTypes}.

Similar to the limitations in approaches regarding implementing the RTBF, one research gap in the literature regarding security and privacy of data on blockchains is the low number of solutions that focus on public keys. A majority of the solutions were proposed to assure the GDPR compliance of personal data in transactions and they tend to overlook public keys. In Figure~\ref{fig:PrivacyDataTypes}, we have demonstrated the solutions proposed, where yellow boxes represent the solutions for transaction data and green ones represent the techniques that considers the privacy of both transaction data and public keys. Similar to Figure~\ref{fig:RTBF}, red hexagons indicate problems raised in the literature for the proposed solutions.

We consider the categorisation presented here a major contribution of this study, as it differentiates the techniques based on their focus (transactional data, meta data or both) and demonstrates both the strength and limitations of the current progress in this research area.

\subsection{Legal Roles and Responsibilities (RQ3)}
\label{subsec:DiscussionlegalRolesResponsibilities}

The third research question (RQ3) aims to investigate how legal roles and responsibilities of different stakeholders of public blockchain networks have been considered in the literature, e.g., who are considered data controllers and processors. This is one of the most common debates in the literature in the context of GDPR compliance of public blockchains. However, it is not possible to report a consensus among the researchers in the sense of identification of data controllers. The ambiguities in this context even yield some researchers to claim that there is no controller in public blockchains, and therefore, the privacy and data protection law is not applicable~\citep{fabiano2018blockchain}.

However, a majority of the studies covered in our SLR discussed several different entities and their roles and responsibilities. In total, three main categories are discussed as controllers in the literature: nodes (lightweight nodes, full nodes, miners, all nodes as joint controller); developers; and users. However, none of the categories dominate the discussions. On the other hand, regarding processors, users and nodes on the blockchain are argued to be processors in almost all of the studies.

\begin{figure}[!htb] 
\includegraphics[width=\linewidth]{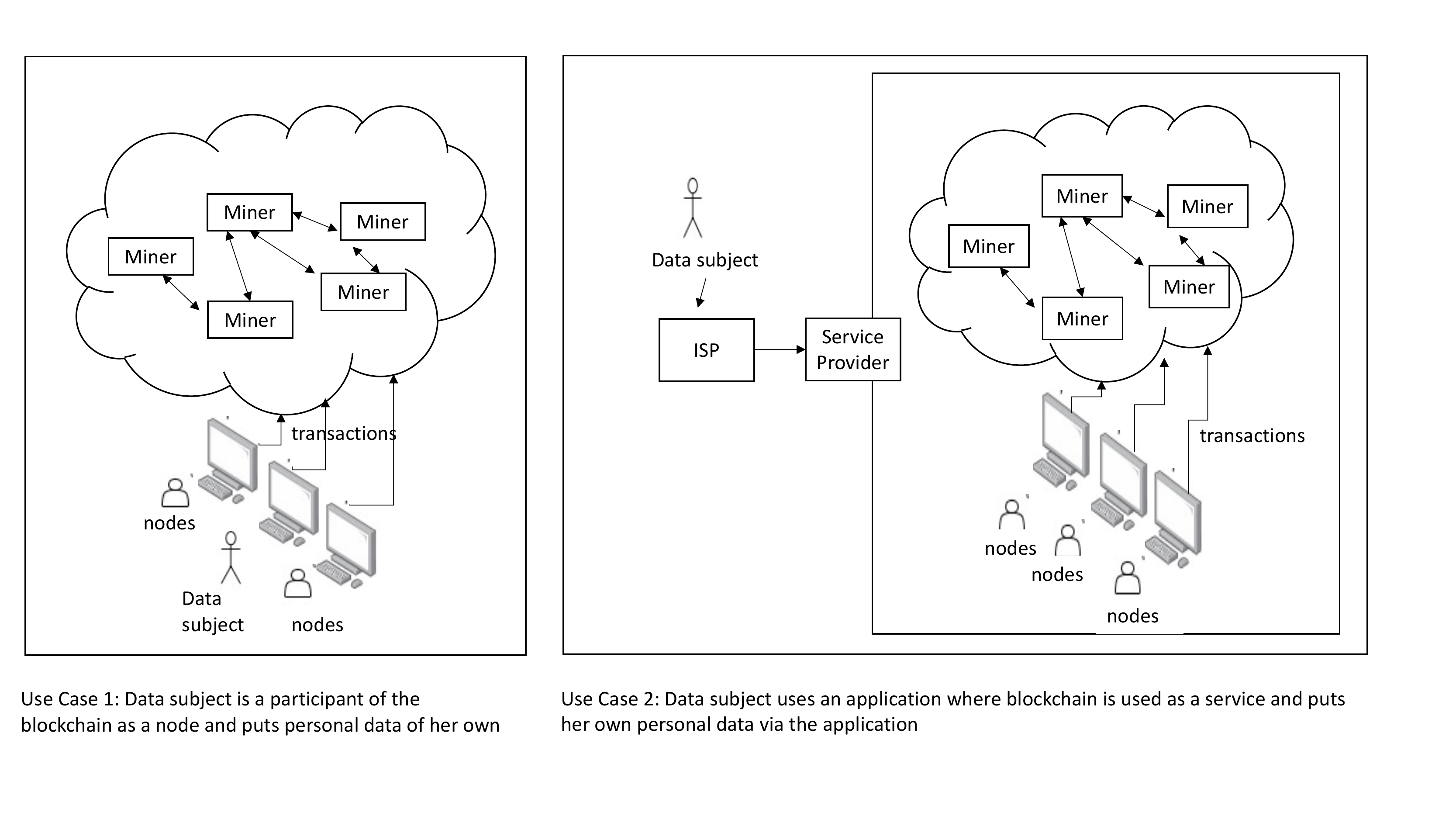}
\caption{Two different use cases of blockchains in terms of roles and responsibilities in the GDPR context}
\label{fig:UCs}
\end{figure}

It is important to highlight here that a vast amount of past research made a over-simplified assumption: personal data is added to the system by their owners (see Use Case 1 in Figure~\ref{fig:UCs}). Past research generally overlooked cases where personal data is added to the system by others on behalf of the data subject, as shown in Use Case 2 in Figure~\ref{fig:UCs}. This largely neglected use case impacts on several aspects including the identification of data controller and the ability of data subjects to exercise their rights given by the GDPR. CNIL slightly covered this use case in one of their reports~\citep{CNIL2018} and explained that a natural person who sells or buys Bitcoin on their own account is not considered as a data controller since s/he is entering personal data outside a professional or commercial activity (in accordance with the principle of domestic exception provided for in Article~2 of the GDPR). On the other hand, if they carry out these transactions on behalf of other natural persons inside their professional or commercial activity, the CNIL report states that they may be considered controllers. This definition is unhelpful to understand the use case where a natural person enters information of others in the course of a purely personal or household activity, with no connection to a professional or commercial activity. The second use case helps indicate that researchers should be more comprehensive when considering the GDPR compliance issue of public bloclchains since they can change many aspects drastically. In addition to the above two use cases, more could be constructed, especially for hybrid blockchain systems where multiple blockchain systems and non-blockchain systems intermingle, leading to even more complicated relationships and behaviours of different entities in the larger ecosystem.

\citeauthor{ibanez2018blockchains} put forward the closest approach to cover this variation and discussed GDPR compliance of blockchains in three most common scenarios of how a data subject interacts with a blockchain~\citep{ibanez2018blockchains}. They evaluated the role assignment, and enumerated applicable strategies for data minimisation and the RBTF and amendments for each scenario individually. Their scenarios are given as follows: an individual interacts directly with a permissionless blockchain; applications that use permissionless blockchains as the backend; and permissioned blockchains. We do not cover the third scenario in this SLR since permissioned blockchains are out of the scope of our work. For the first scenario, \citeauthor{ibanez2018blockchains} argued that it is not possible to identify a controller and the onus of compliance might need to be put on users themselves. For the second scenario, the owners of the intermediary application were argued as data controllers. While this separation is valuable, it falls short of addressing the situation where people put data of others into the blockchain system. This use case reflects a process that completely predates the person being able to give or withdraw consent and exercise other rights. Considering the immutable nature of blockchains, such a use case raises significant privacy concerns. More interestingly, we can also speculate that if a person can or should make irrevocable decisions on their future self's behalf, echoing the importance of the RTBF if the initial decision turns out to be wrong in future. We believe that impossibility of exercising the right to withdraw consent or the RTBF leads to major technical, philosophical, legal, ethical, moral and societal concerns and challenges. Limiting the types of data (e.g., disallowing sensitive personal data or illegal materials) or developing solutions that are not harmful for kids, younger people and other vulnerable people could be argued as responsibilities of service providers, who then should assume the role of data controllers.

\section{Conclusion}
\label{sec:conclusion}

We conducted an SLR of 114 research papers to produce a more in-depth and up-to-date understanding of the current research literature on GDPR compatibility of public blockchain systems. Due to a variety of challenges discussed and solutions proposed for such challenges, as well as the different interpretations regarding complying issues and overlapping concepts, it is challenging to categorise discussions in the literature in a systematic way. However, our study recommends six broad categories to consider while evaluating the GDPR compatibility of public blockchain systems: categories of personal data in blockchains; roles and responsibilities in blockchains; technical solutions proposed for protection of personal data; discussions regarding how to exercise data subject rights guaranteed by the GDPR; discussions regarding lawfulness, fairness and transparency; and discussions regarding GDPR principles and other related elements.

Our contributions are four-fold. Firstly, the variety of personal data on a blockchain has been recognised in our study, which is generally overlooked in similar past studies and we identified the challenges and proposed solutions for each category accordingly. Secondly, we identified and reported limitations and consequences of solutions proposed in the literature as well as contradicting opinions, allowing to get a better idea about the current status of the research in this area. Thirdly, we provided perspectives from both the network and application domain perspectives and categorised discussions accordingly, which have not been covered comprehensively in past studies. Finally, we include a broad scope of GDPR elements in our study and provided a much deeper and concise representation of the literature, identifying not only the conflicts (negative aspects) but also the compliance (positive) aspects between public blockchains and the GDPR.

While our study was comprehensive in means of identifying studies regarding GDPR compatibility aspects of public blockchain systems, we would like to acknowledge some limitations. The first limitation is about the databases we used. Due to limited search facilities, Google Scholar can provide the high number of candidate data items if searched into fulltext, however due to the large number of items and the low hit rate we decided to search into titles only, which might have led us to miss some relevant studies. Despite this limitation, as we also used two additional scientific databases (Scopus and WoS), two most widely used databases for SLRs with a very comprehensive coverage of research papers, we believe that the research papers we covered still represent a very good representation of the current research. Secondly, as stated before, we set our scope to be limited to public blockchains only because it is relatively straightforward to comply with the GDPR for private and consortium blockchains. Therefore, our SLR does not cover solutions proposed for GDPR compatibility that are based on or for private and consortium blockchains only. To address the second limitation, a new SLR can be conducted to focus on private and consortium blockchains, and compare the results with what are reported in this paper for public blockchains.

\bibliographystyle{elsarticle-harv} 
\bibliography{main}

\begin{thebibliography}{135}
\expandafter\ifx\csname natexlab\endcsname\relax\def\natexlab#1{#1}\fi
\providecommand{\url}[1]{\texttt{#1}}
\providecommand{\href}[2]{#2}
\providecommand{\path}[1]{#1}
\providecommand{\DOIprefix}{doi:}
\providecommand{\ArXivprefix}{arXiv:}
\providecommand{\URLprefix}{URL: }
\providecommand{\Pubmedprefix}{pmid:}
\providecommand{\doi}[1]{\href{http://dx.doi.org/#1}{\path{#1}}}
\providecommand{\Pubmed}[1]{\href{pmid:#1}{\path{#1}}}
\providecommand{\bibinfo}[2]{#2}
\ifx\xfnm\relax \def\xfnm[#1]{\unskip,\space#1}\fi
\bibitem[{Ahmed et~al.(2020)Ahmed, Yildirim, Nowostawski, Abomhara, Ramachandra
  and Elezaj}]{ahmed2020towards}
\bibinfo{author}{Ahmed, J.}, \bibinfo{author}{Yildirim, S.},
  \bibinfo{author}{Nowostawski, M.}, \bibinfo{author}{Abomhara, M.},
  \bibinfo{author}{Ramachandra, R.}, \bibinfo{author}{Elezaj, O.},
  \bibinfo{year}{2020}.
\newblock \bibinfo{title}{Towards blockchain-based {GDPR}-compliant online
  social networks: Challenges, opportunities and way forward}, in:
  \bibinfo{booktitle}{Advances in Information and Communication: Proceedings of
  the 2020 Future of Information and Communication Conference (FICC), Volume
  1}, \bibinfo{publisher}{Springer}. pp. \bibinfo{pages}{113--129}.
\newblock \DOIprefix\doi{10.1007/978-3-030-39445-5_10}.
\bibitem[{Al-Abdullah et~al.(2020)Al-Abdullah, Alsmadi, {AlAbdullah} and
  Farkas}]{al2020designing}
\bibinfo{author}{Al-Abdullah, M.}, \bibinfo{author}{Alsmadi, I.},
  \bibinfo{author}{{AlAbdullah}, R.}, \bibinfo{author}{Farkas, B.},
  \bibinfo{year}{2020}.
\newblock \bibinfo{title}{Designing privacy-friendly data repositories: a
  framework for a blockchain that follows the {GDPR}}.
\newblock \bibinfo{journal}{Digital Policy, Regulation and Governance}
  \bibinfo{volume}{22}, \bibinfo{pages}{389--411}.
\newblock \DOIprefix\doi{10.1108/DPRG-04-2020-0050}.
\bibitem[{Alessi et~al.(2019)Alessi, Camillò, Giangreco, Matera, Pino and
  Storelli}]{alessi2019decentralized}
\bibinfo{author}{Alessi, M.}, \bibinfo{author}{Camillò, A.},
  \bibinfo{author}{Giangreco, E.}, \bibinfo{author}{Matera, M.},
  \bibinfo{author}{Pino, S.}, \bibinfo{author}{Storelli, D.},
  \bibinfo{year}{2019}.
\newblock \bibinfo{title}{A decentralized personal data store based on
  {Ethereum}: Towards {GDPR} compliance}.
\newblock \bibinfo{journal}{Journal of Communications Software and Systems}
  \bibinfo{volume}{15}, \bibinfo{pages}{79--88}.
\newblock \DOIprefix\doi{10.24138/jcomss.v15i2.696}.
\bibitem[{Alkouz et~al.(2019)Alkouz, HaiYasien, Alarabeyyat, Samara and
  Al-Saleh}]{alkouz2019eppr}
\bibinfo{author}{Alkouz, A.}, \bibinfo{author}{HaiYasien, A.},
  \bibinfo{author}{Alarabeyyat, A.}, \bibinfo{author}{Samara, K.},
  \bibinfo{author}{Al-Saleh, M.}, \bibinfo{year}{2019}.
\newblock \bibinfo{title}{{EPPR}: Using blockchain for sharing educational
  records}, in: \bibinfo{booktitle}{2019 Sixth HCT Information Technology
  Trends (ITT)}, \bibinfo{publisher}{IEEE}. pp. \bibinfo{pages}{234--239}.
\newblock \DOIprefix\doi{10.1109/ITT48889.2019.9075126}.
\bibitem[{Arisi and Guarda(2020)}]{arisi2020blockchain}
\bibinfo{author}{Arisi, M.}, \bibinfo{author}{Guarda, P.},
  \bibinfo{year}{2020}.
\newblock \bibinfo{title}{Blockchain and {eHealth}: seeking compliance with the
  {General Data Protection Regulation}}.
\newblock \bibinfo{journal}{BioLaw Journal: Rivista di BioDiritto} ,
  \bibinfo{pages}{477--496}\DOIprefix\doi{10.15168/2284-4503-673}.
\bibitem[{{Article~29 Data Protection Working
  Party}(2014)}]{Article29WP2014OpinionAnonymisation}
\bibinfo{author}{{Article~29 Data Protection Working Party}},
  \bibinfo{year}{2014}.
\newblock \bibinfo{title}{{Opinion} 05/2014 on Anonymisation Techniques}.
\newblock \bibinfo{type}{Official Opinion}. European Commission.
\newblock \URLprefix
  \url{https://ec.europa.eu/justice/article-29/documentation/opinion-recommendation/files/2014/wp216_en.pdf}.
\bibitem[{Ateniese et~al.(2017)Ateniese, Magri, Venturi and
  Andrade}]{ateniese2017redactable}
\bibinfo{author}{Ateniese, G.}, \bibinfo{author}{Magri, B.},
  \bibinfo{author}{Venturi, D.}, \bibinfo{author}{Andrade, E.},
  \bibinfo{year}{2017}.
\newblock \bibinfo{title}{Redactable blockchain -- or -- rewriting history in
  {Bitcoin} and friends}, in: \bibinfo{booktitle}{Proceedings of the 2017 IEEE
  European Symposium on Security and Privacy}, \bibinfo{publisher}{IEEE}. pp.
  \bibinfo{pages}{111--126}.
\newblock \DOIprefix\doi{10.1109/EuroSP.2017.37}.
\bibitem[{Barati et~al.(2021)Barati, Aujla, Llanos, Duodu, Rana, Carr and
  Rajan}]{BaratiPrivacy2021}
\bibinfo{author}{Barati, M.}, \bibinfo{author}{Aujla, G.S.},
  \bibinfo{author}{Llanos, J.T.}, \bibinfo{author}{Duodu, K.A.},
  \bibinfo{author}{Rana, O.F.}, \bibinfo{author}{Carr, M.},
  \bibinfo{author}{Rajan, R.}, \bibinfo{year}{2021}.
\newblock \bibinfo{title}{Privacy-aware cloud auditing for gdpr compliance
  verification in online healthcare}.
\newblock \bibinfo{journal}{IEEE Transactions on Industrial Informatics}
  \bibinfo{volume}{18}, \bibinfo{pages}{4808--4819}.
\newblock \DOIprefix\doi{10.1109/TII.2021.3100152}.
\bibitem[{Barati et~al.(2019)Barati, Petri and Rana}]{barati2019developing}
\bibinfo{author}{Barati, M.}, \bibinfo{author}{Petri, I.},
  \bibinfo{author}{Rana, O.F.}, \bibinfo{year}{2019}.
\newblock \bibinfo{title}{Developing {GDPR} compliant user data policies for
  {Internet of Things}}, in: \bibinfo{booktitle}{Proceedings of the 12th
  IEEE/ACM International Conference on Utility and Cloud Computing},
  \bibinfo{publisher}{ACM}. pp. \bibinfo{pages}{133--141}.
\newblock \DOIprefix\doi{10.1145/3344341.3368812}.
\bibitem[{Barati and Rana(2019)}]{barati2019enhancing}
\bibinfo{author}{Barati, M.}, \bibinfo{author}{Rana, O.}, \bibinfo{year}{2019}.
\newblock \bibinfo{title}{Enhancing user privacy in {IoT}: Integration of
  {GDPR} and blockchain}, in: \bibinfo{booktitle}{Blockchain and Trustworthy
  Systems: First International Conference, BlockSys 2019, Guangzhou, China,
  December 7–8, 2019, Proceedings}, \bibinfo{publisher}{Springer}. pp.
  \bibinfo{pages}{322--335}.
\newblock \DOIprefix\doi{10.1007/978-981-15-2777-7_26}.
\bibitem[{Barati and Rana(2021)}]{barati2021privacy}
\bibinfo{author}{Barati, M.}, \bibinfo{author}{Rana, O.}, \bibinfo{year}{2021}.
\newblock \bibinfo{title}{Privacy-aware cloud ecosystems: Architecture and
  performance}.
\newblock \bibinfo{journal}{Concurrency and Computation: Practice and
  Experience} \bibinfo{volume}{33}.
\newblock \DOIprefix\doi{10.1002/cpe.5852}.
\bibitem[{Barati et~al.(2020)Barati, Rana, Petri and
  Theodorakopoulos}]{barati2020gdpr}
\bibinfo{author}{Barati, M.}, \bibinfo{author}{Rana, O.},
  \bibinfo{author}{Petri, I.}, \bibinfo{author}{Theodorakopoulos, G.},
  \bibinfo{year}{2020}.
\newblock \bibinfo{title}{{GDPR} compliance verification in {Internet of
  Things}}.
\newblock \bibinfo{journal}{IEEE Access} \bibinfo{volume}{8},
  \bibinfo{pages}{119697--119709}.
\newblock \DOIprefix\doi{10.1109/ACCESS.2020.3005509}.
\bibitem[{Berberich and Steiner(2016)}]{berberich2016blockchain}
\bibinfo{author}{Berberich, M.}, \bibinfo{author}{Steiner, M.},
  \bibinfo{year}{2016}.
\newblock \bibinfo{title}{Practitioner's corner $\bullet$ blockchain technology
  and the {GDPR} -- how to reconcile privacy and distributed ledgers?}
\newblock \bibinfo{journal}{Eur. Data Prot. L. Rev.} \bibinfo{volume}{2},
  \bibinfo{pages}{422--426}.
\newblock \DOIprefix\doi{10.21552/EDPL/2016/3/21}.
\bibitem[{Bernabe et~al.(2019)Bernabe, Canovas, Hernandez-Ramos, Moreno and
  Skarmeta}]{bernabe2019privacy}
\bibinfo{author}{Bernabe, J.B.}, \bibinfo{author}{Canovas, J.L.},
  \bibinfo{author}{Hernandez-Ramos, J.L.}, \bibinfo{author}{Moreno, R.T.},
  \bibinfo{author}{Skarmeta, A.}, \bibinfo{year}{2019}.
\newblock \bibinfo{title}{Privacy-preserving solutions for blockchain: Review
  and challenges}.
\newblock \bibinfo{journal}{IEEE Access} \bibinfo{volume}{7},
  \bibinfo{pages}{164908--164940}.
\newblock \DOIprefix\doi{10.1109/ACCESS.2019.2950872}.
\bibitem[{Biryukov et~al.(2014)Biryukov, Khovratovich and
  Pustogarov}]{biryukov2014deanonymisation}
\bibinfo{author}{Biryukov, A.}, \bibinfo{author}{Khovratovich, D.},
  \bibinfo{author}{Pustogarov, I.}, \bibinfo{year}{2014}.
\newblock \bibinfo{title}{Deanonymisation of clients in {Bitcoin} {P2P}
  network}, in: \bibinfo{booktitle}{Proceedings of the 2014 ACM SIGSAC
  Conference on Computer and Communications Security},
  \bibinfo{publisher}{ACM}. pp. \bibinfo{pages}{15--29}.
\newblock \DOIprefix\doi{10.1145/2660267.2660379}.
\bibitem[{{Bitcoin Wiki}(2018)}]{bitcoinwiki_lightweight}
\bibinfo{author}{{Bitcoin Wiki}}, \bibinfo{year}{2018}.
\newblock \bibinfo{title}{Lightweight node}.
\newblock \URLprefix \url{https://en.bitcoin.it/wiki/Lightweight_node}.
\bibitem[{{Bitcoin Wiki}(2022)}]{bitcoinwiki_fullnodes}
\bibinfo{author}{{Bitcoin Wiki}}, \bibinfo{year}{2022}.
\newblock \bibinfo{title}{Full node}.
\newblock \URLprefix \url{https://en.bitcoin.it/wiki/Full_node}.
\bibitem[{Bowles et~al.(2021)Bowles, Webber, Blackledge and
  Vermeulen}]{bowles2021blockchain}
\bibinfo{author}{Bowles, J.}, \bibinfo{author}{Webber, T.},
  \bibinfo{author}{Blackledge, E.}, \bibinfo{author}{Vermeulen, A.},
  \bibinfo{year}{2021}.
\newblock \bibinfo{title}{A blockchain-based healthcare platform for secure
  personalised data sharing}, in: \bibinfo{booktitle}{Public Health and
  Informatics: Proceedings of MIE 2021}, \bibinfo{publisher}{IOS Press}. pp.
  \bibinfo{pages}{208--212}.
\newblock \DOIprefix\doi{10.3233/SHTI210150}.
\bibitem[{Buocz et~al.(2019)Buocz, Ehrke-Rabel, H{\"o}dl and
  Eisenberger}]{buocz2019bitcoin}
\bibinfo{author}{Buocz, T.}, \bibinfo{author}{Ehrke-Rabel, T.},
  \bibinfo{author}{H{\"o}dl, E.}, \bibinfo{author}{Eisenberger, I.},
  \bibinfo{year}{2019}.
\newblock \bibinfo{title}{{Bitcoin} and the {GDPR}: Allocating responsibility
  in distributed networks}.
\newblock \bibinfo{journal}{Computer Law \& Security Review}
  \bibinfo{volume}{35}, \bibinfo{pages}{182--198}.
\newblock \DOIprefix\doi{10.1016/j.clsr.2018.12.003}.
\bibitem[{Campanile et~al.(2021)Campanile, Cantiello, Iacono, Marulli and
  Mastroianni}]{campanile2021risk}
\bibinfo{author}{Campanile, L.}, \bibinfo{author}{Cantiello, P.},
  \bibinfo{author}{Iacono, M.}, \bibinfo{author}{Marulli, F.},
  \bibinfo{author}{Mastroianni, M.}, \bibinfo{year}{2021}.
\newblock \bibinfo{title}{Risk analysis of a {GDPR}-compliant deletion
  technique for consortium blockchains based on pseudonymization}, in:
  \bibinfo{booktitle}{Computational Science and Its Applications -- ICCSA 2021:
  21st International Conference, Cagliari, Italy, September 13--16, 2021,
  Proceedings, Part VIII}, \bibinfo{publisher}{Springer}. pp.
  \bibinfo{pages}{3--14}.
\bibitem[{Campanile et~al.(2020)Campanile, Iacono, Levis, Marulli and
  Mastroianni}]{campanile2020privacy}
\bibinfo{author}{Campanile, L.}, \bibinfo{author}{Iacono, M.},
  \bibinfo{author}{Levis, A.H.}, \bibinfo{author}{Marulli, F.},
  \bibinfo{author}{Mastroianni, M.}, \bibinfo{year}{2020}.
\newblock \bibinfo{title}{Privacy regulations, smart roads, blockchain, and
  liability insurance: Putting technologies to work}.
\newblock \bibinfo{journal}{IEEE Security \& Privacy} \bibinfo{volume}{19},
  \bibinfo{pages}{34--43}.
\newblock \DOIprefix\doi{10.1109/MSEC.2020.3012059}.
\bibitem[{Casino et~al.(2019)Casino, Politou, Alepis and
  Patsakis}]{casino2019immutability}
\bibinfo{author}{Casino, F.}, \bibinfo{author}{Politou, E.},
  \bibinfo{author}{Alepis, E.}, \bibinfo{author}{Patsakis, C.},
  \bibinfo{year}{2019}.
\newblock \bibinfo{title}{Immutability and decentralized storage: An analysis
  of emerging threats}.
\newblock \bibinfo{journal}{IEEE Access} \bibinfo{volume}{8},
  \bibinfo{pages}{4737--4744}.
\newblock \DOIprefix\doi{10.1109/ACCESS.2019.2962017}.
\bibitem[{Chavez et~al.(2020)Chavez, Kendzierskyj, Jahankhani and
  Hosseinian}]{chavez2020securing}
\bibinfo{author}{Chavez, N.}, \bibinfo{author}{Kendzierskyj, S.},
  \bibinfo{author}{Jahankhani, H.}, \bibinfo{author}{Hosseinian, A.},
  \bibinfo{year}{2020}.
\newblock \bibinfo{title}{Securing transparency and governance of organ supply
  chain through blockchain}, in: \bibinfo{booktitle}{Policing in the Era of AI
  and Smart Societies}. \bibinfo{publisher}{Springer}, pp.
  \bibinfo{pages}{97--118}.
\newblock \DOIprefix\doi{10.1007/978-3-030-50613-1_4}.
\bibitem[{{CNIL}(2018)}]{CNIL2018}
\bibinfo{author}{{CNIL}}, \bibinfo{year}{2018}.
\newblock \bibinfo{title}{Solutions for a responsible use of the blockchain in
  the context of personal data}.
\newblock \bibinfo{type}{Technical Report}.
\newblock \URLprefix
  \url{https://www.cnil.fr/en/blockchain-and-gdpr-solutions-responsible-use-blockchain-context-personal-data}.
\bibitem[{D'Agostini(2017)}]{dagostini2017gdpr}
\bibinfo{author}{D'Agostini, N.}, \bibinfo{year}{2017}.
\newblock \bibinfo{title}{Gdpr and blockchain: What does this mean for in-house
  counsel?}
\newblock \bibinfo{journal}{International In-house Counsel Journal}
  \bibinfo{volume}{11}.
\newblock \URLprefix
  \url{https://www.iicj.net/subscribersonly/18october/iicj4oct-privacy-NataliaDAgostini-icare.-australia.pdf}.
\bibitem[{Damian et~al.(2019)Damian, Lazăr, Vișoiu, Romanescu and
  Alboaie}]{damian2019applying}
\bibinfo{author}{Damian, C.}, \bibinfo{author}{Lazăr, I.},
  \bibinfo{author}{Vișoiu, D.G.}, \bibinfo{author}{Romanescu, S.},
  \bibinfo{author}{Alboaie, L.}, \bibinfo{year}{2019}.
\newblock \bibinfo{title}{Applying blockchain technologies in funding of
  electrical engineering industry applications}, in:
  \bibinfo{booktitle}{Proceedings of the 2019 International Conference on
  Electromechanical and Energy Systems}, \bibinfo{publisher}{IEEE}.
\newblock \DOIprefix\doi{10.1109/SIELMEN.2019.8905896}.
\bibitem[{Daoui et~al.(2019)Daoui, Fleinert-Jensen and
  Lemperiere}]{daoui2019gdpr}
\bibinfo{author}{Daoui, S.}, \bibinfo{author}{Fleinert-Jensen, T.},
  \bibinfo{author}{Lemperiere, M.}, \bibinfo{year}{2019}.
\newblock \bibinfo{title}{{GDPR}, blockchain and the {French Data Protection
  Authority}: Many answers but some remaining questions}.
\newblock \bibinfo{journal}{Stanford Journal of Blockchain Law \& Policy}
  \bibinfo{volume}{2}, \bibinfo{pages}{240--251}.
\newblock \URLprefix
  \url{https://stanford-jblp.pubpub.org/pub/gdpr-blockchain-france}.
\bibitem[{Daud{\'e}n-Esmel et~al.(2021)Daud{\'e}n-Esmel, Castell{\`a}-Roca,
  Viejo and Domingo-Ferrer}]{dauden2021lightweight}
\bibinfo{author}{Daud{\'e}n-Esmel, C.}, \bibinfo{author}{Castell{\`a}-Roca,
  J.}, \bibinfo{author}{Viejo, A.}, \bibinfo{author}{Domingo-Ferrer, J.},
  \bibinfo{year}{2021}.
\newblock \bibinfo{title}{Lightweight blockchain-based platform for
  {GDPR}-compliant personal data management}, in:
  \bibinfo{booktitle}{Proceedings of the 2021 IEEE 5th International Conference
  on Cryptography, Security and Privacy}, \bibinfo{publisher}{IEEE}. pp.
  \bibinfo{pages}{68--73}.
\newblock \DOIprefix\doi{10.1109/CSP51677.2021.9357602}.
\bibitem[{{de~la~Cruz}(2019)}]{cruz2019privacy}
\bibinfo{author}{{de~la~Cruz}, R.}, \bibinfo{year}{2019}.
\newblock \bibinfo{title}{Privacy laws in the blockchain environment}.
\newblock \bibinfo{journal}{Annals of Emerging Technologies in Computing
  (AETiC)} \bibinfo{volume}{3}, \bibinfo{pages}{34--44}.
\newblock \DOIprefix\doi{10.33166/AETiC.2019.05.005}.
\bibitem[{{de~Sousa} and Pinto(2018)}]{deSousa2018feasibility}
\bibinfo{author}{{de~Sousa}, H.R.}, \bibinfo{author}{Pinto, A.},
  \bibinfo{year}{2018}.
\newblock \bibinfo{title}{On the feasibility of blockchain for online surveys
  with reputation and informed consent support}, in:
  \bibinfo{booktitle}{Ambient Intelligence -- Software and Applications --, 9th
  International Symposium on Ambient Intelligence},
  \bibinfo{publisher}{Springer}. pp. \bibinfo{pages}{314--322}.
\newblock \DOIprefix\doi{10.1007/978-3-030-01746-0_37}.
\bibitem[{{de~Sousa} and Pinto(2019)}]{deSousa2019blockchain}
\bibinfo{author}{{de~Sousa}, H.R.}, \bibinfo{author}{Pinto, A.},
  \bibinfo{year}{2019}.
\newblock \bibinfo{title}{Blockchain based informed consent with reputation
  support}, in: \bibinfo{booktitle}{Blockchain and Applications: International
  Congress}, \bibinfo{publisher}{Springer}. pp. \bibinfo{pages}{54--61}.
\newblock \DOIprefix\doi{10.1007/978-3-030-23813-1_7}.
\bibitem[{Dejanovic et~al.(2019)Dejanovic, Marjanovic, Lendak and
  Erdeljan}]{dejanovic2019using}
\bibinfo{author}{Dejanovic, S.}, \bibinfo{author}{Marjanovic, J.},
  \bibinfo{author}{Lendak, I.}, \bibinfo{author}{Erdeljan, A.},
  \bibinfo{year}{2019}.
\newblock \bibinfo{title}{Using blockchain to decentralize and protect user
  privacy in compliance with {GDPR}}, in: \bibinfo{booktitle}{ICIST 2019
  Proceedings}, \bibinfo{publisher}{Information Society of Serbia}. pp.
  \bibinfo{pages}{171--173}.
\newblock \URLprefix
  \url{http://www.eventiotic.com/eventiotic/files/Papers/URL/a26839e2-6bcd-4194-b481-5cc3e09d892a.pdf}.
\bibitem[{Dekhuijzen(2019)}]{dekhuijzen2019call}
\bibinfo{author}{Dekhuijzen, A.E.}, \bibinfo{year}{2019}.
\newblock \bibinfo{title}{Call for action on the {EDPB} to provide guidance
  concerning {GDPR} and blockchain: Is public blockchain sustainable under the
  {GDPR}?}
\newblock \bibinfo{journal}{Computer Law Review International}
  \bibinfo{volume}{20}, \bibinfo{pages}{33--36}.
\newblock \DOIprefix\doi{10.9785/cri-2019-200202}.
\bibitem[{Duarte(2019)}]{duarte2019introduction}
\bibinfo{author}{Duarte, D.G.}, \bibinfo{year}{2019}.
\newblock \bibinfo{title}{An introduction to {Blockchain} technology from a
  legal perspective and its tensions with the {GDPR}}.
\newblock \bibinfo{journal}{Cyberlaw Journal of the Cyberlaw Research Centre of
  the University of Lisbon School of Law} \DOIprefix\doi{10.2139/ssrn.3545331}.
\bibitem[{Dutta et~al.(2020)Dutta, Das, Dey and
  Bhattacharya}]{dutta2020blockchain}
\bibinfo{author}{Dutta, R.}, \bibinfo{author}{Das, A.}, \bibinfo{author}{Dey,
  A.}, \bibinfo{author}{Bhattacharya, S.}, \bibinfo{year}{2020}.
\newblock \bibinfo{title}{Blockchain vs {GDPR} in collaborative data
  governance}, in: \bibinfo{booktitle}{Cooperative Design, Visualization, and
  Engineering: 17th International Conference, CDVE 2020, Bangkok, Thailand,
  October 25--28, 2020, Proceedings}, \bibinfo{publisher}{Springer}. pp.
  \bibinfo{pages}{81--92}.
\newblock \DOIprefix\doi{10.1007/978-3-030-60816-3}.
\bibitem[{Eichler et~al.(2018)Eichler, Jongerius, McMullen, Naegele, Steininger
  and Wagner}]{eichler2018blockchain}
\bibinfo{author}{Eichler, N.}, \bibinfo{author}{Jongerius, S.},
  \bibinfo{author}{McMullen, G.}, \bibinfo{author}{Naegele, O.},
  \bibinfo{author}{Steininger, L.}, \bibinfo{author}{Wagner, K.},
  \bibinfo{year}{2018}.
\newblock \bibinfo{title}{Blockchain, data protection, and the {GDPR}}.
\newblock \bibinfo{type}{Technical Report} \bibinfo{number}{VR 36105 B
  27/661/52176}. German Blockchain Association (Bundesblock).
\newblock \URLprefix
  \url{https://www.crowdfundinsider.com/wp-content/uploads/2018/06/GDPR_Position_Paper_v1.0.pdf}.
\bibitem[{El-Gazzar and Stendal(2020)}]{el2020examining}
\bibinfo{author}{El-Gazzar, R.}, \bibinfo{author}{Stendal, K.},
  \bibinfo{year}{2020}.
\newblock \bibinfo{title}{Examining how {GDPR} challenges emerging
  technologies}.
\newblock \bibinfo{journal}{Journal of Information Policy}
  \bibinfo{volume}{10}, \bibinfo{pages}{237--275}.
\newblock \DOIprefix\doi{10.5325/jinfopoli.10.2020.0237}.
\bibitem[{Erbguth(2019)}]{erbguth2019five}
\bibinfo{author}{Erbguth, J.}, \bibinfo{year}{2019}.
\newblock \bibinfo{title}{Five ways to {GDPR}-compliant use of blockchains}.
\newblock \bibinfo{journal}{European Data Protection Law Review}
  \bibinfo{volume}{5}, \bibinfo{pages}{427--433}.
\newblock \DOIprefix\doi{10.21552/edpl/2019/3/19}.
\bibitem[{{European Parliament}(2015)}]{EU_DPD1995}
\bibinfo{author}{{European Parliament}}, \bibinfo{year}{2015}.
\newblock \bibinfo{title}{{Directive} 95/46/{EC} of the {European Parliament}
  and of the {Council} of 24 {October} 1995 on the protection of individuals
  with regard to the processing of personal data and on the free movement of
  such data}.
\newblock \bibinfo{howpublished}{Official Journal of the European Union No L
  281/31}.
\newblock \URLprefix
  \url{https://eur-lex.europa.eu/legal-content/EN/TXT/?uri=celex\%3A31995L0046}.
\bibitem[{{European Parliament}(2016)}]{GDPR}
\bibinfo{author}{{European Parliament}}, \bibinfo{year}{2016}.
\newblock \bibinfo{title}{{Regulation} ({EU}) (2016) 2016/679 of the {European
  Parliament} and of the {Council} of 27 {April} on the protection of natural
  persons with regard to the processing of personal data and on the free
  movement of such data, and repealing {Directive} 95/46/{EC} ({General Data
  Protection Regulation})}.
\newblock \bibinfo{howpublished}{Official Journal of the European Union 59(L
  119)}.
\newblock \URLprefix \url{https://eur-lex.europa.eu/eli/reg/2016/679/oj}.
\bibitem[{Faber et~al.(2019)Faber, Michelet, Weidmann, Mukkamala and
  Vatrapu}]{faber2019bpdims}
\bibinfo{author}{Faber, B.}, \bibinfo{author}{Michelet, G.C.},
  \bibinfo{author}{Weidmann, N.}, \bibinfo{author}{Mukkamala, R.R.},
  \bibinfo{author}{Vatrapu, R.}, \bibinfo{year}{2019}.
\newblock \bibinfo{title}{{BPDIMS}: A blockchain-based personal data and
  identity management system}, in: \bibinfo{booktitle}{Proceedings of the 52nd
  Hawaii International Conference on System Sciences},
  \bibinfo{publisher}{University of Hawai`i at Mānoa, USA}. pp.
  \bibinfo{pages}{6855--6864}.
\newblock \URLprefix \url{http://hdl.handle.net/10125/60121}.
\bibitem[{Fabiano(2017a)}]{Fabiano2018a}
\bibinfo{author}{Fabiano, N.}, \bibinfo{year}{2017}a.
\newblock \bibinfo{title}{{Internet of Things} and blockchain: Legal issues and
  privacy. the challenge for a privacy standard}, in:
  \bibinfo{booktitle}{Proceedings of the 2017 IEEE International Conference on
  Internet of Things, IEEE Green Computing and Communications, IEEE Cyber,
  Physical and Social Computing, IEEE Smart Data}, \bibinfo{publisher}{IEEE}.
  pp. \bibinfo{pages}{727--734}.
\newblock \DOIprefix\doi{10.1109/iThings-GreenCom-CPSCom-SmartData.2017.112}.
\bibitem[{Fabiano(2017b)}]{fabiano2017IoT}
\bibinfo{author}{Fabiano, N.}, \bibinfo{year}{2017}b.
\newblock \bibinfo{title}{The {Internet of Things} ecosystem: The blockchain
  and privacy issues. the challenge for a global privacy standard}, in:
  \bibinfo{booktitle}{Proceedings of the 2017 International Conference on
  Internet of Things for the Global Community}, \bibinfo{publisher}{IEEE}.
\newblock \DOIprefix\doi{10.1109/IoTGC.2017.8008970}.
\bibitem[{Fabiano(2018)}]{fabiano2018blockchain}
\bibinfo{author}{Fabiano, N.}, \bibinfo{year}{2018}.
\newblock \bibinfo{title}{Blockchain and data protection: The value of personal
  data}.
\newblock \bibinfo{journal}{Journal of Systemics, Cybernetics and Informatics}
  \bibinfo{volume}{6}, \bibinfo{pages}{112--115}.
\newblock \URLprefix
  \url{http://www.iiisci.org/journal/sci/FullText.asp?var=&id=ZA165NO18}.
\bibitem[{Farshid et~al.(2019)Farshid, Reitz and
  Ro{\ss}bach}]{farshid2019design}
\bibinfo{author}{Farshid, S.}, \bibinfo{author}{Reitz, A.},
  \bibinfo{author}{Ro{\ss}bach, P.}, \bibinfo{year}{2019}.
\newblock \bibinfo{title}{Design of a forgetting blockchain: A possible way to
  accomplish {GDPR} compatibility}, in: \bibinfo{booktitle}{Proceedings of the
  52nd Hawaii International Conference on System Sciences},
  \bibinfo{publisher}{University of Hawai`i at Mānoa, USA}. pp.
  \bibinfo{pages}{7087--7095}.
\newblock \URLprefix \url{https://hdl.handle.net/10125/60145}.
\bibitem[{Ferrari et~al.(2018)Ferrari, Quintais, Giannopoulou and
  Bodo}]{Ferrari2018}
\bibinfo{author}{Ferrari, V.}, \bibinfo{author}{Quintais, J.P.},
  \bibinfo{author}{Giannopoulou, A.}, \bibinfo{author}{Bodo, B.},
  \bibinfo{year}{2018}.
\newblock \bibinfo{title}{{EU Blockchain Observatory and Forum} {Workshop} on
  {GDPR}, data policy and compliance}.
\newblock \bibinfo{type}{Research Nodes} \bibinfo{number}{2018/1}. Blockchain
  \& Society Policy Research Lab, Institute for Information Law, University of
  Amsterdam.
\newblock \URLprefix
  \url{https://hdl.handle.net/11245.1/2c8359fb-8ced-4c17-8168-78e18bc0db53}.
\bibitem[{Finck(2018)}]{Finck2018a}
\bibinfo{author}{Finck, M.}, \bibinfo{year}{2018}.
\newblock \bibinfo{title}{Blockchains and data protection in the {European
  Union}}.
\newblock \bibinfo{journal}{European Data Protection Law Review}
  \bibinfo{volume}{4}, \bibinfo{pages}{17--35}.
\newblock \DOIprefix\doi{10.21552/edpl/2018/1/6}.
\bibitem[{G{\'e}raud et~al.(2017)G{\'e}raud, Naccache and
  Ro{\c{s}}ie}]{geraud2017twisting}
\bibinfo{author}{G{\'e}raud, R.}, \bibinfo{author}{Naccache, D.},
  \bibinfo{author}{Ro{\c{s}}ie, R.}, \bibinfo{year}{2017}.
\newblock \bibinfo{title}{Twisting lattice and graph techniques to compress
  transactional ledgers}, in: \bibinfo{booktitle}{Security and Privacy in
  Communication Systems: 13th International Conference, SecureComm 2017,
  Niagara Falls, ON, Canada, October 22–25, 2017, Proceedings},
  \bibinfo{publisher}{Springer}. pp. \bibinfo{pages}{108--127}.
\newblock \DOIprefix\doi{10.1007/978-3-319-78813-5_6}.
\bibitem[{Giannopoulou(2020)}]{giannopoulou2020data}
\bibinfo{author}{Giannopoulou, A.}, \bibinfo{year}{2020}.
\newblock \bibinfo{title}{Data protection compliance challenges for
  self-sovereign identity}, in: \bibinfo{booktitle}{Blockchain and
  Applications: 2nd International Congress}, \bibinfo{publisher}{Springer}. pp.
  \bibinfo{pages}{91--100}.
\newblock \DOIprefix\doi{10.1007/978-3-030-52535-4_10}.
\bibitem[{Giannopoulou(2021)}]{giannopoulou2021putting}
\bibinfo{author}{Giannopoulou, A.}, \bibinfo{year}{2021}.
\newblock \bibinfo{title}{Putting data protection by design on the blockchain}.
\newblock \bibinfo{journal}{European Data Protection Law Review}
  \bibinfo{volume}{7}, \bibinfo{pages}{388--399}.
\newblock \DOIprefix\doi{10.21552/edpl/2021/3/7}.
\bibitem[{Giannopoulou and Ferrari(2018)}]{giannopoulou2018distributed}
\bibinfo{author}{Giannopoulou, A.}, \bibinfo{author}{Ferrari, V.},
  \bibinfo{year}{2018}.
\newblock \bibinfo{title}{Distributed data protection and liability on
  blockchains}, in: \bibinfo{booktitle}{Internet Science: INSCI 2018
  International Workshops, St.~Petersburg, Russia, October 24--26, 2018,
  Revised Selected Papers}, \bibinfo{publisher}{Springer}. pp.
  \bibinfo{pages}{203--211}.
\newblock \DOIprefix\doi{10.1007/978-3-030-17705-8_17}.
\bibitem[{Giordanengo(2019)}]{giordanengo2019possible}
\bibinfo{author}{Giordanengo, A.}, \bibinfo{year}{2019}.
\newblock \bibinfo{title}{Possible usages of smart contracts (blockchain) in
  healthcare and why no one is using them}, in: \bibinfo{booktitle}{MEDINFO
  2019: Health and Wellbeing e-Networks for All}. \bibinfo{publisher}{IOS
  Press}, pp. \bibinfo{pages}{596--600}.
\newblock \DOIprefix\doi{10.3233/SHTI190292}.
\bibitem[{Giordano(2021)}]{giordano2021blockchain}
\bibinfo{author}{Giordano, M.T.}, \bibinfo{year}{2021}.
\newblock \bibinfo{title}{Blockchain and the {GDPR}: New challenges for privacy
  and security}, in: \bibinfo{booktitle}{Blockchain, Law and Governance}.
  \bibinfo{publisher}{Springer}, pp. \bibinfo{pages}{275--286}.
\newblock \DOIprefix\doi{10.1007/978-3-030-52722-8_20}.
\bibitem[{Guggenmos et~al.(2020)Guggenmos, Lockl, Rieger, Wenninger and
  Fridgen}]{guggenmos2020develop}
\bibinfo{author}{Guggenmos, F.}, \bibinfo{author}{Lockl, J.},
  \bibinfo{author}{Rieger, A.}, \bibinfo{author}{Wenninger, A.},
  \bibinfo{author}{Fridgen, G.}, \bibinfo{year}{2020}.
\newblock \bibinfo{title}{How to develop a {GDPR}-compliant blockchain solution
  for cross-organizational workflow management: Evidence from the {German}
  asylum procedure}, in: \bibinfo{booktitle}{Proceedings of the 53rd Hawaii
  International Conference on System Sciences}, \bibinfo{publisher}{University
  of Hawai`i at Mānoa, USA}. pp. \bibinfo{pages}{4023--4032}.
\newblock \URLprefix \url{http://hdl.handle.net/10125/64234}.
\bibitem[{Haque et~al.(2021)Haque, Islam, Hyrynsalmi, Naqvi and
  Smolander}]{haque2021gdpr}
\bibinfo{author}{Haque, A.B.}, \bibinfo{author}{Islam, A.K.M.N.},
  \bibinfo{author}{Hyrynsalmi, S.}, \bibinfo{author}{Naqvi, B.},
  \bibinfo{author}{Smolander, K.}, \bibinfo{year}{2021}.
\newblock \bibinfo{title}{{GDPR} compliant blockchains--a systematic literature
  review}.
\newblock \bibinfo{journal}{IEEE Access} \bibinfo{volume}{9},
  \bibinfo{pages}{50593--50606}.
\newblock \DOIprefix\doi{10.1109/ACCESS.2021.3069877}.
\bibitem[{Hasselgren et~al.(2020)Hasselgren, Wan, Horn, Kralevska, Gligoroski
  and Faxvaag}]{hasselgren2020}
\bibinfo{author}{Hasselgren, A.}, \bibinfo{author}{Wan, P.K.},
  \bibinfo{author}{Horn, M.}, \bibinfo{author}{Kralevska, K.},
  \bibinfo{author}{Gligoroski, D.}, \bibinfo{author}{Faxvaag, A.},
  \bibinfo{year}{2020}.
\newblock \bibinfo{title}{{GDPR} compliance for blockchain applications in
  healthcare}.
\newblock \bibinfo{howpublished}{arXiv:2009.12913 [cs.CR]}.
\newblock \DOIprefix\doi{10.48550/arxiv.2009.12913}.
\bibitem[{Heiss et~al.(2020)Heiss, Ulbricht and Eberhardt}]{heiss2020put}
\bibinfo{author}{Heiss, J.}, \bibinfo{author}{Ulbricht, M.R.},
  \bibinfo{author}{Eberhardt, J.}, \bibinfo{year}{2020}.
\newblock \bibinfo{title}{Put your money where your mouth is -- towards
  blockchain-based consent violation detection}, in:
  \bibinfo{booktitle}{Proceedings of the 2020 International Conference on
  Blockchain and Cryptocurrency}, \bibinfo{publisher}{IEEE}.
\newblock \DOIprefix\doi{10.1109/ICBC48266.2020.9169455}.
\bibitem[{Herian(2018)}]{herian2018regulating}
\bibinfo{author}{Herian, R.}, \bibinfo{year}{2018}.
\newblock \bibinfo{title}{Regulating disruption: Blockchain, {GDPR}, and
  questions of data sovereignty}.
\newblock \bibinfo{journal}{Journal of Internet Law} \bibinfo{volume}{22},
  \bibinfo{pages}{1,8--16}.
\newblock \URLprefix \url{https://www.proquest.com/docview/2089334432}.
\bibitem[{Hillmann et~al.(2020)Hillmann, Kn{\"u}pfer, Heiland and
  Karcher}]{hillmann2020selective}
\bibinfo{author}{Hillmann, P.}, \bibinfo{author}{Kn{\"u}pfer, M.},
  \bibinfo{author}{Heiland, E.}, \bibinfo{author}{Karcher, A.},
  \bibinfo{year}{2020}.
\newblock \bibinfo{title}{Selective deletion in a blockchain}, in:
  \bibinfo{booktitle}{Proceedings of the 40th IEEE International Conference on
  Distributed Computing Systems}, \bibinfo{publisher}{IEEE}. pp.
  \bibinfo{pages}{1249--1256}.
\newblock \DOIprefix\doi{10.1109/ICDCS47774.2020.00160}.
\bibitem[{Hofman et~al.(2019)Hofman, Lemieux, Joo and
  Batista}]{hofman2019margin}
\bibinfo{author}{Hofman, D.}, \bibinfo{author}{Lemieux, V.L.},
  \bibinfo{author}{Joo, A.}, \bibinfo{author}{Batista, D.A.},
  \bibinfo{year}{2019}.
\newblock \bibinfo{title}{``the margin between the edge of the world and
  infinite possibility'': Blockchain, {GDPR} and information governance}.
\newblock \bibinfo{journal}{Records Management Journal} \bibinfo{volume}{29},
  \bibinfo{pages}{240--257}.
\newblock \DOIprefix\doi{10.1108/RMJ-12-2018-0045}.
\bibitem[{{Hungarian National Authority for Data Protection for Data Protection
  and Freedom of Information}(2018)}]{hunagrianDPO}
\bibinfo{author}{{Hungarian National Authority for Data Protection for Data
  Protection and Freedom of Information}}, \bibinfo{year}{2018}.
\newblock \bibinfo{title}{The opinion of the {Hungarian National Authority for
  Data Protection and Freedom of Information} on blockchain technology in the
  context of data protection}.
\newblock \URLprefix
  \url{https://www.naih.hu/files/Blockchain-Opinion-2018-01-29.pdf}.
\bibitem[{Ibáñez et~al.(2018)Ibáñez, O'Hara and
  Simperl}]{ibanez2018blockchains}
\bibinfo{author}{Ibáñez, L.D.}, \bibinfo{author}{O'Hara, K.},
  \bibinfo{author}{Simperl, E.}, \bibinfo{year}{2018}.
\newblock \bibinfo{title}{On Blockchains and the {General Data Protection
  Regulation}}.
\newblock \bibinfo{type}{Project Report}.
\newblock \DOIprefix\doi{https://eprints.soton.ac.uk/422879/}.
\bibitem[{{ICO, UK}(a)}]{ICOLegitimateUse}
\bibinfo{author}{{ICO, UK}}, a.
\newblock \bibinfo{title}{Legitimate interests}.
\newblock \URLprefix
  \url{https://ico.org.uk/for-organisations/guide-to-data-protection/guide-to-the-general-data-protection-regulation-gdpr/lawful-basis-for-processing/legitimate-interests/}.
\bibitem[{{ICO, UK}(b)}]{ICOPortability}
\bibinfo{author}{{ICO, UK}}, b.
\newblock \bibinfo{title}{Right to data portability}.
\newblock \URLprefix
  \url{https://ico.org.uk/for-organisations/guide-to-data-protection/guide-to-the-general-data-protection-regulation-gdpr/individual-rights/right-to-data-portability/}.
\bibitem[{{ICO, UK}(c)}]{ICO_Encryption}
\bibinfo{author}{{ICO, UK}}, c.
\newblock \bibinfo{title}{What is encryption?}
\newblock \URLprefix
  \url{https://ico.org.uk/for-organisations/guide-to-data-protection/guide-to-the-general-data-protection-regulation-gdpr/encryption/what-is-encryption/}.
\bibitem[{{ICO, UK}(2022)}]{ICO_Fairness}
\bibinfo{author}{{ICO, UK}}, \bibinfo{year}{2022}.
\newblock \bibinfo{title}{Principle (a): Lawfulness, fairness and
  transparency}.
\newblock \URLprefix
  \url{https://ico.org.uk/for-organisations/guide-to-data-protection/guide-to-the-general-data-protection-regulation-gdpr/principles/lawfulness-fairness-and-transparency/}.
\bibitem[{{Information Commissioner's Office (ICO), UK}()}]{UK_GDPR}
\bibinfo{author}{{Information Commissioner's Office (ICO), UK}}, .
\newblock \bibinfo{title}{The {UK} {GDPR}}.
\newblock \URLprefix
  \url{https://ico.org.uk/for-organisations/dp-at-the-end-of-the-transition-period/data-protection-and-the-eu-in-detail/the-uk-gdpr/}.
\bibitem[{Jaccard and Tharin(2018)}]{jaccard2018gdpr}
\bibinfo{author}{Jaccard, G.}, \bibinfo{author}{Tharin, A.},
  \bibinfo{year}{2018}.
\newblock \bibinfo{title}{{GDPR} \& blockchain: the {Swiss} take}.
\newblock \bibinfo{journal}{Jusletter IT} \URLprefix
  \url{https://lawded.ch/wp-content/uploads/2019/07/Jusletter-IT_gdpr-blockchain-t_5aebbf8be4_en.pdf}.
\bibitem[{Kadena and Holicza(2018)}]{kadena2018security}
\bibinfo{author}{Kadena, E.}, \bibinfo{author}{Holicza, P.},
  \bibinfo{year}{2018}.
\newblock \bibinfo{title}{Security issues in the blockchain(ed) world}, in:
  \bibinfo{booktitle}{Proceedings of the 18th IEEE International Symposium on
  Computational Intelligence and Informatics}, \bibinfo{publisher}{IEEE}. pp.
  \bibinfo{pages}{211--216}.
\newblock \DOIprefix\doi{10.1109/CINTI.2018.8928212}.
\bibitem[{Kaiser et~al.(2018)Kaiser, Steger, Dorri, Festl, Stocker, Fellmann
  and Kanhere}]{kaiser2018towards}
\bibinfo{author}{Kaiser, C.}, \bibinfo{author}{Steger, M.},
  \bibinfo{author}{Dorri, A.}, \bibinfo{author}{Festl, A.},
  \bibinfo{author}{Stocker, A.}, \bibinfo{author}{Fellmann, M.},
  \bibinfo{author}{Kanhere, S.}, \bibinfo{year}{2018}.
\newblock \bibinfo{title}{Towards a privacy-preserving way of vehicle data
  sharing -- a case for blockchain technology?}, in:
  \bibinfo{booktitle}{Advanced Microsystems for Automotive Applications 2018:
  Smart Systems for Clean, Safe and Shared Road Vehicles (Proceedings of the
  AMAA 2018 Conference)}, \bibinfo{publisher}{Springer}. pp.
  \bibinfo{pages}{111--122}.
\newblock \DOIprefix\doi{10.1007/978-3-319-99762-9_10}.
\bibitem[{Karasek-Wojciechowicz(2021)}]{karasek2021reconciliation}
\bibinfo{author}{Karasek-Wojciechowicz, I.}, \bibinfo{year}{2021}.
\newblock \bibinfo{title}{Reconciliation of anti-money laundering instruments
  and {European} data protection requirements in permissionless blockchain
  spaces}.
\newblock \bibinfo{journal}{Journal of Cybersecurity} \bibinfo{volume}{7}.
\newblock \DOIprefix\doi{10.1093/cybsec/tyab004}.
\bibitem[{Kolan et~al.(2020)Kolan, Tjoa and Kieseberg}]{kolan2020medical}
\bibinfo{author}{Kolan, A.}, \bibinfo{author}{Tjoa, S.},
  \bibinfo{author}{Kieseberg, P.}, \bibinfo{year}{2020}.
\newblock \bibinfo{title}{Medical blockchains and privacy in {Austria} -
  technical and legal aspects}, in: \bibinfo{booktitle}{Proceedings of the 2020
  International Conference on Software Security and Assurance},
  \bibinfo{publisher}{IEEE}.
\newblock \DOIprefix\doi{10.1109/ICSSA51305.2020.00009}.
\bibitem[{Kondova and Erbguth(2020)}]{kondova2020self}
\bibinfo{author}{Kondova, G.}, \bibinfo{author}{Erbguth, J.},
  \bibinfo{year}{2020}.
\newblock \bibinfo{title}{Self-sovereign identity on public blockchains and the
  {GDPR}}, in: \bibinfo{booktitle}{Proceedings of the 35th Annual ACM Symposium
  on Applied Computing}, \bibinfo{publisher}{ACM}. pp.
  \bibinfo{pages}{342--345}.
\newblock \DOIprefix\doi{10.1145/3341105.3374066}.
\bibitem[{Koscina et~al.(2021)Koscina, Lombard-Platet and
  Negri-Ribalta}]{koscina2021blockchain}
\bibinfo{author}{Koscina, M.}, \bibinfo{author}{Lombard-Platet, M.},
  \bibinfo{author}{Negri-Ribalta, C.}, \bibinfo{year}{2021}.
\newblock \bibinfo{title}{A blockchain-based marketplace platform for circular
  economy}, in: \bibinfo{booktitle}{Proceedings of the 36th Annual ACM
  Symposium on Applied Computing}, \bibinfo{publisher}{ACM}. pp.
  \bibinfo{pages}{1746--1749}.
\bibitem[{Kusber et~al.(2020)Kusber, Schwalm, Shamburger and
  Korte}]{Kusber2020}
\bibinfo{author}{Kusber, T.}, \bibinfo{author}{Schwalm, S.},
  \bibinfo{author}{Shamburger, K.}, \bibinfo{author}{Korte, U.},
  \bibinfo{year}{2020}.
\newblock \bibinfo{title}{Criteria for trustworthy digital transactions -
  blockchain/{DLT} between {eI-DAS}, {GDPR}, data and evidence preservation},
  in: \bibinfo{booktitle}{Open Identity Summit 2020},
  \bibinfo{publisher}{Gesellschaft für Informatik e.V.}. pp.
  \bibinfo{pages}{49--60}.
\newblock \DOIprefix\doi{10.18420/ois2020_04}.
\bibitem[{Labancz(2019)}]{labancz4conflict}
\bibinfo{author}{Labancz, A.}, \bibinfo{year}{2019}.
\newblock \bibinfo{title}{The conflict of blockchain and the {EU} {General Data
  Protection Regulation} in the area of business law}, in:
  \bibinfo{booktitle}{Law 4.0 -- Challenges of the Digital Age}.
  \bibinfo{publisher}{Széchenyi István University}, pp.
  \bibinfo{pages}{76--81}.
\bibitem[{Lee et~al.(2019)Lee, Yang, Onik and Kim}]{lee2019modifiable}
\bibinfo{author}{Lee, N.Y.}, \bibinfo{author}{Yang, J.}, \bibinfo{author}{Onik,
  M.M.H.}, \bibinfo{author}{Kim, C.S.}, \bibinfo{year}{2019}.
\newblock \bibinfo{title}{Modifiable public blockchains using truncated hashing
  and sidechains}.
\newblock \bibinfo{journal}{IEEE Access} \bibinfo{volume}{7},
  \bibinfo{pages}{173571--173582}.
\newblock \DOIprefix\doi{10.1109/ACCESS.2019.2956628}.
\bibitem[{Liberati et~al.(2009)Liberati, Altman, Tetzlaff, Mulrow, G{\o}tzsche,
  Ioannidis, Clarke, Devereaux, Kleijnen and Moher}]{liberati2009prisma}
\bibinfo{author}{Liberati, A.}, \bibinfo{author}{Altman, D.G.},
  \bibinfo{author}{Tetzlaff, J.}, \bibinfo{author}{Mulrow, C.},
  \bibinfo{author}{G{\o}tzsche, P.C.}, \bibinfo{author}{Ioannidis, J.P.A.},
  \bibinfo{author}{Clarke, M.}, \bibinfo{author}{Devereaux, P.J.},
  \bibinfo{author}{Kleijnen, J.}, \bibinfo{author}{Moher, D.},
  \bibinfo{year}{2009}.
\newblock \bibinfo{title}{The {PRISMA} statement for reporting systematic
  reviews and meta-analyses of studies that evaluate health care interventions:
  explanation and elaboration}.
\newblock \bibinfo{journal}{Journal of Clinical Epidemiology}
  \bibinfo{volume}{62}.
\newblock \DOIprefix\doi{10.1136/bmj.b2700}.
\bibitem[{Lima(2018)}]{lima2018blockchain}
\bibinfo{author}{Lima, C.}, \bibinfo{year}{2018}.
\newblock \bibinfo{title}{Blockchain {GDPR} Privacy by Design: How
  Decentralized Blockchain Internet will Comply with {GDPR} Data Privacy}.
\newblock \bibinfo{type}{Technical Report}.
\newblock \URLprefix
  \url{https://blockchain.ieee.org/images/files/pdf/blockchain-gdpr-privacy-by-design.pdf}.
\bibitem[{Loukil et~al.(2020)Loukil, Ghedira-Guegan and
  Benharkat}]{loukil2020patriot}
\bibinfo{author}{Loukil, F.}, \bibinfo{author}{Ghedira-Guegan, C.},
  \bibinfo{author}{Benharkat, A.N.}, \bibinfo{year}{2020}.
\newblock \bibinfo{title}{{PATRIoT}: A data sharing platform for {IoT} using a
  service-oriented approach based on blockchain}, in:
  \bibinfo{booktitle}{Service-Oriented Computing: 18th International
  Conference, ICSOC 2020, Dubai, United Arab Emirates, December 14–17, 2020,
  Proceedings}, \bibinfo{publisher}{Springer}. pp. \bibinfo{pages}{121--129}.
\newblock \DOIprefix\doi{10.1007/978-3-030-65310-1_10}.
\bibitem[{Lyons et~al.(2018)Lyons, Courcelas and
  Timsit}]{EUBlockchainObservatoryForum2018report}
\bibinfo{author}{Lyons, T.}, \bibinfo{author}{Courcelas, L.},
  \bibinfo{author}{Timsit, K.}, \bibinfo{year}{2018}.
\newblock \bibinfo{title}{Blockchain and the {GDPR}: a thematic report prepared
  by the {European Union Blockchain Observatory and Forum}}.
\newblock \bibinfo{type}{Thematic Report}. European Union Blockchain
  Observatory and Forum.
\newblock \URLprefix
  \url{https://www.eublockchainforum.eu/sites/default/files/reports/20181016_report_gdpr.pdf}.
\bibitem[{Ma et~al.(2018)Ma, Guo, Wang, Xiao, Xu, Dai, Cheng, Yi and
  Wang}]{ma2018nudging}
\bibinfo{author}{Ma, S.}, \bibinfo{author}{Guo, C.}, \bibinfo{author}{Wang,
  H.}, \bibinfo{author}{Xiao, H.}, \bibinfo{author}{Xu, B.},
  \bibinfo{author}{Dai, H.N.}, \bibinfo{author}{Cheng, S.},
  \bibinfo{author}{Yi, R.}, \bibinfo{author}{Wang, T.}, \bibinfo{year}{2018}.
\newblock \bibinfo{title}{Nudging data privacy management of open banking based
  on blockchain}, in: \bibinfo{booktitle}{Proceedings of the 2018 15th
  International Symposium on Pervasive Systems, Algorithms and Networks},
  \bibinfo{publisher}{IEEE}. pp. \bibinfo{pages}{72--79}.
\newblock \DOIprefix\doi{10.1109/I-SPAN.2018.00021}.
\bibitem[{Makhdoom et~al.(2020)Makhdoom, Zhou, Abolhasan, Lipman and
  Ni}]{makhdoom2020privysharing}
\bibinfo{author}{Makhdoom, I.}, \bibinfo{author}{Zhou, I.},
  \bibinfo{author}{Abolhasan, M.}, \bibinfo{author}{Lipman, J.},
  \bibinfo{author}{Ni, W.}, \bibinfo{year}{2020}.
\newblock \bibinfo{title}{{PrivySharing}: A blockchain-based framework for
  privacy-preserving and secure data sharing in smart cities}.
\newblock \bibinfo{journal}{Computers \& Security} \bibinfo{volume}{88}.
\newblock \DOIprefix\doi{10.1016/j.cose.2019.101653}.
\bibitem[{Mannan et~al.(2019)Mannan, Sethuram and Younge}]{mannan2019gdpr}
\bibinfo{author}{Mannan, R.}, \bibinfo{author}{Sethuram, R.},
  \bibinfo{author}{Younge, L.}, \bibinfo{year}{2019}.
\newblock \bibinfo{title}{Practitioner's corner $\bullet$ {GDPR} and
  blockchain: A compliance approach}.
\newblock \bibinfo{journal}{European Data Protection Law Review}
  \bibinfo{volume}{5}, \bibinfo{pages}{421--426}.
\newblock \DOIprefix\doi{10.21552/edpl/2019/3/18}.
\bibitem[{Marikyan et~al.(2021)Marikyan, Llanos, Barati, Aujla, Li, Adu-Duodu,
  Tahir, Rana, Papagiannidis, Ranjan et~al.}]{marikyan2021privacy}
\bibinfo{author}{Marikyan, D.}, \bibinfo{author}{Llanos, J.},
  \bibinfo{author}{Barati, M.}, \bibinfo{author}{Aujla, G.},
  \bibinfo{author}{Li, Y.}, \bibinfo{author}{Adu-Duodu, K.},
  \bibinfo{author}{Tahir, S.}, \bibinfo{author}{Rana, O.},
  \bibinfo{author}{Papagiannidis, S.}, \bibinfo{author}{Ranjan, R.}, et~al.,
  \bibinfo{year}{2021}.
\newblock \bibinfo{title}{Privacy \& cloud services: Are we there yet?}, in:
  \bibinfo{booktitle}{Proceedings of the 2021 IEEE International Conference on
  Service-Oriented System Engineering}, \bibinfo{publisher}{IEEE}. pp.
  \bibinfo{pages}{11--19}.
\newblock \DOIprefix\doi{10.1109/SOSE52839.2021.00006}.
\bibitem[{Martin-Bariteau(2018)}]{Martin-Bariteau2018}
\bibinfo{author}{Martin-Bariteau, F.}, \bibinfo{year}{2018}.
\newblock \bibinfo{title}{Blockchain and the {European Union} {General Data
  Protection Regulation}: The {CNIL}'s Perspective}.
\newblock \bibinfo{type}{Working Paper}. Blckchn.ca.
\newblock \DOIprefix\doi{10.2139/ssrn.3275783}.
\bibitem[{Meiklejohn et~al.(2013)Meiklejohn, Pomarole, Jordan, Levchenko,
  McCoy, Voelker and Savage}]{meiklejohn2013fistful}
\bibinfo{author}{Meiklejohn, S.}, \bibinfo{author}{Pomarole, M.},
  \bibinfo{author}{Jordan, G.}, \bibinfo{author}{Levchenko, K.},
  \bibinfo{author}{McCoy, D.}, \bibinfo{author}{Voelker, G.M.},
  \bibinfo{author}{Savage, S.}, \bibinfo{year}{2013}.
\newblock \bibinfo{title}{A fistful of {Bitcoins}: Characterizing payments
  among men with no names}, in: \bibinfo{booktitle}{Proceedings of the 2013
  Internet Measurement Conference}, \bibinfo{publisher}{ACM}. pp.
  \bibinfo{pages}{127--140}.
\newblock \DOIprefix\doi{10.1145/2504730.2504747}.
\bibitem[{Millard(2018)}]{millard2018blockchain}
\bibinfo{author}{Millard, C.}, \bibinfo{year}{2018}.
\newblock \bibinfo{title}{Blockchain and law: Incompatible codes?}
\newblock \bibinfo{journal}{Computer Law \& Security Review}
  \bibinfo{volume}{34}, \bibinfo{pages}{843--846}.
\newblock \DOIprefix\doi{10.1016/j.clsr.2018.06.006}.
\bibitem[{Moerel(2018)}]{moerel2018blockchain}
\bibinfo{author}{Moerel, L.}, \bibinfo{year}{2018}.
\newblock \bibinfo{title}{Blockchain \& data protection ... and why they are
  not on a collision course}.
\newblock \bibinfo{journal}{European Review of Private Law}
  \bibinfo{volume}{26}, \bibinfo{pages}{825--851}.
\newblock \DOIprefix\doi{10.54648/erpl2018057}.
\bibitem[{Molina et~al.(2020)Molina, Betarte and Luna}]{molina2020}
\bibinfo{author}{Molina, F.}, \bibinfo{author}{Betarte, G.},
  \bibinfo{author}{Luna, C.}, \bibinfo{year}{2020}.
\newblock \bibinfo{title}{A blockchain based and {GDPR}-compliant design of a
  system for digital education certificates}.
\newblock \bibinfo{howpublished}{arXiv:2010.12980 [cs.SE]}.
\newblock \DOIprefix\doi{10.48550/arxiv.2010.12980}.
\bibitem[{Molina et~al.(2021)Molina, Betarte and Luna}]{molina2021design}
\bibinfo{author}{Molina, F.}, \bibinfo{author}{Betarte, G.},
  \bibinfo{author}{Luna, C.}, \bibinfo{year}{2021}.
\newblock \bibinfo{title}{Design principles for constructing {GDPR}-compliant
  blockchain solutions}, in: \bibinfo{booktitle}{Proceedings of the 2021 4th
  IEEE/ACM International Workshop on Emerging Trends in Software Engineering
  for Blockchain}, \bibinfo{publisher}{IEEE}.
\newblock \DOIprefix\doi{10.1109/WETSEB52558.2021.00008}.
\bibitem[{Moslavac(2020)}]{moslavac2020consent}
\bibinfo{author}{Moslavac, B.}, \bibinfo{year}{2020}.
\newblock \bibinfo{title}{Consent by {GDPR} vs.\ blockchain}.
\newblock \bibinfo{journal}{Revista Acad{\^e}mica Escola Superior do
  Minist{\'e}rio P{\'u}blico do Cear{\'a}} \bibinfo{volume}{12},
  \bibinfo{pages}{149--166}.
\newblock \URLprefix
  \url{https://www.bib.irb.hr/1076576/download/1076576.ARTIGO-149-166.pdf}.
\bibitem[{Naik and Jenkins(2020)}]{naik2020your}
\bibinfo{author}{Naik, N.}, \bibinfo{author}{Jenkins, P.},
  \bibinfo{year}{2020}.
\newblock \bibinfo{title}{Your identity is yours: Take back control of your
  identity using {GDPR} compatible self-sovereign identity}, in:
  \bibinfo{booktitle}{Proceedings of the 2020 7th International Conference on
  Behavioural and Social Computing}, \bibinfo{publisher}{IEEE}.
\newblock \DOIprefix\doi{10.1109/BESC51023.2020.9348298}.
\bibitem[{Neisse et~al.(2018)Neisse, Steri and Fovino}]{Neisse2018}
\bibinfo{author}{Neisse, R.}, \bibinfo{author}{Steri, G.},
  \bibinfo{author}{Fovino, I.N.}, \bibinfo{year}{2018}.
\newblock \bibinfo{title}{Blockchain-based identity management and data usage
  control (extended abstract)}, in: \bibinfo{booktitle}{Privacy and Identity
  Management. The Smart Revolution: 12th IFIP WG 9.2, 9.5, 9.6/11.7, 11.6/SIG
  9.2.2 International Summer School, Ispra, Italy, September 4-8, 2017, Revised
  Selected Papers}, \bibinfo{publisher}{Springer}. pp.
  \bibinfo{pages}{237--239}.
\newblock \DOIprefix\doi{10.1007/978-3-319-92925-5_15}.
\bibitem[{Neisse et~al.(2017)Neisse, Steri and
  Nai-Fovino}]{neisse2017blockchain}
\bibinfo{author}{Neisse, R.}, \bibinfo{author}{Steri, G.},
  \bibinfo{author}{Nai-Fovino, I.}, \bibinfo{year}{2017}.
\newblock \bibinfo{title}{A blockchain-based approach for data accountability
  and provenance tracking}, in: \bibinfo{booktitle}{Proceedings of the 12th
  International Conference on Availability, Reliability and Security},
  \bibinfo{publisher}{ACM}.
\newblock \DOIprefix\doi{10.1145/3098954.3098958}.
\bibitem[{Pagallo et~al.(2018)Pagallo, Bassi, Crepaldi and
  Durante}]{pagallo2018chronicle}
\bibinfo{author}{Pagallo, U.}, \bibinfo{author}{Bassi, E.},
  \bibinfo{author}{Crepaldi, M.}, \bibinfo{author}{Durante, M.},
  \bibinfo{year}{2018}.
\newblock \bibinfo{title}{Chronicle of a clash foretold: Blockchains and the
  {GDPR}'s right to erasure}, in: \bibinfo{booktitle}{Legal Knowledge and
  Information Systems}, \bibinfo{publisher}{IOS Press}. pp.
  \bibinfo{pages}{81--90}.
\newblock \DOIprefix\doi{10.3233/978-1-61499-935-5-81}.
\bibitem[{Pedersen et~al.(2019)Pedersen, Risius and
  Beck}]{Pedersen2019BlockchainDecisionPath}
\bibinfo{author}{Pedersen, A.B.}, \bibinfo{author}{Risius, M.},
  \bibinfo{author}{Beck, R.}, \bibinfo{year}{2019}.
\newblock \bibinfo{title}{A ten-step decision path to determine when to use
  blockchain technologies}.
\newblock \bibinfo{journal}{MIS Quarterly Executive} \bibinfo{volume}{18},
  \bibinfo{pages}{99--115}.
\newblock \URLprefix \url{https://aisel.aisnet.org/misqe/vol18/iss2/3/}.
\bibitem[{Peel(2019)}]{peel2019gdpr}
\bibinfo{author}{Peel, J.}, \bibinfo{year}{2019}.
\newblock \bibinfo{title}{The {GDPR}: The biggest threat to the implementation
  of blockchain technology in global supply chains}.
\newblock \bibinfo{journal}{UMKC Law Review} \bibinfo{volume}{88},
  \bibinfo{pages}{497--517}.
\newblock \URLprefix
  \url{https://www.europarl.europa.eu/RegData/etudes/STUD/2019/634445/EPRS_STU(2019)634445_EN.pdf}.
\bibitem[{Pei et~al.(2020)Pei, Li, Wu, Sun and Cao}]{pei2020udpp}
\bibinfo{author}{Pei, X.}, \bibinfo{author}{Li, X.}, \bibinfo{author}{Wu, X.},
  \bibinfo{author}{Sun, L.}, \bibinfo{author}{Cao, Y.}, \bibinfo{year}{2020}.
\newblock \bibinfo{title}{{UDPP}: Blockchain based open platform as a privacy
  enabler}, in: \bibinfo{booktitle}{Proceedings of the 2020 10th Annual
  Computing and Communication Workshop and Conference},
  \bibinfo{publisher}{IEEE}. pp. \bibinfo{pages}{500--505}.
\newblock \DOIprefix\doi{10.1109/CCWC47524.2020.9031142}.
\bibitem[{Poelman and Iqbal(2021)}]{poelman2021investigating}
\bibinfo{author}{Poelman, M.}, \bibinfo{author}{Iqbal, S.},
  \bibinfo{year}{2021}.
\newblock \bibinfo{title}{Investigating the compliance of the {GDPR}:
  Processing personal data on a blockchain}, in:
  \bibinfo{booktitle}{Proceedings of the IEEE 5th International Conference on
  Cryptography, Security and Privacy}, \bibinfo{publisher}{IEEE}. pp.
  \bibinfo{pages}{38--44}.
\newblock \DOIprefix\doi{10.1109/CSP51677.2021.9357590}.
\bibitem[{Politou et~al.(2019)Politou, Casino, Alepis and
  Patsakis}]{Politou2019}
\bibinfo{author}{Politou, E.}, \bibinfo{author}{Casino, F.},
  \bibinfo{author}{Alepis, E.}, \bibinfo{author}{Patsakis, C.},
  \bibinfo{year}{2019}.
\newblock \bibinfo{title}{Blockchain mutability: Challenges and proposed
  solutions}.
\newblock \bibinfo{journal}{IEEE Transactions on Emerging Topics in Computing}
  \bibinfo{volume}{9}.
\newblock \DOIprefix\doi{10.1109/TETC.2019.2949510}.
\bibitem[{Puthal et~al.(2018)Puthal, Malik, Mohanty, Kougianos and
  Yang}]{puthal2018blockchain}
\bibinfo{author}{Puthal, D.}, \bibinfo{author}{Malik, N.},
  \bibinfo{author}{Mohanty, S.P.}, \bibinfo{author}{Kougianos, E.},
  \bibinfo{author}{Yang, C.}, \bibinfo{year}{2018}.
\newblock \bibinfo{title}{The blockchain as a decentralized security framework
  [future directions]}.
\newblock \bibinfo{journal}{IEEE Consumer Electronics Magazine}
  \bibinfo{volume}{7}, \bibinfo{pages}{18--21}.
\newblock \DOIprefix\doi{10.1109/MCE.2017.2776459}.
\bibitem[{Quiniou(2019)}]{quiniou2019blockchain}
\bibinfo{author}{Quiniou, M.}, \bibinfo{year}{2019}.
\newblock \bibinfo{title}{Blockchain: The Advent of Disintermediation}.
\newblock \bibinfo{publisher}{ISTE Ltd}.
\newblock \DOIprefix\doi{10.1002/9781119629573}.
\bibitem[{Ramos and Silva(2019)}]{ramos2019privacy}
\bibinfo{author}{Ramos, L.F.M.}, \bibinfo{author}{Silva, J.M.C.},
  \bibinfo{year}{2019}.
\newblock \bibinfo{title}{Privacy and data protection concerns regarding the
  use of blockchains in smart cities}, in: \bibinfo{booktitle}{Proceedings of
  the 12th International Conference on Theory and Practice of Electronic
  Governance}, \bibinfo{publisher}{ACM}. pp. \bibinfo{pages}{342--347}.
\newblock \DOIprefix\doi{10.1145/3326365.3326410}.
\bibitem[{Rampone(2018)}]{rampone2018data}
\bibinfo{author}{Rampone, F.}, \bibinfo{year}{2018}.
\newblock \bibinfo{title}{Data protection in the blockchain environment: {GDPR}
  is not a hurdle to permissionless {DLT} solutions}.
\newblock \bibinfo{journal}{Ciberspazio e diritto} \bibinfo{volume}{19},
  \bibinfo{pages}{457--478}.
\newblock \URLprefix
  \url{https://www.mucchieditore.it/index.php?option=com_virtuemart&view=productdetails&virtuemart_product_id=2794}.
\bibitem[{Rantos et~al.(2018)Rantos, Drosatos, Demertzis, Ilioudis and
  Papanikolaou}]{Rantos2018}
\bibinfo{author}{Rantos, K.}, \bibinfo{author}{Drosatos, G.},
  \bibinfo{author}{Demertzis, K.}, \bibinfo{author}{Ilioudis, C.},
  \bibinfo{author}{Papanikolaou, A.}, \bibinfo{year}{2018}.
\newblock \bibinfo{title}{Blockchain-based consents management for personal
  data processing in the {IoT} ecosystem}, in: \bibinfo{booktitle}{Proceedings
  of the 15th International Joint Conference on e-Business and
  Telecommunications}, \bibinfo{publisher}{SciTePress}. pp.
  \bibinfo{pages}{572--577}.
\newblock \DOIprefix\doi{10.5220/0006911005720577}.
\bibitem[{Riva(2020)}]{riva2020happens}
\bibinfo{author}{Riva, G.M.}, \bibinfo{year}{2020}.
\newblock \bibinfo{title}{What happens in blockchain stays in blockchain. a
  legal solution to conflicts between digital ledgers and privacy rights}.
\newblock \bibinfo{journal}{Frontiers in Blockchain}
  \DOIprefix\doi{10.3389/fbloc.2020.00036}.
\bibitem[{Rotondi et~al.(2019)Rotondi, Saltarella, Giordano and
  Pellecchia}]{rotondi2019distributed}
\bibinfo{author}{Rotondi, D.}, \bibinfo{author}{Saltarella, M.},
  \bibinfo{author}{Giordano, G.}, \bibinfo{author}{Pellecchia, F.},
  \bibinfo{year}{2019}.
\newblock \bibinfo{title}{Distributed ledger technology and {European Union}
  {General Data Protection Regulation} compliance in a flexible working
  context}.
\newblock \bibinfo{journal}{Internet Technology Letters} \bibinfo{volume}{2}.
\newblock \DOIprefix\doi{10.1002/itl2.127}.
\bibitem[{Sarier(2021a)}]{sarier2021comments}
\bibinfo{author}{Sarier, N.D.}, \bibinfo{year}{2021}a.
\newblock \bibinfo{title}{Comments on biometric-based non-transferable
  credentials and their application in blockchain-based identity management}.
\newblock \bibinfo{journal}{Computers \& Security} \bibinfo{volume}{105}.
\newblock \DOIprefix\doi{10.1016/j.cose.2021.102243}.
\bibitem[{Sarier(2021b)}]{sarier2021efficient}
\bibinfo{author}{Sarier, N.D.}, \bibinfo{year}{2021}b.
\newblock \bibinfo{title}{Efficient biometric-based identity management on the
  {Blockchain} for smart industrial applications}.
\newblock \bibinfo{journal}{Pervasive and Mobile Computing}
  \bibinfo{volume}{71}.
\newblock \DOIprefix\doi{10.1016/j.pmcj.2020.101322}.
\bibitem[{Schellekens(2020)}]{schellekens2020conceptualizations}
\bibinfo{author}{Schellekens, M.}, \bibinfo{year}{2020}.
\newblock \bibinfo{title}{Conceptualizations of the controller in
  permissionless blockchains}.
\newblock \bibinfo{journal}{Journal of Intellectual Property, Information
  Technology and E-Commerce Law} \bibinfo{volume}{11},
  \bibinfo{pages}{215--227}.
\newblock \DOIprefix\doi{https://www.jipitec.eu/issues/jipitec-11-2-2020/5099}.
\bibitem[{Schellinger et~al.(2022)Schellinger, Urbach, V{\"o}lter and
  Sedlmeir}]{schellinger2022yes}
\bibinfo{author}{Schellinger, B.}, \bibinfo{author}{Urbach, N.},
  \bibinfo{author}{V{\"o}lter, F.}, \bibinfo{author}{Sedlmeir, J.},
  \bibinfo{year}{2022}.
\newblock \bibinfo{title}{Yes, {I} do: Marrying blockchain applications with
  {GDPR}}, in: \bibinfo{booktitle}{Proceedings of the 55th Hawaii International
  Conference on System Sciences}, \bibinfo{publisher}{University of Hawai`i at
  Mānoa, USA}. pp. \bibinfo{pages}{4631--4640}.
\newblock \URLprefix \url{http://hdl.handle.net/10125/79900}.
\bibitem[{Schmelz et~al.(2018)Schmelz, Fischer, Niemeier, Zhu and
  Grechenig}]{schmelz2018towards}
\bibinfo{author}{Schmelz, D.}, \bibinfo{author}{Fischer, G.},
  \bibinfo{author}{Niemeier, P.}, \bibinfo{author}{Zhu, L.},
  \bibinfo{author}{Grechenig, T.}, \bibinfo{year}{2018}.
\newblock \bibinfo{title}{Towards using public blockchain in
  information-centric networks: Challenges imposed by the {European Union}'s
  {General Data Protection Regulation}}, in: \bibinfo{booktitle}{IEEE
  International Conference on Hot Information-Centric Networking},
  \bibinfo{publisher}{IEEE}. pp. \bibinfo{pages}{223--228}.
\newblock \DOIprefix\doi{10.1109/HOTICN.2018.8606000}.
\bibitem[{Schmelz et~al.(2020)Schmelz, Pinter, Brottrager, Niemeier, Lamber and
  Grechenig}]{schmelz2020securing}
\bibinfo{author}{Schmelz, D.}, \bibinfo{author}{Pinter, K.},
  \bibinfo{author}{Brottrager, J.}, \bibinfo{author}{Niemeier, P.},
  \bibinfo{author}{Lamber, R.}, \bibinfo{author}{Grechenig, T.},
  \bibinfo{year}{2020}.
\newblock \bibinfo{title}{Securing the rights of data subjects with blockchain
  technology}, in: \bibinfo{booktitle}{Proceedings of the 2020 3rd
  International Conference on Information and Computer Technologies},
  \bibinfo{publisher}{IEEE}. pp. \bibinfo{pages}{284--288}.
\newblock \DOIprefix\doi{10.1109/ICICT50521.2020.00050}.
\bibitem[{Schwerin(2018)}]{Schwerin2018}
\bibinfo{author}{Schwerin, S.}, \bibinfo{year}{2018}.
\newblock \bibinfo{title}{Blockchain and privacy protection in the case of the
  {European General Data Protection Regulation} ({GDPR}): A {Delphi} study}.
\newblock \bibinfo{journal}{The Journal of The British Blockchain Association}
  \bibinfo{volume}{1}.
\newblock \DOIprefix\doi{10.31585/jbba-1-1-(4)2018}.
\bibitem[{Shahaab et~al.(2020)Shahaab, Maude, Hewage and
  Khan}]{shahaab2020managing}
\bibinfo{author}{Shahaab, A.}, \bibinfo{author}{Maude, R.},
  \bibinfo{author}{Hewage, C.}, \bibinfo{author}{Khan, I.},
  \bibinfo{year}{2020}.
\newblock \bibinfo{title}{Managing gender change information on immutable
  blockchain in context of {GDPR}}.
\newblock \bibinfo{journal}{The Journal of The British Blockchain Association}
  \bibinfo{volume}{3}, \bibinfo{pages}{23--28}.
\newblock \DOIprefix\doi{10.31585/jbba-3-1-(3)2020}.
\bibitem[{Sharma et~al.(2020)Sharma, Halder and Singh}]{sharma2020blockchain}
\bibinfo{author}{Sharma, B.}, \bibinfo{author}{Halder, R.},
  \bibinfo{author}{Singh, J.}, \bibinfo{year}{2020}.
\newblock \bibinfo{title}{Blockchain-based interoperable healthcare using
  zero-knowledge proofs and proxy re-encryption}, in:
  \bibinfo{booktitle}{Proceedings of the 2020 12th International Conference on
  Communication Systems \& Networks}, \bibinfo{publisher}{IEEE}.
\newblock \DOIprefix\doi{10.1109/COMSNETS48256.2020.9027413}.
\bibitem[{Silva and Garcia(2021)}]{silva2021our}
\bibinfo{author}{Silva, W.}, \bibinfo{author}{Garcia, A.C.B.},
  \bibinfo{year}{2021}.
\newblock \bibinfo{title}{Where is our data? a blockchain-based information
  chain of custody model for privacy improvement}, in:
  \bibinfo{booktitle}{Proceedings of the 2021 IEEE 24th International
  Conference on Computer Supported Cooperative Work in Design},
  \bibinfo{publisher}{IEEE}. pp. \bibinfo{pages}{329--334}.
\newblock \DOIprefix\doi{10.1109/CSCWD49262.2021.9437727}.
\bibitem[{Sim et~al.(2019)Sim, Chua and Tahir}]{sim2019blockchain}
\bibinfo{author}{Sim, W.L.}, \bibinfo{author}{Chua, H.N.},
  \bibinfo{author}{Tahir, M.}, \bibinfo{year}{2019}.
\newblock \bibinfo{title}{Blockchain for identity management: The implications
  to personal data protection}, in: \bibinfo{booktitle}{Proceedings of the 2019
  IEEE Conference on Application, Information and Network Security},
  \bibinfo{publisher}{IEEE}. pp. \bibinfo{pages}{30--35}.
\newblock \DOIprefix\doi{10.1109/AINS47559.2019.8968708}.
\bibitem[{Stan and Miclea(2019)}]{stan2019new}
\bibinfo{author}{Stan, O.P.}, \bibinfo{author}{Miclea, L.},
  \bibinfo{year}{2019}.
\newblock \bibinfo{title}{New era for technology in healthcare powered by
  {GDPR} and blockchain}, in: \bibinfo{booktitle}{6th International Conference
  on Advancements of Medicine and Health Care through Technology; 17--20
  October 2018, Cluj-Napoca, Romania}, \bibinfo{publisher}{Springer}. pp.
  \bibinfo{pages}{311--317}.
\newblock \DOIprefix\doi{10.1007/978-981-13-6207-1_49}.
\bibitem[{Subramanian et~al.(2020)Subramanian, Cousins, Bouyad, Sheth and
  Conway}]{subramanian2020blockchain}
\bibinfo{author}{Subramanian, H.}, \bibinfo{author}{Cousins, K.},
  \bibinfo{author}{Bouyad, L.B.}, \bibinfo{author}{Sheth, A.},
  \bibinfo{author}{Conway, D.}, \bibinfo{year}{2020}.
\newblock \bibinfo{title}{Blockchain regulations and decentralized
  applications: Panel report from {AMCIS} 2018}.
\newblock \bibinfo{journal}{Communications of the Association for Information
  Systems} \bibinfo{volume}{47}.
\newblock \DOIprefix\doi{10.17705/1CAIS.04709}.
\bibitem[{Suripeddi and Purandare(2021)}]{suripeddi2021blockchain}
\bibinfo{author}{Suripeddi, M.K.S.}, \bibinfo{author}{Purandare, P.},
  \bibinfo{year}{2021}.
\newblock \bibinfo{title}{Blockchain and {GDPR} -- a study on compatibility
  issues of the distributed ledger technology with {GDPR} data processing}.
\newblock \bibinfo{journal}{Journal of Physics: Conference Series}
  \bibinfo{volume}{1964}.
\newblock \DOIprefix\doi{10.1088/1742-6596/1964/4/042005}.
\bibitem[{Tatar et~al.(2020)Tatar, Gokce and Nussbaum}]{tatar2020law}
\bibinfo{author}{Tatar, U.}, \bibinfo{author}{Gokce, Y.},
  \bibinfo{author}{Nussbaum, B.}, \bibinfo{year}{2020}.
\newblock \bibinfo{title}{Law versus technology: Blockchain, {GDPR}, and tough
  tradeoffs}.
\newblock \bibinfo{journal}{Computer Law \& Security Review}
  \bibinfo{volume}{38}, \bibinfo{pages}{105454}.
\newblock \DOIprefix\doi{10.1016/j.clsr.2020.105454}.
\bibitem[{Teperdjian(2020)}]{teperdjian2020puzzle}
\bibinfo{author}{Teperdjian, R.}, \bibinfo{year}{2020}.
\newblock \bibinfo{title}{The puzzle of squaring blockchain with the {General
  Data Protection Regulation}}.
\newblock \bibinfo{journal}{Jurimetrics Journal} \bibinfo{volume}{60},
  \bibinfo{pages}{253--313}.
\newblock \URLprefix
  \url{https://www.americanbar.org/digital-asset-abstract.html/content/dam/aba/publications/Jurimetrics/spring2020/teperdjian.pdf}.
\bibitem[{{Van~Eecke} and Haie(2018)}]{van2018blockchain}
\bibinfo{author}{{Van~Eecke}, P.}, \bibinfo{author}{Haie, A.G.},
  \bibinfo{year}{2018}.
\newblock \bibinfo{title}{Practitioner's corner $\bullet$ blockchain and the
  {GDPR}: The {EU Blockchain Observatory} report}.
\newblock \bibinfo{journal}{European Data Protection Law Review}
  \bibinfo{volume}{4}, \bibinfo{pages}{531--534}.
\newblock \DOIprefix\doi{10.21552/edpl/2018/4/18}.
\bibitem[{Walters(2019)}]{walters2019privacy}
\bibinfo{author}{Walters, N.}, \bibinfo{year}{2019}.
\newblock \bibinfo{title}{Privacy law issues in blockchains: An analysis of
  {PIPEDA}, the {GDPR}, and proposals for compliance}.
\newblock \bibinfo{journal}{Canadian Journal of Law and Technology}
  \bibinfo{volume}{17}, \bibinfo{pages}{276--305}.
\newblock
  \DOIprefix\doi{https://digitalcommons.schulichlaw.dal.ca/cjlt/vol17/iss2/5/}.
\bibitem[{Wilford et~al.(2021)Wilford, McBride, Brooks, Eke, Akintoye, Owoseni,
  Leach, Flick, Malcolm and Stacey}]{wilford2021digital}
\bibinfo{author}{Wilford, S.H.}, \bibinfo{author}{McBride, N.},
  \bibinfo{author}{Brooks, L.}, \bibinfo{author}{Eke, D.O.},
  \bibinfo{author}{Akintoye, S.}, \bibinfo{author}{Owoseni, A.},
  \bibinfo{author}{Leach, T.}, \bibinfo{author}{Flick, C.},
  \bibinfo{author}{Malcolm, F.}, \bibinfo{author}{Stacey, M.},
  \bibinfo{year}{2021}.
\newblock \bibinfo{title}{The digital network of networks: Regulatory risk and
  policy challenges of vaccine passports}.
\newblock \bibinfo{journal}{European Journal of Risk Regulation}
  \bibinfo{volume}{12}, \bibinfo{pages}{393--403}.
\newblock \DOIprefix\doi{10.1017/err.2021.35}.
\bibitem[{Wirth and Kolain(2018)}]{wirth2018privacy}
\bibinfo{author}{Wirth, C.}, \bibinfo{author}{Kolain, M.},
  \bibinfo{year}{2018}.
\newblock \bibinfo{title}{Privacy by blockchain design: a blockchain-enabled
  {GDPR}-compliant approach for handling personal data}, in:
  \bibinfo{booktitle}{Proceedings of 1st ERCIM Blockchain Workshop 2018},
  \bibinfo{publisher}{EUSSET}.
\newblock \DOIprefix\doi{10.18420/blockchain2018_03}.
\bibitem[{Wrigley(2019)}]{wrigley2019people}
\bibinfo{author}{Wrigley, S.}, \bibinfo{year}{2019}.
\newblock \bibinfo{title}{``when people just click'': Addressing the
  difficulties of controller/processor agreements online}, in:
  \bibinfo{booktitle}{Legal Tech, Smart Contracts and Blockchain}.
  \bibinfo{publisher}{Springer}, pp. \bibinfo{pages}{221--252}.
\newblock \DOIprefix\doi{10.1007/978-981-13-6086-2_9}.
\bibitem[{Wulff(2020)}]{wulff2020right}
\bibinfo{author}{Wulff, C.M.}, \bibinfo{year}{2020}.
\newblock \bibinfo{title}{The right to be forgotten in post-{Google} {Spain}
  case law: an example of legal interpretivism in action?}
\newblock \bibinfo{journal}{Comparative Law Review} \bibinfo{volume}{26},
  \bibinfo{pages}{255--279}.
\newblock \DOIprefix\doi{10.12775/CLR.2020.010}.
\bibitem[{Zemler(2019)}]{zemler2019concepts}
\bibinfo{author}{Zemler, F.}, \bibinfo{year}{2019}.
\newblock \bibinfo{title}{Concepts for {GDPR}-compliant processing of personal
  data on blockchain: A literature review}.
\newblock \bibinfo{journal}{Anwendungen und Konzepte der Wirtschaftsinformatik}
  \bibinfo{volume}{10}, \bibinfo{pages}{96--107}.
\newblock \URLprefix
  \url{https://ojs-hslu.ch/ojs3211/index.php/akwi/issue/view/10}.
\bibitem[{Zemler and Westner(2019)}]{zemler2019blockchain}
\bibinfo{author}{Zemler, F.}, \bibinfo{author}{Westner, M.},
  \bibinfo{year}{2019}.
\newblock \bibinfo{title}{Blockchain and {GDPR}: Application scenarios and
  compliance requirements}, in: \bibinfo{booktitle}{Proceedings of the 2019
  Portland International Conference on Management of Engineering and
  Technology}, \bibinfo{publisher}{IEEE}.
\newblock \DOIprefix\doi{10.23919/PICMET.2019.8893923}.
\bibitem[{Zheng et~al.(2018)Zheng, Mukkamala, Vatrapu and
  Ordieres-Mere}]{zheng2018blockchain}
\bibinfo{author}{Zheng, X.}, \bibinfo{author}{Mukkamala, R.R.},
  \bibinfo{author}{Vatrapu, R.}, \bibinfo{author}{Ordieres-Mere, J.},
  \bibinfo{year}{2018}.
\newblock \bibinfo{title}{Blockchain-based personal health data sharing system
  using cloud storage}, in: \bibinfo{booktitle}{Proceedings of the 2018 IEEE
  20th International Conference on e-Health Networking, Applications and
  Services}, \bibinfo{publisher}{IEEE}.
\newblock \DOIprefix\doi{10.1109/HealthCom.2018.8531125}.
\bibitem[{Zichichi et~al.(2020)Zichichi, Ferretti and
  D'Angelo}]{zichichi2020efficiency}
\bibinfo{author}{Zichichi, M.}, \bibinfo{author}{Ferretti, S.},
  \bibinfo{author}{D'Angelo, G.}, \bibinfo{year}{2020}.
\newblock \bibinfo{title}{On the efficiency of decentralized file storage for
  personal information management systems}, in: \bibinfo{booktitle}{Proceedings
  of the 2020 IEEE Symposium on Computers and Communications},
  \bibinfo{publisher}{IEEE}.
\bibitem[{Zieglmeier and Daiqui(2021)}]{zieglmeier2021gdpr}
\bibinfo{author}{Zieglmeier, V.}, \bibinfo{author}{Daiqui, G.L.},
  \bibinfo{year}{2021}.
\newblock \bibinfo{title}{{GDPR}-compliant use of blockchain for secure usage
  logs}, in: \bibinfo{booktitle}{Proceedings of EASE 2021: Evaluation and
  Assessment in Software Engineering}, \bibinfo{publisher}{ACM}. pp.
  \bibinfo{pages}{313--320}.
\newblock \DOIprefix\doi{10.1145/3463274.3463349}.

\end{thebibliography}

\end{document}